\documentclass[reprint, twoside]{revtex4-2}
\usepackage{subfig}
\usepackage{comment}
\usepackage{hyperref}
\usepackage{tikz}
\usepackage{braket}
\usepackage{amsmath}
\usepackage{bbold}
\usepackage[english]{babel}
\usepackage[utf8]{inputenc}
\usepackage[T1]{fontenc}

\begin{document}

\title{Optimization of Quantum-Repeater Networks using Stochastic Automatic Differentiation}
\author{Guus Avis}
\email{guusavis@hotmail.com}
\affiliation{Manning College of Information and Computer Sciences, University of Massachusetts Amherst, Amherst, Massachusetts 01003, USA}
\author{Stefan Krastanov}
\affiliation{Manning College of Information and Computer Sciences, University of Massachusetts Amherst, Amherst, Massachusetts 01003, USA}
\date{\today}

\begin{abstract}
Quantum repeaters are envisioned to enable long-distance entanglement distribution.
Analysis of quantum-repeater networks could hasten their realization by informing design decisions and research priorities.
Determining derivatives of network properties is crucial towards that end, facilitating optimizations and revealing parameter sensitivity.
Doing so, however, is difficult because the networks are discretely random.
Here we use a recently developed technique, stochastic automatic differentiation, to automatically extract derivatives from discrete Monte Carlo simulations of repeater networks.
With these derivatives, we optimize rate-fidelity tradeoffs in a repeater chain, determine the chain's sensitivity with respect to the coherence times of different nodes, and finally choose the locations of quantum repeaters in a two-dimensional plane to optimize the guaranteed quality of service between four end nodes.
In particular, the technique enabled us to discover how the best achievable quality of service, the minimal number of repeaters required to improve a network, and the number of repeaters required to saturate the network scale with the physical size of the network.
\end{abstract}

\maketitle

\section{Introduction}

Future quantum networks promise to distribute quantum entanglement between their end nodes.
This entanglement can be used for a myriad of applications such as secure communication through quantum-key distribution~\cite{bennett1992a, ekert1991, bennett2014}, secure delegated quantum computing~\cite{broadbent2009, fitzsimons2017, leichtle2021}, distributed sensing~\cite{eldredge2018, proctor2018}, clock synchronization~\cite{komar2014} and distributed quantum computing~\cite{grover1997, nickerson2014, cirac1999}.
Large-scale quantum networks, and eventually a quantum internet~\cite{wehner2018}, are envisioned to use auxiliary nodes called quantum repeaters to create entanglement between distant end nodes~\cite{azuma2015, briegel1998, duan2001, dur1999, jiang2009}.
Recent breakthroughs include the operation of the first three-node quantum network~\cite{pompili2021}, heralded entanglement generation over deployed fiber at metropolitan distances~\cite{knaut2024, stolk2024}, and proof-of-principle quantum-repeater experiments~\cite{bhaskar2020, langenfeld2021}.
However, realizing practical quantum networks is still a major challenge, requiring developments in physics, engineering, and computer science.
\\

Theoretical analysis of quantum networks is crucial to their speedy development.
Important questions include how to design network protocols that alleviate hardware requirements~\cite{goodenough2021, vardoyan2023, krastanov2019, rozpedek2018, jiang2007a, haldar2025, haldar2024a, li2021a, pouryousef2023, pathumsoot2021}, what hardware improvements are the most crucial~\cite{avis2022a, silva2024, silva2021, vandam2024, mora2024}, and how best to allocate resources in a network~\cite{rabbie2022, islam2023, pouryousef2024a, silva2024, sripotchanart2024, avis2024, chehimi2023, dhara2021a, jiang2007a, luong2016, pathumsoot2021, rozpedek2018a, rozpedek2021, wo2023, yao2021, dai2021a, yao2021}.
When addressing such questions it is useful to calculate derivatives of network properties.
For instance, knowing the derivative of a performance benchmark with respect to the parameters of a network protocol can enable the optimal configuration of that protocol using a gradient-descent algorithm.
Moreover, how ``crucial'' a hardware improvement is can be quantified using the derivative of performance benchmarks with respect to hardware parameters.
When analytical expressions are available, standard differentiation techniques such as symbolic differentiation can be used.
However, due to the probabilistic nature of quantum processes, quantum networks display complicated discrete random behavior that is difficult to capture in analytical models except in specific or idealized scenarios~\cite{rozpedek2018a, kamin2023, goodenough2025, coopmans2022, sangouard2007, guha2015, shchukin2019, deandrade2024b}.
For this reason, many studies of quantum networks have turned to numerical Monte Carlo simulations.
Particularly popular are discrete-event simulators~\cite{coopmans2021, wu2021b, wallnofer2024, chen2023, satoh2022}.
\\

A popular method for obtaining non-symbolic derivatives is numerical differentiation, particularly finite differences~\cite{leveque2007}.
This method, which we discuss in more detail in Section~\ref{sec:finitediff}, provides an unbiased estimator of the derivative only in a limit when the variance diverges.
Roughly speaking, one obtains either inaccurate results with small error bars (precise), or accurate results with large error bars (imprecise).
A slew of more modern methods has been developed~\cite{glynn1990, grathwohl2018, heidergott2008, jang2017, kingma2022a, kleijnen1996, mohamed2020, pflug1989, tucker2017, williams1992, glasserman1991}.
However, when the Monte Carlo simulation involves discrete randomness, they all either introduce biases or have large variances.
This is problematic as quantum networks are discretely random.
Recently, however, a new method has been developed that we will refer to here as \textit{stochastic automatic differentiation} (stochastic AD)~\cite{arya2023}.
If a computer program samples the (discrete) random variable $X(p)$, stochastic AD can be used to construct another computer program that samples $X'(p)$ such that
\begin{equation} \label{eq:X'}
\mathbb E[X'(p)] = \frac{d \mathbb E[X(p)]}{dp},
\end{equation}
where the variance of $X'(p)$ is bounded.
Roughly speaking, stochastic AD achieves this feat by not estimating the value of $\mathbb E[X(p + \epsilon)]$ and $\mathbb E[X(p)]$ independently, as one would do in finite differences, but instead uses shared randomness between the two evaluations.
Moreover, the programs created by stochastic AD are \textit{composable}, meaning that when a random variable $X(p)$ is a function of random variables $X_1(p)$ and $X_2(p)$, a program that samples $X'(p)$ can be constructed from programs that sample $X'_1(p)$ and $X'_2(p)$.
This enables the technique to be used as a form of \textit{automatic differentiation} (AD).
While AD is a well established technique, traditional versions of AD either do not work or are inefficient in the context of discrete randomness, which is crucial for simulations of quantum networks.
We will further explain AD and stochastic AD in Section \ref{sec:autodiff}.
\\

Being able to calculate derivatives for discrete random variables is particularly important in quantum mechanics, as any measurement in a finite-dimensional Hilbert space is discretely random.
Using AD to differentiate through discretely random quantum evolution has been done before in the context of model-aware reinforcement learning, using the REINFORCE gradient estimator~\cite{williams1992}.
Examples include reinforcement learning for optimal control of quantum systems~\cite{porotti2023} and for quantum metrology~\cite{belliardo2024, belliardo2024a}.
While REINFORCE is general purpose and unbiased, it is known for having a large variance, which often necessitates using tricks for variance reduction.
In contrast, stochastic AD, which we use in this paper, does not require such tricks and provides a much smaller variance.
A comparison between REINFORCE and stochastic AD for a discrete random walk is included in \cite{arya2023}, demonstrating that stochastic AD provides a significantly smaller variance even when variance-reduction tricks are used for REINFORCE.
To the best of our knowledge, AD has never been used in any form to differentiate through the dynamics of quantum-repeater networks.
In this paper, we apply stochastic AD to obtain derivatives from Monte Carlo simulations of quantum-repeater networks and demonstrate how this enables their optimization and analysis.
\\

This paper is structured as follows.
First, in Section \ref{sec:autodiff}, we further explain what (stochastic) AD is and how it works.
We remark though that understanding stochastic AD is not needed to understand the results obtained using stochastic AD.
In Section \ref{sec:model}, we explain the quantum-network model that is employed throughout the rest of the paper.
Then, in Section \ref{sec:results}, we give four different demonstrations of the usefulness of stochastic AD for quantum networks.
In Section \ref{sec:finitediff}, we explicitly compare stochastic AD to finite differences.
In Section \ref{sec:optimizing_bright_state_params}, we use gradient descent enabled by stochastic AD to optimize rate-fidelity tradeoffs on the elementary links of a quantum-repeater chain.
In Section \ref{sec:sensitivity_analysis}, we perform a sensitivity analysis with respect to the coherence times at the different nodes in the same repeater chain.
Finally, we present our most extensive result in Section \ref{sec:repeater_placement}, where we investigate the optimal placement of quantum repeaters in a 2D plane.

\section{(Stochastic) automatic differentiation}
\label{sec:autodiff}

AD is a technique that goes back to the 1950s~\cite{nolan1953} and the 1960s~\cite{wengert1964} that is nowadays particularly popular in the context of machine learning~\cite{baydin2018}.
Computer programs usually evaluate functions by composing simpler programs that evaluate simpler functions, e.g., $f(x) = g(h(x), k(x))$.
For deterministic functions, the derivative of the larger function can itself be written as a composition of the derivatives of the smaller functions using the chain rule, e.g.,
\begin{equation}
\frac{df}{dx} = \frac{\partial g}{\partial h} \frac {dh}{dx} + \frac{\partial g}{\partial k} \frac {dk}{dx}.
\end{equation}
In practice, every program is eventually composed of only a relatively small number of programs that evaluate simple functions (e.g., addition and multiplication), and hence by only writing programs that evaluate the derivatives of those simple functions and teaching a computer about composability, every program can be ``automatically'' differentiated.
The two main types of AD are forward accumulation and backward accumulation.
These differ in the order in which the different functions in the composition are evaluated; forward accumulation starts at the top function and works its way down, while backward accumulation starts at the lowest-level functions and works its way to the top.
If $f$ is a function $\mathcal R^n \to \mathcal R^m$, forward accumulation is more efficient for $n \ll m$ while backward accumulation is more efficient for $n \gg m$.
\\

Perhaps the simplest form of forward accumulation is based on \textit{dual numbers} of the type $x + \delta \epsilon$, where $x$ and $\delta$ are real numbers and $\epsilon$ can be thought of as an infinitesimal perturbation that has the special property $\epsilon^2 = 0$~\cite{baydin2018}.
When feeding such a number into a function $f(x)$, it follows from the Taylor expansion of $f$ that
\begin{equation}
f(x + \delta \epsilon) = f(x) + f'(x) \delta \epsilon.
\end{equation}
Forward accumulation can be implemented by creating a special data structure that represents a dual number and specifying how simple functions should act on dual numbers so that they adhere to the above relation correctly.
Then, the derivative $f'(x)$ is found for any function $f(x)$ that is composed of those simple functions by evaluating $f(x + \epsilon)$ and extracting the coefficient in front of $\epsilon$.
\\

Stochastic AD~\cite{arya2023} works by generalizing dual numbers to \textit{stochastic triples} of the form
\begin{equation}
x + \delta \epsilon + \left(\Delta \text{ with probability } w \epsilon \right).
\end{equation}
Instead of having just a smooth, infinitesimal perturbation $\delta \epsilon$, it also contains a discrete perturbation of finite size that only occurs with infinitesimal probability.
Now, let $C_X(p)$ be a computer program that samples the random variable $X(p)$.
Then, the program can be implemented such that $C_X(p+\epsilon)$ randomly returns a stochastic triple $x + \delta \epsilon + (\Delta \text{ with probability } w \epsilon)$, where $x$, $\delta$, $\Delta$ and $w$ are random variables that may be strongly correlated and have the properties $X(p) = x$ and
\begin{equation}
X'(p) = \delta + w \Delta,
\end{equation}
such that $X'(p)$ satisfies Eq.~\eqref{eq:X'} and has bounded variance.
Hence, the derivative of $\mathbb E[X(p)]$ can be estimated without bias and without diverging error bar by executing $C_X(p + \epsilon)$ several times, calculating $\delta + w \Delta$ for each obtained sample, and then taking the average of that quantity.
If it has been correctly implemented how to sample $C_{x_i}(p + \epsilon)$ for some simple random variables $x_i$ (e.g., geometric and binomial) and how simple deterministic functions should act on stochastic triples,
$C_X(p + \epsilon)$ can be sampled automatically for any $X$ that is composed of those simple random variables and deterministic functions.
Stochastic AD has been implemented as a library in the Julia programming language under the name StochasticAD.jl~\cite{arya2024}, which is what we use to obtain derivatives from quantum-network simulations.
\\

\section{Model for quantum repeaters}
\label{sec:model}

In this paper, we study quantum networks in which end nodes perform an entanglement-based quantum-key-distribution protocol known as BBM92~\cite{bennett1992a}, facilitated by quantum repeaters.
If the quantum network is a repeater chain, the repeaters execute a swap-ASAP protocol (defined below) to create entanglement between the end nodes.
After end-to-end entanglement has been achieved, both end nodes measure their qubits in either the Pauli X or Z basis according to BBM92 to produce a secret key.
If the quantum network is not a repeater chain, we will assume that a pair of end nodes that wants to produce a secret key reserves a path through the network to effectively create a repeater chain again (see Section \ref{sec:repeater_placement}).
Entanglement generation between neighboring nodes in a repeater chain is a discrete random process; the number of attempts required to obtain a single success is geometrically distributed.
Therefore, it would be very difficult to obtain derivatives (e.g., of the rate at which a secret key can be created as a function of the repeater positions) from a typical Monte Carlo simulation of such a repeater chain.
Using stochastic AD, however, we can obtain derivatives and perform gradient-based optimizations with ease, as demonstrated in Section~\ref{sec:results}.
In the rest of this section, we describe our model for quantum-repeater chains in more detail.
\\

Each pair of neighboring nodes in the chain is connected by an elementary link that can be used to generate elementary-link entanglement.
In a chain of $N$ nodes (of which $N - 2$ are repeaters), there are $N - 1$ elementary links.
Generating elementary-link entanglement is a sequence of probabilistic attempts.
Each attempt at elementary link $i$ succeeds only with probability $p_i$ and takes an amount of time $\Delta t_i$.
We assume that the attempt duration is limited by the transmission of optical pulses through fiber, giving
\begin{equation}
\Delta t_i = \frac {L_i} c,
\end{equation}
where $L_i$ is the length of elementary link $i$ and $c$ is the speed of light in fiber, which we assume to equal $200,000$ km/s (approximately two thirds the speed of light in vacuum).
When an attempt is successful, at the end of the corresponding time slot, the two-qubit state $\rho_i$ is created between the two nodes.
Ideally this is the Bell state, $\rho_i = \ket{\phi^+}\bra{\phi^+}$ with $\ket{\phi^+} = \tfrac 1 {\sqrt 2} \left( \ket{00} + \ket{11} \right)$, but more realistically there will be noise and $\rho_i$ will be some mixed state instead.
We here assume all elementary-link states are Werner states \cite{werner1989, dur1999} of the form
\begin{equation}
\rho_i = w_i \ket{\phi^+}\bra{\phi^+} + (1 - w_i) \frac {\mathbb 1} 4,
\end{equation}
where $w_i$ is the Werner parameter of elementary link $i$.
The fidelity of elementary link $i$ is given by
\begin{equation}
F_i \equiv \braket{\phi^+| \rho_i | \phi^+} = \frac {1 + 3 w_i} {4}.
\end{equation}
We note that a Werner state can be thought of as the result of depolarizing noise on a perfect Bell state.
This model has been chosen for its conceptual simplicity, as the contribution of this paper lies in the application of stochastic AD and not in the specific repeater model.
\\

When a repeater has successfully generated entanglement with one neighbor but not the other, it will store its entangled qubit in quantum memory while continuing attempts at entanglement generation with the other neighbor.
We assume each repeater has only a single qubit per elementary link, so the elementary link on which entanglement has already been generated falls idle.
If qubit $k$ is stored at node $n$ with coherence time $T_n$ for an amount of time $t$, we assume it undergoes depolarizing noise according to
\begin{equation}
\rho \mapsto e^{-\frac t {T_n}} \rho + (1 - e^{- \frac t {T_n}}) \frac {\mathbb 1}{2} \otimes \text{Tr}_k{\rho},
\end{equation}
where $\text{Tr}_k$ is the partial trace over qubit $k$.
We note that in the Werner-state model, this effectively evolves the Werner parameter of the two-qubit state that qubit $k$ is a part of as
\begin{equation}
w_i \mapsto e^{- \frac t {T_n}} w_i.
\end{equation}
Then, as soon as the second elementary link succeeds, the swap-ASAP protocol prescribes that the repeater immediately performs entanglement swapping~\cite{avis2022a, avis2024, coopmans2021, goodenough2025, haldar2025, inesta2023, kamin2023}.
Entanglement swapping is a measurement in the Bell basis followed by single-qubit corrections that combines two entangled states into a longer-distance entangled state~\cite{bennett1993a}.
If the repeater shares Werner states with nodes $A$ and $B$ with parameters $w_A$ and $w_B$ respectively, entanglement swapping results in a Werner state with parameter $w = w_A w_B$ shared between nodes $A$ and $B$, and an empty register at the repeater.
When every repeater has performed entanglement swapping, the end nodes of the repeater chain will hold an entangled state.
In order to keep the model simple it does not include classical communication between nodes, for example, to inform end nodes that entanglement swapping has taken place.
\\

Here, we discern two different variations of the swap-ASAP protocol.
First, there is the single-shot protocol.
In this protocol, each repeater remains idle after entanglement swapping until end-to-end entanglement has been successfully created.
Only then are the elementary links restarted, and the next round of entanglement commences with a clean slate.
Second, there is the multi-shot or concurrent protocol.
In this protocol, elementary links are restarted as soon as possible and repeaters keep swapping as soon as possible.
It can then happen that one repeater has already performed entanglement swapping twice before even the first end-to-end state has been created, as perhaps the elementary links that connect that repeater are very fast while there is another elementary link elsewhere in the chain that is very slow.
While the multi-shot protocol may seem to make more efficient use of resources, it may be susceptible to congestion and lead to increased qubit storage times and hence noise.
One could potentially use cutoff mechanisms~\cite{rozpedek2018a, santra2019, li2021a} to protect against this noise, but we do not consider such mechanisms here (see Section~\ref{sec:conclusion} for further discussion).
We point out that while the single-shot protocol has been extensively studied and even analytical results have been obtained for the behavior of such chains~\cite{goodenough2025, kamin2023}, we are not aware of any such results for the multi-shot protocol.
Throughout the rest of the paper, unless stated otherwise, we consider the multi-shot protocol as it allows us to demonstrate stochastic AD in a regime that is currently not analytically tractable.
For both the single-shot and multi-shot protocol, it is difficult to obtain derivatives from a typical Monte Carlo simulation without stochastic AD.
\\

Finally, an end-to-end entangled state is created a time $T_\text{ent}$ after the previous state and with Werner parameter $w$.
Using the BBM92 protocol for quantum-key distribution \cite{bennett1992a}, the states can be used to distribute a secret key between the end nodes.
Asymptotically the secret-key rate (SKR), which is the average number of secret bits shared per time unit, is given by~\cite{shor2000}
\begin{equation} \label{eq:skr}
\text{SKR} = \frac 1 {\mathbb E[T_\text{ent}]} \max(1 - 2h(\text{QBER}), 0),
\end{equation}
where $h(x) = -x\log_2(x) - (1-x)\log_2(1-x)$ is the binary entropy function.
The quantum-bit error rate (QBER) is the probability that two Pauli measurements that ideally would be correlated give different results, and is given by
\begin{equation}
\text{QBER} = \frac{1 - \mathbb E[w]} 2
\end{equation}
for Werner states.
The secret-key rate then allows us to combine the rate of entanglement distribution and the quality of entangled states into a single assessment of the utility of the repeater chain.
\\

In order to estimate the secret-key rate, we have developed the open-source Julia package QuantumNetworkRecipes.jl~\cite{quantum_network_recipes.jl}.
We use this package to run simple Monte Carlo simulations in which the number of attempts required for each elementary link to succeed is determined by sampling geometric distributions.
$T_\text{ent}$ is then sampled by calculating the time when all required elementary-link states are in place and the last entanglement swap can be performed.
The final Werner parameter $w$, on the other hand, is sampled by calculating qubit storage times from the differences between the times when adjacent elementary links are successful.
The simulations have been set up to be compatible with stochastic AD as implemented in StochasticAD.jl~\cite{arya2024} and hence can be used to not only estimate $\mathbb E[T_\text{ent}]$ and $\mathbb E[w]$ but also their derivatives.
These estimates can then in turn be used to estimate the secret-key rate and its derivative.
This allows us not only to perform gradient-based optimization techniques (see Sections~\ref{sec:finitediff}, \ref{sec:optimizing_bright_state_params} and \ref{sec:repeater_placement}), but also to perform sensitivity analysis (see Section~\ref{sec:sensitivity_analysis}).
In Appendix~\ref{app:benchmarking} we evaluate the computational cost of applying stochastic AD to the simulations.

\section{Results}
\label{sec:results}

In this section, we demonstrate the usefulness of stochastic AD to study and optimize quantum networks.
First, in Section \ref{sec:finitediff}, we compare stochastic AD to finite differences.
Then, in Section \ref{sec:optimizing_bright_state_params}, we optimize the secret-key rate of a repeater chain by tuning the rate-fidelity tradeoffs of the elementary links.
In Section~\ref{sec:sensitivity_analysis}, we perform a sensitivity analysis to identify the most crucial coherence times in that same repeater chain.
Finally, in Section~\ref{sec:repeater_placement}, we optimize the placement of quantum repeaters in a 2D plane to get the best guaranteed quality of service between four end nodes.

\subsection{Comparison to finite differences}
\label{sec:finitediff}

To demonstrate the advantages of using stochastic AD to extract derivatives from  Monte Carlo simulations of quantum networks, we here compare it to finite differences.
Finite differences defines three basic types of derivative estimators: the forward difference, the backward difference, and the central difference.
The central difference of a function $f(x)$ approximates its derivative as
\begin{equation}
f'(x) \approx \frac{f(x + \frac 1 2 \epsilon) - f(x - \frac 1 2 \epsilon)} \epsilon
\end{equation}
for some value of $\epsilon$.
The forward difference and backward difference are defined similarly, but with the function arguments shifted by $\tfrac 1 2 \epsilon$ forwards and backwards, respectively.
The standard error in the central difference is given by
\begin{equation}
\frac 1 {|\epsilon|} \sqrt{ \Delta_f(x + \frac 1 2 \epsilon) ^2 + \Delta_f(x - \frac 1 2 \epsilon)^2},
\end{equation}
where $\Delta_f(x)$ is the standard error in the evaluation of $f$ at value $x$.
This, then, poses a problem: the estimator will only be unbiased (i.e., converge to the true value of the derivative) in the limit $\epsilon \to 0$.
But in that same limit, the standard error diverges.
The same problem holds for the forward difference and the backward difference.
Therefore, when using finite differences, one has to compromise between accuracy and precision by choosing a suitable finite value for $\epsilon$.
In contrast, when using stochastic AD, the estimator is always unbiased and has a bounded standard error.
\\

To illustrate this point, we compare the derivative of the secret-key rate (defined in Eq.~\eqref{eq:skr}) with respect to the position of a quantum repeater that is placed between two end nodes separated by 100 km using both stochastic AD and finite differences.
Specifically, we compare stochastic AD to the central difference, as we found it is more accurate here than the forward and backward differences.
We assume both elementary links create perfectly entangled states ($F_1 = F_2 = 1$) and each node has a coherence time of $T_1 = T_2 = T_3 = 0.13$ s.
Moreover, the success probability for elementary link $i$ with length $L_i$ is
\begin{equation}
p_i = 10 ^ {- \frac{\gamma_i}{10} L_i}.
\end{equation}
Here, $\gamma_1 = 0.2 \, \text{km}^{-1}$, corresponding to 0.2~dB/km attenuation losses for the first elementary link, and $\gamma_2 = 0.6 \, \text{km}^{-1}$, corresponding to 0.6~dB/km for the second.
Of particular interest is finding the repeater location that optimizes the secret-key rate.
Because the two different elementary links have different attenuation coefficients, the optimal position will not be in the center, providing a simple but nontrivial (if somewhat contrived) optimization problem.
The optimal position can be found by identifying where the derivative is zero.
\\

\begin{figure}
\includegraphics[width=\columnwidth]{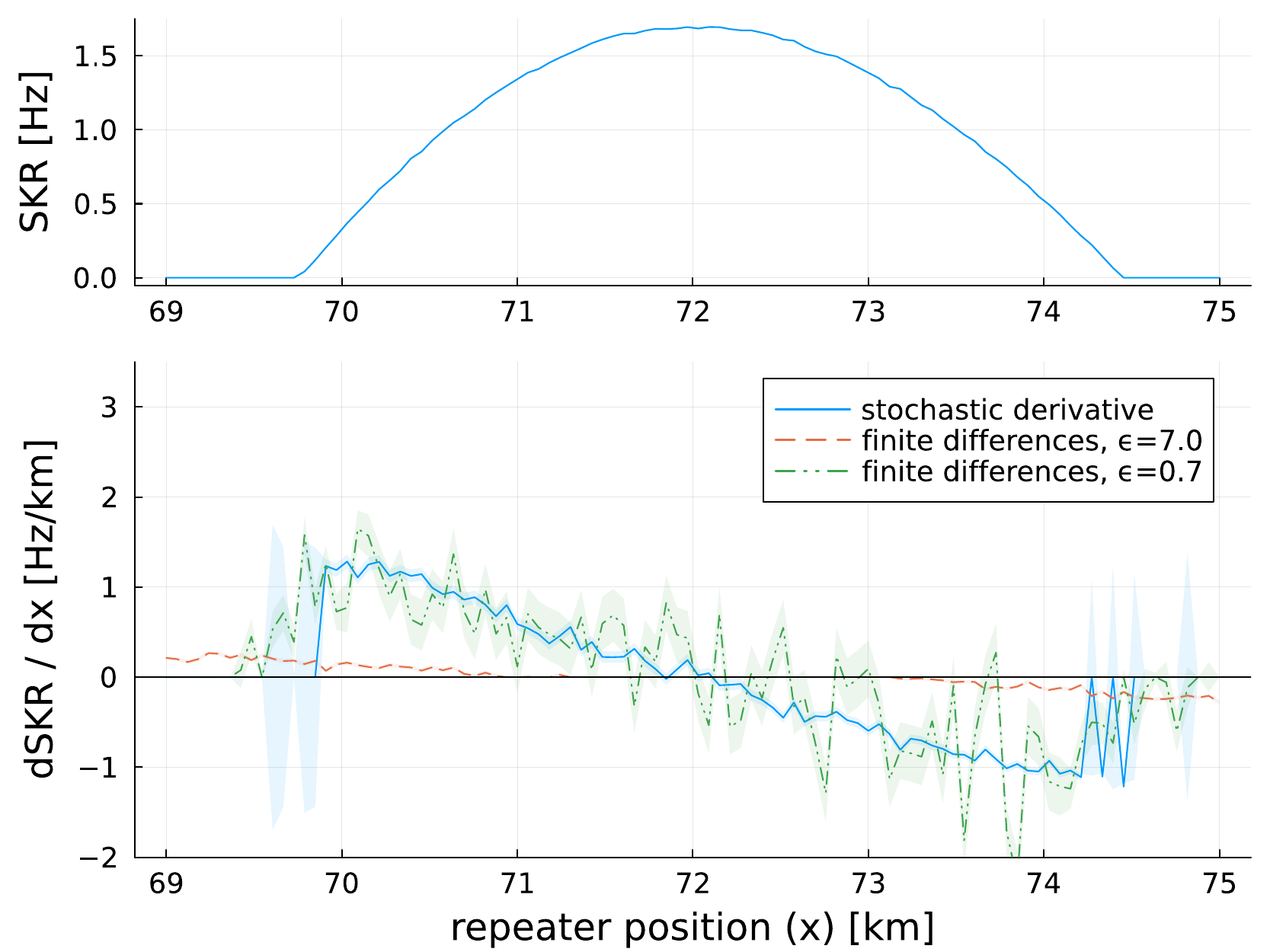}
\caption{
    Comparison between stochastic AD and finite differences.
    Shown are the secret-key rate (SKR, defined in Eq. \eqref{eq:skr}) and its derivative over 100~km using a quantum repeater with a coherence time of 0.13~s that suffers 0.2~dB/km attenuation losses on one side and 0.6~dB/km on the other, as a function of the repeater's position.
    Each data point for the secret-key rate and its derivative is based on $10^7$ and $10^4$ samples respectively.
    Shaded regions indicate standard errors.
}
\label{fig:compare_to_finitediff}
\end{figure}

The comparison is shown in Figure~\ref{fig:compare_to_finitediff}.
For finite differences, it can be seen that choosing a small value for $\epsilon$ makes it very hard to pinpoint the exact zero of the function due to the large error bars.
On the other hand, a large value of $\epsilon$ results in a sharply defined line, but it is zero over a range of more than 2 km leaving the exact optimum again ambiguous.
In general, finite differences becomes unreliable when a function has important features at a scale that is similar to or smaller than $\epsilon$.
A good compromise requires choosing a suitable value for $\epsilon$, which may be different not only per function but also per region of the same function, and also depends on the number of samples taken for each data point.
While choosing a good $\epsilon$ heuristically could perhaps be feasible in this pedagogical example, that is not typically true for complicated high-dimensional functions.
In contrast, stochastic AD provides an unbiased value of the derivative with a relatively small standard error without the need for finetuning a tradeoff, making it easy to identify the optimum for any function.
\\

\subsection{Optimizing a repeater chain}
\label{sec:optimizing_bright_state_params}

Quantum networks typically have tunable parameters that control tradeoffs between the rate at which entangled states are produced, and their fidelity.
Examples of such parameters are the bright-state parameter in single-click entanglement generation~\cite{cabrillo1999, bose1999, humphreys2018, pompili2021, stolk2024} (discussed below), the time window for photon detection~\cite{schupp2021, krutyanskiy2023, hensen2015, stolk2024}, the coincidence time window for double-click entanglement generation~\cite{schupp2021, krutyanskiy2023, avis2022a}, the mean photon number of pulses used to drive SPDC sources~\cite{knaut2024, krovi2016} and cutoff times that determine how much decoherence entangled states are allowed to suffer before they are discarded~\cite{rozpedek2018a, santra2019, li2021a, pompili2021, langenfeld2021, haldar2024a}.
In order to get the best performance out of a network, or to reduce the hardware requirements for making the network feasible in the first place, it is important to determine the best values of these parameters for particular quantum-network applications.
This can be phrased as the problem of optimizing a network utility function.
In this section, we use stochastic AD to solve such optimization problems.
\\

For concreteness, in this section, we optimize the bright-state parameter of single-click entanglement generation.
When entanglement is generated using the single-click protocol, emitters at both nodes sharing the elementary link are prepared in the state
\begin{equation}
\sqrt{1 - \alpha} \ket{\text{dark}} + \sqrt{\alpha} \ket{\text{bright}},
\end{equation}
where $\alpha$ is the bright-state parameter.
After that the emitter is excited, resulting in photon emission only if the emitter is in the bright state, thereby creating entanglement between the emitter and the absence/presence of a photon.
Photons emitted by both nodes are then directed to a midpoint station where an entangling measurement is performed that effectively counts the total number of photons that arrive.
If only a single photon is measured, in the absence of photon loss, this projects the emitters on one of the states $\tfrac 1 {\sqrt 2} \left(\ket{\text{bright, dark}} \pm \ket{\text{dark, bright}}\right)$.
However, when two photons are emitted but one is lost, this will also result in only a single photon being measured, and hence the entanglement will be noisy.
The noise can be suppressed by making $\alpha$ small, but this comes at the cost of a reduced success rate.
Hence, it tunes a tradeoff between rate and fidelity.
\\

Using the repeater-chain model introduced in Section~\ref{sec:model},
we here address the problem of choosing the bright-state parameters for each of the $N - 1$ elementary links such that the secret-key rate, defined in Eq.~\eqref{eq:skr}, is optimized.
A simplified version of this optimization was phrased as a mathematical program by Vardoyan et al.~\cite{vardoyan2023} that allows for convex optimization for a specific class of utility functions, a class that was extended further in \cite{kar2025}.
Optimization of bright-state parameters under more complicated conditions was performed using genetic algorithms in~\cite{avis2022a, mora2024}.
Here we follow Vardoyan et al. by making the typical simplifying assumption that each elementary link produces Werner states.
In particular, when $\alpha_i$ is the bright-state parameter of elementary link $i$, states on that link are created with Werner parameter
\begin{equation}
w_i = 1 - \frac 3 4 \alpha_i
\end{equation}
and with success probability
\begin{equation}
p_i = 2 \alpha_i 10^{- \frac \gamma {10} L_i},
\end{equation}
where we assume $\gamma = 0.2 \, \text{km}^{-1}$ (corresponding to 0.2~dB/km attenuation losses) and $L_i$ is the length.
Unlike Vardoyan et al. though, we consider a dynamical model of the repeater chain where each node only has a single qubit per elementary link and qubits undergo decoherence while waiting for entanglement swapping, which greatly complicates determining the end-to-end rate and state.
\\

\begin{figure*}[!tbp]
    \centering
    \subfloat[] {\includegraphics[width=.32\textwidth]{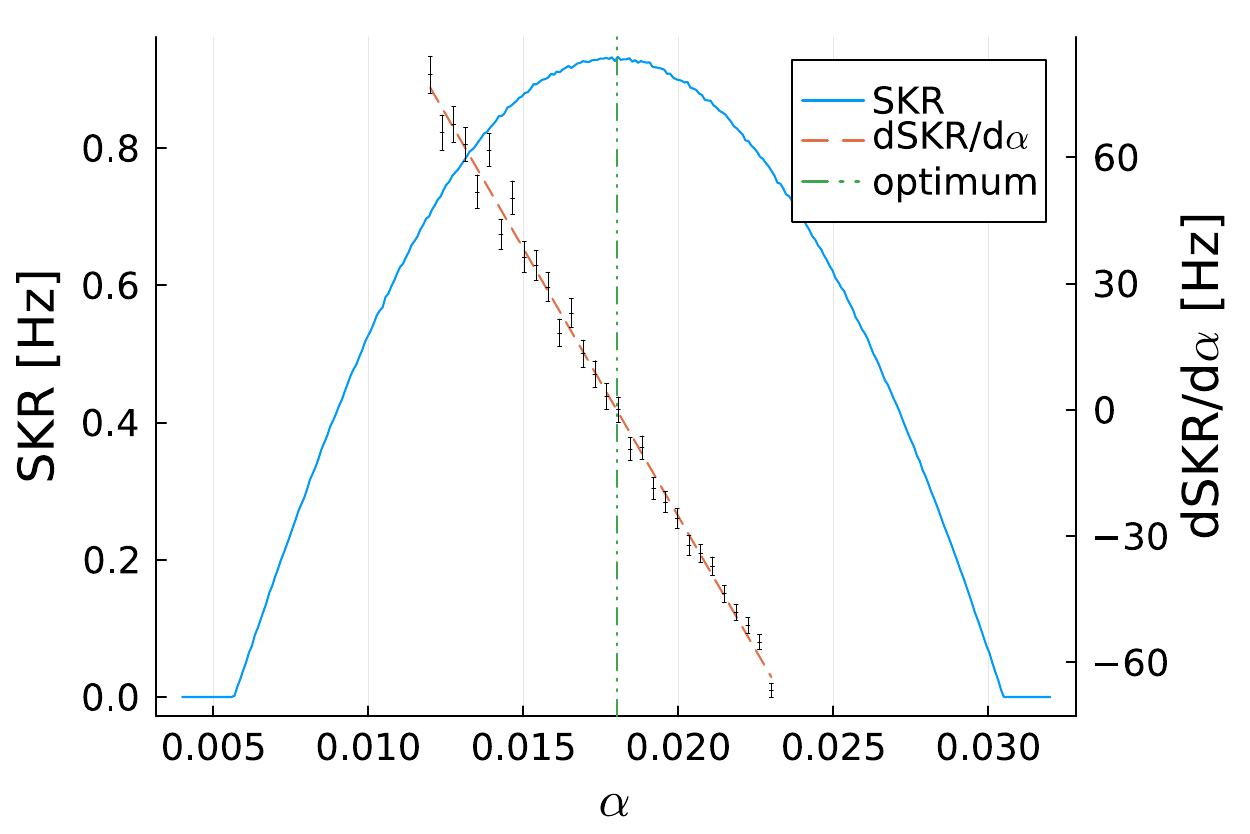}}
    \hfill
    \subfloat[] {\includegraphics[width=.32\textwidth]{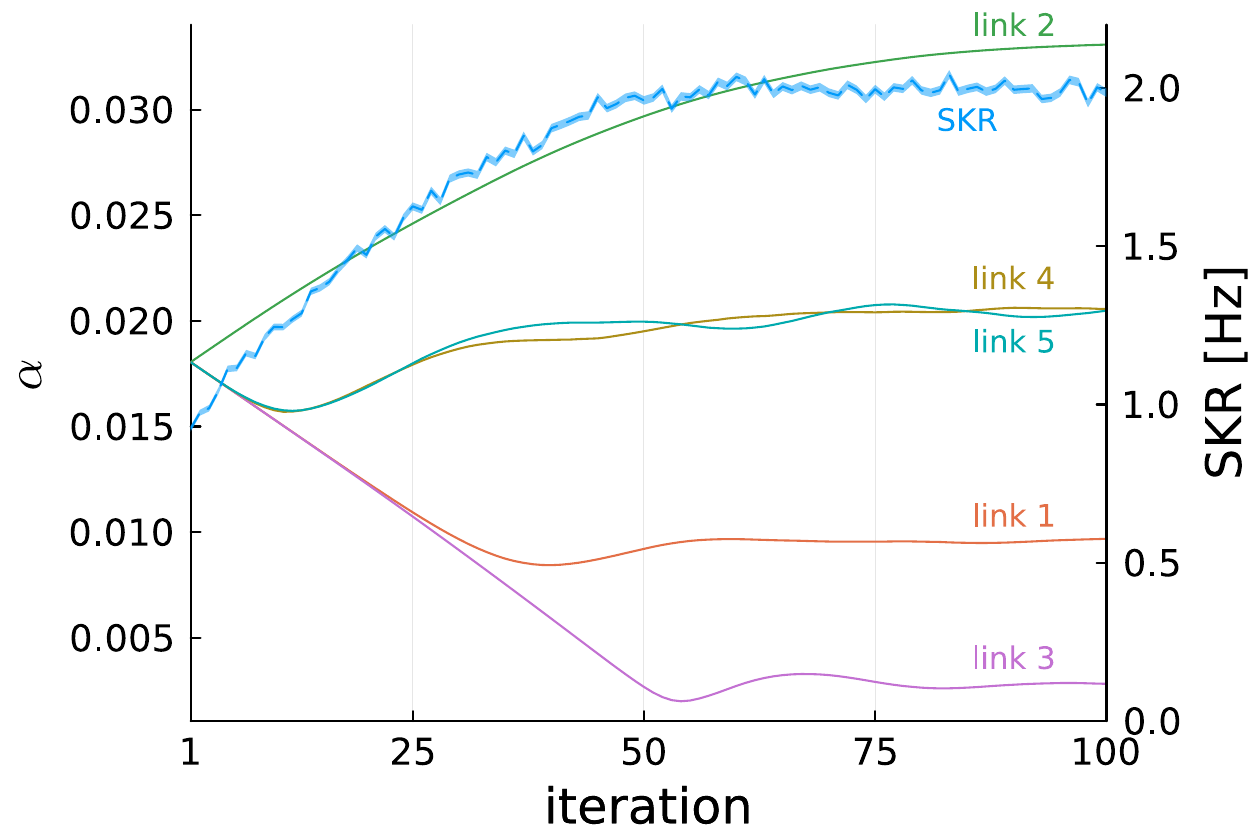}}
    \hfill
    \subfloat[] {\includegraphics[width=.32\textwidth]{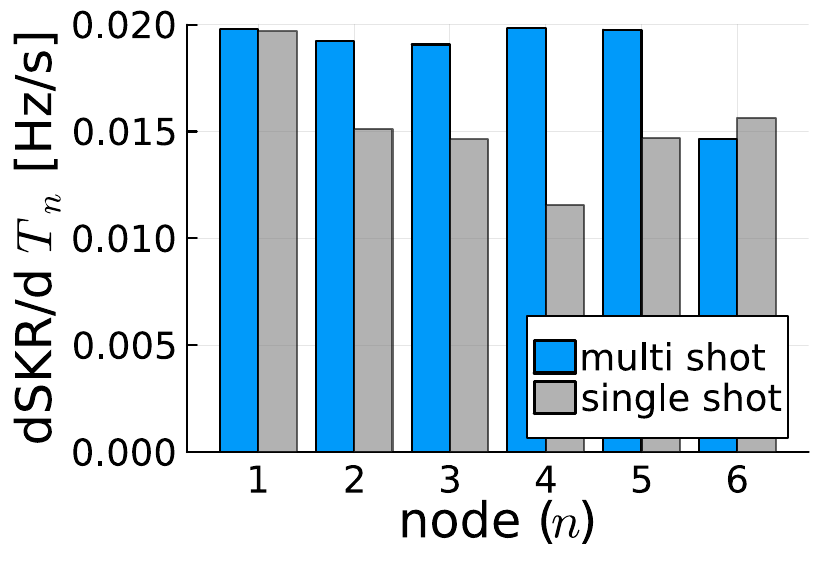}}
    \\
    \subfloat[] {\includegraphics[width=\textwidth]{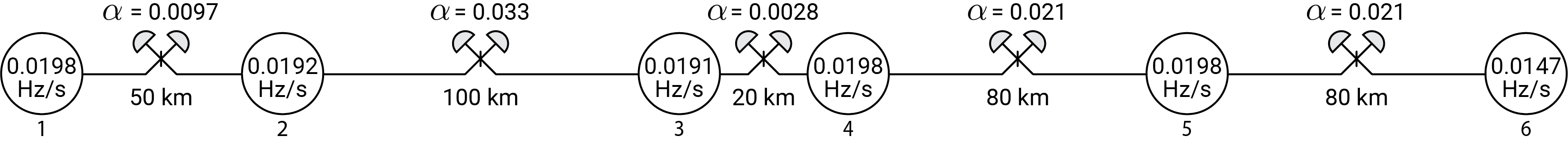}}
    \caption{
    Optimization and sensitivity analysis of a repeater chain.
    \textbf{(a)} The optimal value of the bright-state parameter ($\alpha$) in case the same value is used for every elementary link is determined using a linear fit.
    The secret-key rate (SKR, defined in Eq. \eqref{eq:skr}) is estimated using $10^6$ samples per data point for illustration and the derivative is estimated using $10^4$ samples for each of the 30 data points to determine the fit.
    \textbf{(b)} Gradient descent is used to optimize the bright-state parameters of the five elementary links individually, using 100 total iterations of $10^4$ samples each.
    The solution found at (a) is used as the initial guess.
    The secret-key rate is estimated after the optimization finishes, with the shaded region indicating the standard error.
    \textbf{(c)} The sensitivity of the secret-key rate with respect to the coherence times of the six different nodes in the chain when using the optimal values of the bright-state parameters, for both the multi-shot and single-shot entanglement-generation protocols.
    The derivatives are estimated using $10^5$ samples each.
    Standard errors are all smaller than $10^{-4}$ Hz/s and are not shown.
    \textbf{(d)} The repeater-chain setup, with optimal bright-state parameters and coherence-time sensitivities inscribed.
    }
    \label{fig:repeater_chain}
\end{figure*}

Here, we optimize the bright-state parameters of the asymmetric repeater chain displayed in Figure~\ref{fig:repeater_chain} (d), where the coherence time of every node is $T_n = 10$ s.
We use the Monte Carlo simulation described in Section~\ref{sec:model} to estimate the secret-key rate as a function of the bright-state parameters, and we use stochastic AD to estimate its derivatives.
First, we set all five bright-state parameters to be equal and estimate their optimal value using a linear fit of the derivative, as shown in Figure~\ref{fig:repeater_chain} (a).
Then we use a gradient-descent algorithm with momentum called Adam~\cite{kingma2017} to optimize over the full five-dimensional space, using the optimal value when they are all the same as the initial guess.
As shown in Figure~\ref{fig:repeater_chain} (b), optimizing the five bright-state parameters individually leads to very different values for each of them, allowing the secret-key rate to be doubled from $\sim 1$ Hz to $\sim 2$ Hz compared to using the same value for all.
The full two-step optimization took less than a minute using a single core of an Intel i7 13th-gen processor.
% 57.1 s using a single core of a i7-1360P processor in a Dell XPS 13 laptop 
Further optimization results and an investigation of how the results would change if the single-shot protocol was used instead, can be found in Appendix~\ref{app:shots}.

\subsection{Sensitivity analysis}
\label{sec:sensitivity_analysis}

The derivatives obtainable through stochastic AD are useful beyond just optimization.
In particular, they can be used to perform sensitivity analysis.
If a parameter is changed a little bit, how will that affect the network as a whole?
Sensitivity analysis can be used to identify the most crucial hardware parameters in which to invest during research and development.
Moreover, when deploying a quantum network, sensitivity analysis can help decide how much to spend on each individual component.
For instance, if there are only enough funds for one top-of-the-line quantum repeater while all other repeaters must be older models, at which network node should it be placed?
Finally, sensitivity analysis can also be used to identify how often different network components should be scheduled for maintenance.
Components associated with parameters that the network is insensitive to can be allowed to deteriorate a bit, while other components may need to be kept in mint condition.
\\

Here, we revisit the repeater chain studied in Section~\ref{sec:optimizing_bright_state_params} and determine the derivatives of the secret-key rate with respect to the coherence times of the different nodes in the network.
The results can be seen in Figure~\ref{fig:repeater_chain} (c) for both the multi-shot protocol that is used throughout the rest of the paper and the single-shot protocol (see Section~\ref{sec:model} for details).
In both cases, the bright-state parameters are set to the values that were found to be optimal for that particular protocol.
For the optimized parameters of the single-shot protocol, see Appendix~\ref{app:shots}.
For the multi-shot case, we see that the sensitivity is almost the same for all nodes except the rightmost end node (node 6), which has a smaller sensitivity.
On the other hand, for the single-shot case, there are much larger differences in the sensitivities.
This demonstrates that the relative importance of the different hardware parameters can be highly dependent on the exact network protocol that is executed.
In the single-shot case, we see that if some of the quantum memories could be upgraded, node 1 would be first in line while node 4 would be last.
Moreover, if maintenance is required to counteract drift or degradation of the coherence time~\cite{stephens2022, danageozian2023, klimov2018, matityahu2021}, node 1 would probably require service more often than node 4.
This is yet another example of an important result for the engineering of quantum-repeater networks that would have been difficult to obtain reliably without stochastic AD.
\\

\subsection{Repeater placement}
\label{sec:repeater_placement}

As our final and most complex example of stochastic AD, we apply it to the design of quantum-repeater networks.
In particular, we consider the following question:
Given a set of end nodes that wish to perform quantum communication, how should quantum repeaters and connections between them be allocated?
In case there are two end nodes, the problem reduces to choosing how many repeaters should be placed and how they should be spaced~\cite{avis2024, jiang2007a, silva2024, yao2021, chehimi2023, pathumsoot2021, luong2016, dhara2021a, rozpedek2021, wo2023, rozpedek2018a}.
Repeater allocation with more than two end nodes was first studied in~\cite{rabbie2022}, where the problem was formulated as choosing repeater locations and optical fibers from some preexisting infrastructure such that minimal performance requirements are satisfied and the total network cost (i.e., the number of repeaters) is minimized.
Heuristics for solving a slightly modified version of this formulation were proposed in~\cite{islam2023, sripotchanart2024}.
In~\cite{pouryousef2024a} a similar problem was considered, where instead a network utility function is maximized while the total number of repeaters is constrained.
\\

Here, we consider a different formulation of the repeater-allocation problem that we refer to as the \textit{greenfield scenario}.
Instead of being limited to preexisting infrastructure, repeaters can be placed anywhere in a two-dimensional Euclidean space, and elementary links between any two nodes can be established with a length that is equal to the distance between the two nodes.
Just as in the first formulation, one can either minimize the cost subject to a performance constraint as in~\cite{rabbie2022} or maximize the performance subject to a cost constraint as in~\cite{pouryousef2024a}.
Here we take the second approach, as it will allow us to study explicitly the effect of the number of repeaters on the network performance.
We observe interesting behavior where the minimum number of repeaters required to improve network performance and the number of repeaters at which the network becomes saturated both scale approximately linearly with the physical size of the network.
This result was only made possible by stochastic AD, allowing us to apply gradient-based optimization techniques to Monte Carlo simulations of quantum networks.
\\

While many future quantum networks may be built on top of existing infrastructure to limit costs, the greenfield scenario is the more relevant formulation when no such preexisting infrastructure is available, as might, for example, be the case when constructing sensor networks in remote places~\cite{gottesman2012, khabiboulline2019a, brady2022, eldredge2018, guo2020a, proctor2018, sekatski2020, xia2019, zang2024, zhang2021}.
Moreover, the greenfield scenario is also applicable to mobile repeater nodes that communicate through line-of-sight free-space links and that can be freely (re)configured, e.g., networks of drones~\cite{kumar2022, liu2020b, conrad2023}, balloons~\cite{karakosta-amarantidou2025}, ships~\cite{hill2016, qin2021}, submarines~\cite{gariano2019} or satellites~\cite{yin2020, kumar2022, vergoossen2020, sidhu2021}.
Finally, solving repeater allocation in the greenfield scenario can also be useful when building networks using existing infrastructure.
The greenfield scenario reveals the ultimate performance limit for a given number of repeaters, providing a benchmark against which to judge how limiting a set of restrictions is.
Moreover, the optimization problem is NP-hard on existing infrastructure~\cite{rabbie2022, sripotchanart2024}; solutions from the greenfield scenario could potentially serve to construct heuristics for optimizing repeater allocation on preexisting infrastructure.
\\

As the network utility function that we optimize the repeater placement for, we choose the guaranteed minimal quality of service the network can offer to its users in terms of the secret-key rate (see Eq.~\eqref{eq:skr}).
Writing $\text{SKR}_{i, j}$ for the secret-key rate corresponding to the single best path between end nodes $i$ and $j$, the network utility $\text{SKR}_\text{min}$ is given by
\begin{equation} \label{eq:skr_min}
\text{SKR}_\text{min} = \min_{i, j} \{\text{SKR}_{i, j}\}.
\end{equation}
Note that this choice for the network utility does not account for the possibility of conflicting paths in the network;
it is the minimal guaranteed secret-key rate in case only one pair of end nodes utilizes the network at the same time.
We focus specifically on the placement of $N$ quantum repeaters that need to connect four end nodes that are positioned on the vertices of a square with edges of length $D$.
Below, we will investigate the effects of $N$ and $D$ on the solutions.
For our model (described in Section~\ref{sec:model}), we assume the success probability for elementary link $i$ of length $L_i$ is
\begin{equation}
p_i = 10 ^ {- \frac \gamma {10} L_i},
\end{equation}
with $\gamma$ again $0.2 \, \text{km}^{-1}$, corresponding to 0.2~dB/km attenuation losses.
Furthermore, we assume $F_i = 0.99$ for the elementary-link fidelity of each link and $T_n = 10$ s for the coherence time of each node. 
\\

In order to perform the optimization, we apply a gradient-descent algorithm that incorporates elements of simulated annealing~\cite{delahaye2019}.
In each iteration, a pathfinding algorithm is used to identify the best path for each pair of end nodes (see Appendix~\ref{app:pathfinding} for more details).
Stochastic AD is then applied to the worst of all the best paths to determine the derivative of $\text{SKR}_\text{min}$ with respect to the $x$ and $y$ coordinates of all repeater nodes in the network.
For the next iteration, the nodes are displaced in accordance with the Adam gradient-descent algorithm~\cite{kingma2017}.
However, the learning rate of the algorithm, as well as the accuracy with which the derivative is determined, are controlled by a temperature parameter (see Appendix~\ref{app:annealing} for more details).
The temperature starts at a high value, resulting in repeater nodes that undergo large and mostly random displacements.
Then, according to an exponential schedule, the temperature is reduced epoch-by-epoch, until only small steps precisely in the direction of the nearest local optimum are taken.
The goal of this method is to allow exploration of the parameter space and avoid directly getting stuck in the nearest local optimum, while still enabling good exploitation of the final solution that is found.
This is a random method with no guarantee of finding the global optimum; for each value of $N$ and $D$, we perform the optimization many times (at least 25, on average approximately 47) and only report the best solution.
We have developed the open-source Julia package RepeaterPlacement.jl~\cite{repeater_placement.jl} to perform these optimizations.
\\

\begin{figure}[!tbp]
    \centering 
    \includegraphics[width=\columnwidth]{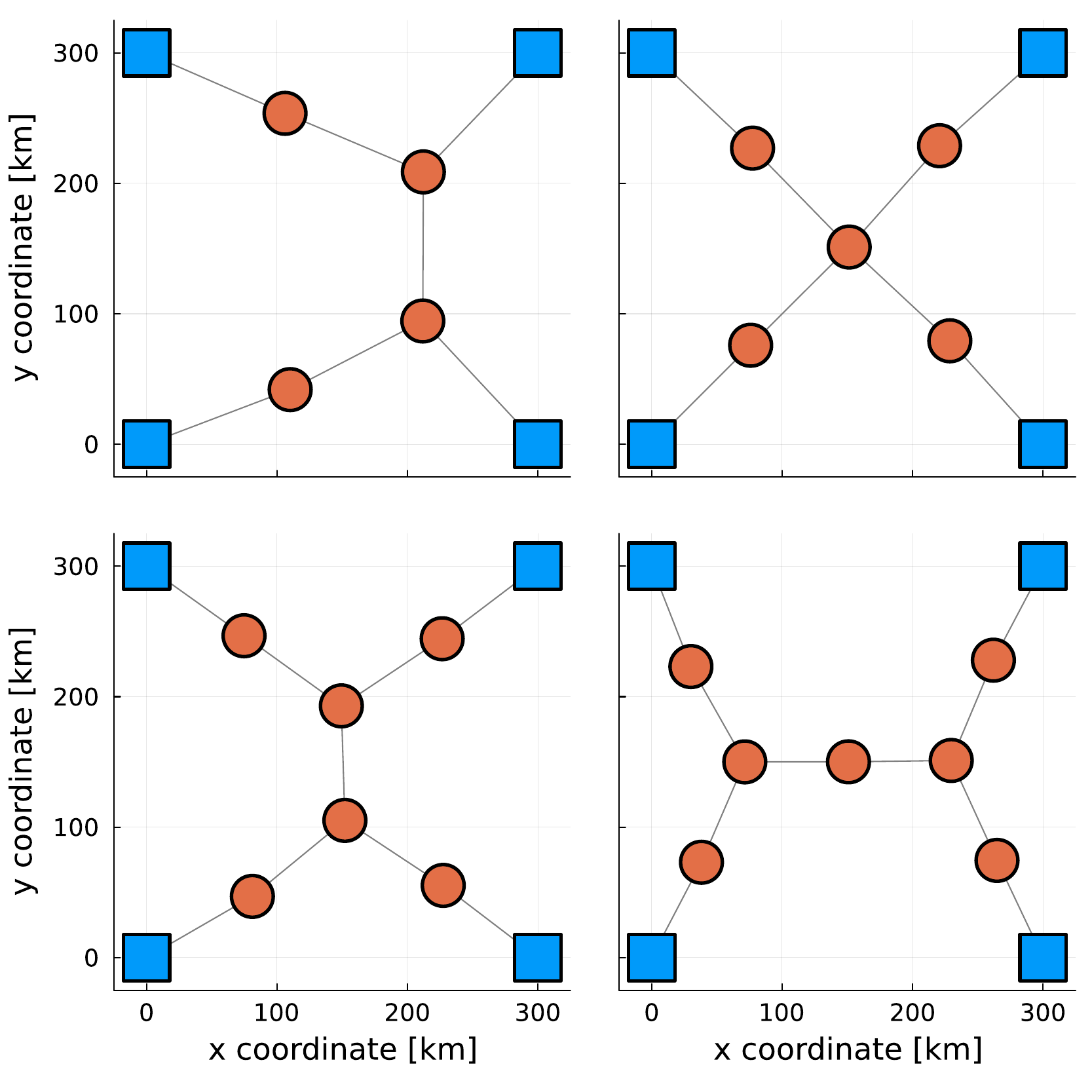}
    \caption{
    Examples of solutions for the optimization of repeater locations.
    End nodes (blue squares) form a square with edges of length $D = 300$ km.
    The total numbers of repeaters (orange circles) are $N = 4, 5, 6, 7$.
    }
\label{fig:placement_solution_examples}
\end{figure}

For examples of optimization solutions, see Figure~\ref{fig:placement_solution_examples}.
One notable feature of these solutions is that they spontaneously break the symmetry of the underlying problem.
That is, the repeaters do not always inherit the symmetry of the end nodes, which is the symmetry of the square.
In particular, it appears that while the reflection symmetries remain intact, rotations by multiples of $90^\circ$ are sometimes broken.
This implies a solution that is degenerate, as the utility function is invariant under these transformations.
This is indeed reflected in the output of the optimizer, which randomly returns one of the degenerate solutions.
More examples can be found in Appendix~\ref{app:more_solutions}.
\\

\begin{figure}[!tbp]
    \centering 
    \includegraphics[width=\columnwidth]{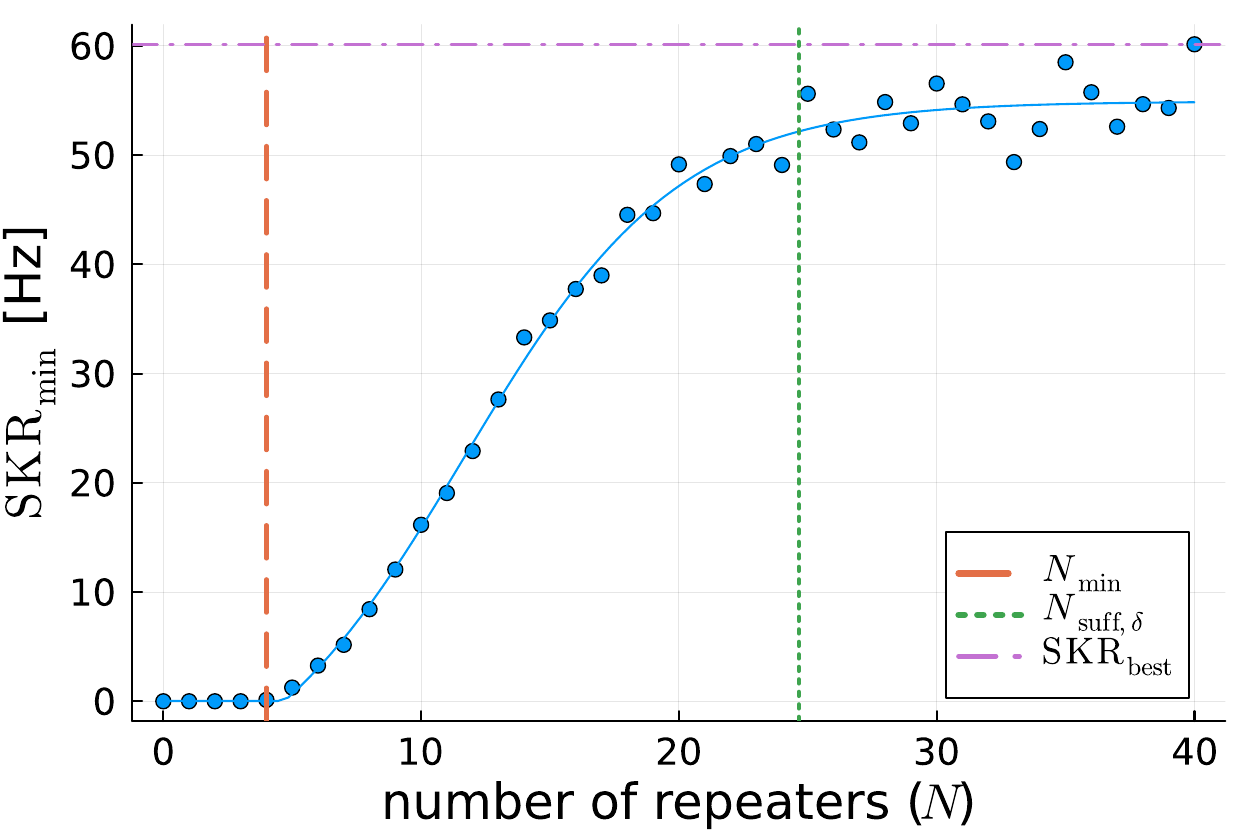}
    \caption{
    Dependence of the optimized network utility ($\text{SKR}_\text{min}$, defined in Eq.~\eqref{eq:skr_min}) on the number of repeaters ($N$).
    End nodes form a square with edges of length $D = 300$ km.
    Shown are the optimization results, a fit function as described by Eq. \eqref{eq:fit_function}, the minimal number of repeaters ($N_\text{min}$, defined in Eq. \eqref{eq:N_min}), the sufficient number of repeaters ($N_{\text{suff, }\delta}$, defined in Eq. \eqref{eq:N_suff}, with $\delta = 0.05$), and the best secret-key rate ($\text{SKR}_\text{best}$, defined in Eq. \eqref{eq:best_skr}).
    The error bars in the data are smaller than the marker size and have been suppressed.
    }
\label{fig:placement_features_example}
\end{figure}

In Figure~\ref{fig:placement_features_example}, we show the optimal value of $\text{SKR}_\text{min}$ that was found for $N = 0, 1, 2, ..., 40$ for fixed $D=300$ km.
Similar figures for different values of $D$ can be found in Appendix~\ref{app:more_solutions}.
There are three numbers of interest that we extract from such figures.
Firstly, it can be seen that for small values of $N$, the secret-key rate is close to zero and is not improved by adding more repeaters.
The reason for this is that when the elementary links of the potential network paths are too long, it is better to perform direct entanglement generation between end nodes than to use repeaters.
When using repeaters, the memory decoherence will be so large that the noise in the final states exceeds what is tolerated by the BBM92 protocol.
In contrast, when using direct transmission there is no memory decoherence, and a nonzero though very small key rate is possible (e.g., $\text{SKR}_\text{min} \approx 7 \cdot 10^{-7}$ for $D = 300$ km and $N = 0, 1, 2, 3$).
Only at some point does adding repeaters improve network performance.
We define this point as the minimal number of repeaters:
\begin{equation} \label{eq:N_min}
N_\text{min}(D) = \min \{N|\text{SKR}_\text{min}(D, N) > \text{SKR}_\text{min}(D, 0)\}.
\end{equation}
In practice, we determine $N_\text{min}(D)$ by finding the smallest $N$ for which $\text{SKR}_\text{min}(D, N) > 1.03 \cdot \text{SKR}_\text{min}(D, 0)$, to account for statistical errors in the estimate of $\text{SKR}_\text{min}$.
For example, $N_\text{min}(300 \text{ km}) = 4$, as shown by the orange dashed vertical line in Figure~\ref{fig:placement_features_example}.
\\

Secondly, we see that for large values of $N$ the secret-key rate is not improved much by adding more repeaters.
The reason for this is that adding repeaters to a path may improve the rate of entanglement generation, but because links are only generated with a fidelity of 0.99 it will come at the cost of increased error rates.
To quantify this, we fit a logistic function of the type
\begin{equation} \label{eq:fit_function}
f(N) = \max \left( \frac {c_1} {1 + e^{- c_2 * (N - c_3)}} + c_4, c_5 \right)
\end{equation}
to the data, where the free parameters $c_i$ are determined through a least-squares procedure.
We then define the sufficient number of repeaters as the smallest value of $N$ that is $\delta$ close to the supremum of the fit:
\begin{equation} \label{eq:N_suff}
N_{\text{suff, }\delta}(D) = \min \left\{N \bigg | \frac{\lim_{M \to \infty} f(M) - f(N)}  {\lim_{M \to \infty} f(M)} < \delta \right \},
\end{equation}
For example, $N_{\text{suff, } 0.05}(300 \text{ km}) \approx 20.7$, as indicated by the green dotted line in Figure~\ref{fig:placement_features_example}.
\\

Finally, we define the best secret-key rate as the best network utility found for a given value of $D$ (which may not necessarily be for the largest $N$ due to the random nature of the optimizations):
\begin{equation} \label{eq:best_skr}
\text{SKR}_\text{best} (D) = \max_N \{ \text{SKR}_\text{min}(D, N) \}.
\end{equation}
For example, $\text{SKR}_\text{best}(300 \text{ km}) \approx 59$ Hz, as shown by the horizontal line in Figure~\ref{fig:placement_features_example}.
\\

\begin{figure}[!tbp]
    \centering 
    \includegraphics[width=\columnwidth]{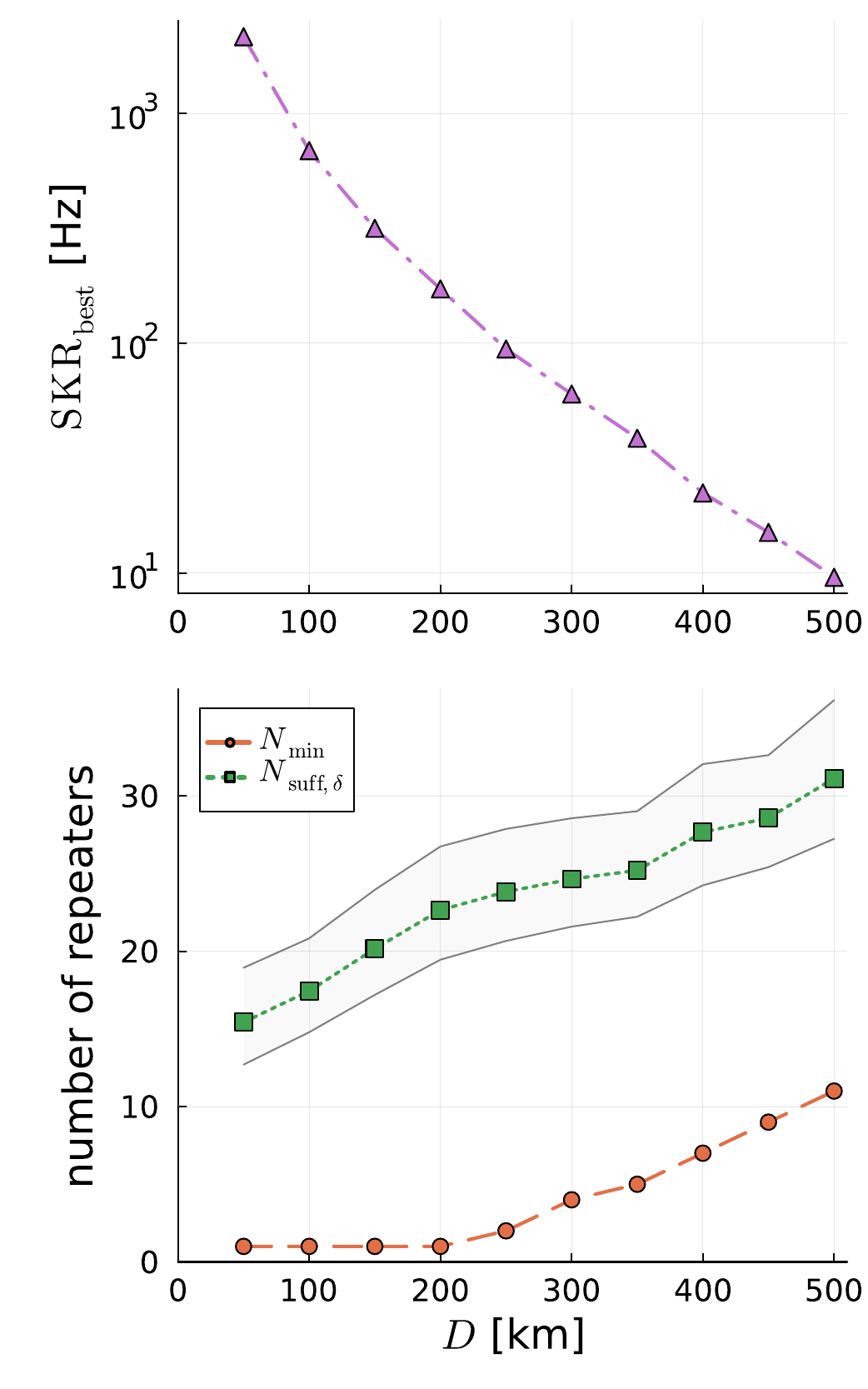}
    \caption{
    Dependence of the best secret-key rate ($\text{SKR}_\text{best}$, defined in Eq. \eqref{eq:best_skr}), the minimum number of repeaters ($N_\text{min}$, defined in Eq. \eqref{eq:N_min}), and the sufficient number of repeaters ($N_{\text{suff, }\delta}$, defined in Eq. \eqref{eq:N_suff}, with $\delta = 0.05$ for the middle line, $\delta = 0.02$ for the upper line and $\delta = 0.1$ for the lower line) on length ($D$) of the edges of the square that the end nodes form.
    }
\label{fig:placement_feature_dependence}
\end{figure}

To understand how the problem depends on the scale of the network, we have investigated the dependence of the minimum number of repeaters, the sufficient number of repeaters, and the best secret-key rate on the parameter $D$.
The results are shown in Figure~\ref{fig:placement_feature_dependence}.
It can be seen that the best secret-key rate falls approximately, but not exactly, exponentially with $D$.
Approximately one order of magnitude is lost each time $D$ is increased by 250 km.
While thousands of Hz of secret-key rate are achievable for $D = 50$ km, not even 10 Hz is achievable for $D = 500$ km.
For the minimum number of repeaters, we see that a single repeater is enough to enhance key rates up to and including $D = 200$ km.
From 250 km onwards the minimum increases approximately linearly with $D$, on average by 3 repeaters per 100 km up to 10 repeaters at $D = 500$ km.
Looking at the sufficient number of repeaters though, we see that in order to get close to the best possible secret-key rate more than 15 repeaters are already required at $D = 50$ km.
This increases approximately linearly by 3.5 repeaters per 100 km with more than 31 repeaters required at $D = 500$ km.
We note that while these findings can inform heuristics for determining how many repeaters to put in a network and where to place them, our optimization technique can be applied to different hardware configurations to guide the deployment of specific quantum networks.

\section{Conclusion} \label{sec:conclusion}

In this paper, we have demonstrated that stochastic AD can be used to obtain derivatives in the study of quantum networks and that these derivatives can consequently be put to good use.
To show that the technique can be utilized towards the optimal deployment of quantum networks, we have first tuned rate-fidelity tradeoffs in a repeater chain, demonstrating an increase of a factor of two in the secret-key rate compared to what would be obtained in a more naive, one-dimensional optimization.
Second, we have introduced the problem of choosing repeater locations in a two-dimensional plane.
We have provided explicit solutions, showing where best to deploy a constrained number of quantum repeaters.
Moreover, we have determined how the best obtainable network utility, the minimal number of repeaters required to obtain a benefit, and the number of repeaters at which the network is saturated scale with the physical dimensions of the network.
Additionally, we have performed a sensitivity analysis, demonstrating how stochastic AD can be used to find the most crucial parameters in a network and informing investment decisions on the levels of research and development, network deployment, and network operation.
All the code used to perform these studies has been made open source~\cite{quantum_network_recipes.jl, repeater_placement.jl} and can be reused easily to study and optimize specific network instances.
\\

The simulations we have used were specifically created to model the swap-ASAP protocol (see Section~\ref{sec:model}) in a way that is compatible with StochasticAD.jl~\cite{arya2024}, the Julia implementation of stochastic AD created by the authors of \cite{arya2023}.
An interesting direction for future research is using stochastic AD to study more complicated network protocols.
To that end, it would be particularly useful if stochastic AD was integrated with a general-purpose quantum-network simulator, e.g., \cite{coopmans2021, wu2021b, wallnofer2024, chen2023, satoh2022}.
Doing so may prove challenging though, as stochastic AD is still a somewhat experimental technique.
For instance, we have written our simulation code~\cite{quantum_network_recipes.jl} such that it can also be used to model certain cutoff protocols known to alleviate the effects of memory decoherence~\cite{rozpedek2018a, santra2019, li2021a} in a way that is compatible with StochasticAD.jl.
However, we have observed that actually applying stochastic AD to these protocols gives rise to performance issues and very large variances.
These problems are currently not well understood, and similar problems could very well arise when integrating stochastic AD with a simulator.
Hopefully, future research will resolve these problems; the authors of StochasticAD.jl have been made aware, and a minimal working example based on the cutoff protocol will be used as a benchmark in future development.
\\

Another direction of future research would be to continue the study of repeater placement in the greenfield scenario, i.e., in case repeaters can be freely placed.
A straightforward extension would be to investigate the influence of parameters other than the scale parameter $D$ and the number of repeaters $N$, e.g., one could reproduce Figure~\ref{fig:placement_feature_dependence} but with the elementary-link fidelity or coherence time on the x-axis.
Other interesting questions include the following:
What is the influence of how the end nodes are distributed on the optimal repeater locations and the resulting network performance?
Do solutions look different when another network-utility function is optimized, for example, based on a multi-party network application such as conference-key agreement~\cite{epping2017, murta2020a}?
What if a different repeater protocol or model is used?
How can the repeater locations and network parameters be optimized concurrently, e.g., the bright-state parameters of single-click entanglement generation?
This last problem may be most natural to formulate as a bilevel optimization~\cite{colson2007}.
Furthermore, we did not consider the concurrency of network traffic.
When taking this into account, the network's control-plane protocol that directs that traffic needs to be considered, potentially also resulting in a bilevel optimization.
\\

\paragraph*{Code availability}
Two open-source Julia packages were developed to make the investigations reported on in this paper possible, QuantumNetworkRecipes.jl~\cite{quantum_network_recipes.jl} and RepeaterPlacement.jl~\cite{repeater_placement.jl}.
All the code used for this work has been bundled in a GitHub repository~\cite{code_for_autodiff_paper}, which has been archived at~\cite{avis2025a}.
\\

\paragraph*{Data availability}
The data presented in this work has been made available at~\cite{avis2025b}.

\begin{acknowledgements}

We thank Gaurav Arya, Kenneth Goodenough, and Gayane Vardoyan for useful discussions.
This work was supported by NSF CQN grant number 1941583, NSF grant number 2346089, and NSF grant number 2402861.
\\

\end{acknowledgements}

\bibliography{StochasticAD}

\begin{thebibliography}{130}%
\makeatletter
\providecommand \@ifxundefined [1]{%
 \@ifx{#1\undefined}
}%
\providecommand \@ifnum [1]{%
 \ifnum #1\expandafter \@firstoftwo
 \else \expandafter \@secondoftwo
 \fi
}%
\providecommand \@ifx [1]{%
 \ifx #1\expandafter \@firstoftwo
 \else \expandafter \@secondoftwo
 \fi
}%
\providecommand \natexlab [1]{#1}%
\providecommand \enquote  [1]{``#1''}%
\providecommand \bibnamefont  [1]{#1}%
\providecommand \bibfnamefont [1]{#1}%
\providecommand \citenamefont [1]{#1}%
\providecommand \href@noop [0]{\@secondoftwo}%
\providecommand \href [0]{\begingroup \@sanitize@url \@href}%
\providecommand \@href[1]{\@@startlink{#1}\@@href}%
\providecommand \@@href[1]{\endgroup#1\@@endlink}%
\providecommand \@sanitize@url [0]{\catcode `\\12\catcode `\$12\catcode
  `\&12\catcode `\#12\catcode `\^12\catcode `\_12\catcode `\%12\relax}%
\providecommand \@@startlink[1]{}%
\providecommand \@@endlink[0]{}%
\providecommand \url  [0]{\begingroup\@sanitize@url \@url }%
\providecommand \@url [1]{\endgroup\@href {#1}{\urlprefix }}%
\providecommand \urlprefix  [0]{URL }%
\providecommand \Eprint [0]{\href }%
\providecommand \doibase [0]{https://doi.org/}%
\providecommand \selectlanguage [0]{\@gobble}%
\providecommand \bibinfo  [0]{\@secondoftwo}%
\providecommand \bibfield  [0]{\@secondoftwo}%
\providecommand \translation [1]{[#1]}%
\providecommand \BibitemOpen [0]{}%
\providecommand \bibitemStop [0]{}%
\providecommand \bibitemNoStop [0]{.\EOS\space}%
\providecommand \EOS [0]{\spacefactor3000\relax}%
\providecommand \BibitemShut  [1]{\csname bibitem#1\endcsname}%
\let\auto@bib@innerbib\@empty
%</preamble>
\bibitem [{\citenamefont {Bennett}\ \emph {et~al.}(1992)\citenamefont
  {Bennett}, \citenamefont {Brassard},\ and\ \citenamefont
  {Mermin}}]{bennett1992a}%
  \BibitemOpen
  \bibfield  {author} {\bibinfo {author} {\bibfnamefont {C.~H.}\ \bibnamefont
  {Bennett}}, \bibinfo {author} {\bibfnamefont {G.}~\bibnamefont {Brassard}},\
  and\ \bibinfo {author} {\bibfnamefont {N.~D.}\ \bibnamefont {Mermin}},\
  }\bibfield  {title} {\bibinfo {title} {Quantum cryptography without
  {{Bell}}'s theorem},\ }\href {https://doi.org/10.1103/PhysRevLett.68.557}
  {\bibfield  {journal} {\bibinfo  {journal} {Phys. Rev. Lett.}\ }\textbf
  {\bibinfo {volume} {68}},\ \bibinfo {pages} {557} (\bibinfo {year}
  {1992})}\BibitemShut {NoStop}%
\bibitem [{\citenamefont {Ekert}(1991)}]{ekert1991}%
  \BibitemOpen
  \bibfield  {author} {\bibinfo {author} {\bibfnamefont {A.~K.}\ \bibnamefont
  {Ekert}},\ }\bibfield  {title} {\bibinfo {title} {Quantum cryptography based
  on {{Bell}}'s theorem},\ }\href {https://doi.org/10.1103/PhysRevLett.67.661}
  {\bibfield  {journal} {\bibinfo  {journal} {Phys. Rev. Lett.}\ }\textbf
  {\bibinfo {volume} {67}},\ \bibinfo {pages} {661} (\bibinfo {year}
  {1991})}\BibitemShut {NoStop}%
\bibitem [{\citenamefont {Bennett}\ and\ \citenamefont
  {Brassard}(2014)}]{bennett2014}%
  \BibitemOpen
  \bibfield  {author} {\bibinfo {author} {\bibfnamefont {C.~H.}\ \bibnamefont
  {Bennett}}\ and\ \bibinfo {author} {\bibfnamefont {G.}~\bibnamefont
  {Brassard}},\ }\bibfield  {title} {\bibinfo {title} {Quantum cryptography:
  {{Public}} key distribution and coin tossing},\ }\href
  {https://doi.org/10.1016/j.tcs.2014.05.025} {\bibfield  {journal} {\bibinfo
  {journal} {Theoretical Computer Science}\ }\bibinfo {series} {Theoretical
  {{Aspects}} of {{Quantum Cryptography}} -- Celebrating 30 Years of
  {{BB84}}},\ \textbf {\bibinfo {volume} {560}},\ \bibinfo {pages} {7}
  (\bibinfo {year} {2014})}\BibitemShut {NoStop}%
\bibitem [{\citenamefont {Broadbent}\ \emph {et~al.}(2009)\citenamefont
  {Broadbent}, \citenamefont {Fitzsimons},\ and\ \citenamefont
  {Kashefi}}]{broadbent2009}%
  \BibitemOpen
  \bibfield  {author} {\bibinfo {author} {\bibfnamefont {A.}~\bibnamefont
  {Broadbent}}, \bibinfo {author} {\bibfnamefont {J.}~\bibnamefont
  {Fitzsimons}},\ and\ \bibinfo {author} {\bibfnamefont {E.}~\bibnamefont
  {Kashefi}},\ }\bibfield  {title} {\bibinfo {title} {Universal {{Blind Quantum
  Computation}}},\ }in\ \href {https://doi.org/10.1109/FOCS.2009.36} {\emph
  {\bibinfo {booktitle} {2009 50th {{Annual IEEE Symposium}} on {{Foundations}}
  of {{Computer Science}}}}}\ (\bibinfo {year} {2009})\ pp.\ \bibinfo {pages}
  {517--526}\BibitemShut {NoStop}%
\bibitem [{\citenamefont {Fitzsimons}\ and\ \citenamefont
  {Kashefi}(2017)}]{fitzsimons2017}%
  \BibitemOpen
  \bibfield  {author} {\bibinfo {author} {\bibfnamefont {J.~F.}\ \bibnamefont
  {Fitzsimons}}\ and\ \bibinfo {author} {\bibfnamefont {E.}~\bibnamefont
  {Kashefi}},\ }\bibfield  {title} {\bibinfo {title} {Unconditionally
  verifiable blind quantum computation},\ }\href
  {https://doi.org/10.1103/PhysRevA.96.012303} {\bibfield  {journal} {\bibinfo
  {journal} {Phys. Rev. A}\ }\textbf {\bibinfo {volume} {96}},\ \bibinfo
  {pages} {012303} (\bibinfo {year} {2017})}\BibitemShut {NoStop}%
\bibitem [{\citenamefont {Leichtle}\ \emph {et~al.}(2021)\citenamefont
  {Leichtle}, \citenamefont {Music}, \citenamefont {Kashefi},\ and\
  \citenamefont {Ollivier}}]{leichtle2021}%
  \BibitemOpen
  \bibfield  {author} {\bibinfo {author} {\bibfnamefont {D.}~\bibnamefont
  {Leichtle}}, \bibinfo {author} {\bibfnamefont {L.}~\bibnamefont {Music}},
  \bibinfo {author} {\bibfnamefont {E.}~\bibnamefont {Kashefi}},\ and\ \bibinfo
  {author} {\bibfnamefont {H.}~\bibnamefont {Ollivier}},\ }\bibfield  {title}
  {\bibinfo {title} {Verifying {{BQP Computations}} on {{Noisy Devices}} with
  {{Minimal Overhead}}},\ }\href {https://doi.org/10.1103/PRXQuantum.2.040302}
  {\bibfield  {journal} {\bibinfo  {journal} {PRX Quantum}\ }\textbf {\bibinfo
  {volume} {2}},\ \bibinfo {pages} {040302} (\bibinfo {year}
  {2021})}\BibitemShut {NoStop}%
\bibitem [{\citenamefont {Eldredge}\ \emph {et~al.}(2018)\citenamefont
  {Eldredge}, \citenamefont {{Foss-Feig}}, \citenamefont {Gross}, \citenamefont
  {Rolston},\ and\ \citenamefont {Gorshkov}}]{eldredge2018}%
  \BibitemOpen
  \bibfield  {author} {\bibinfo {author} {\bibfnamefont {Z.}~\bibnamefont
  {Eldredge}}, \bibinfo {author} {\bibfnamefont {M.}~\bibnamefont
  {{Foss-Feig}}}, \bibinfo {author} {\bibfnamefont {J.~A.}\ \bibnamefont
  {Gross}}, \bibinfo {author} {\bibfnamefont {S.~L.}\ \bibnamefont {Rolston}},\
  and\ \bibinfo {author} {\bibfnamefont {A.~V.}\ \bibnamefont {Gorshkov}},\
  }\bibfield  {title} {\bibinfo {title} {Optimal and secure measurement
  protocols for quantum sensor networks},\ }\href
  {https://doi.org/10.1103/PhysRevA.97.042337} {\bibfield  {journal} {\bibinfo
  {journal} {Phys. Rev. A}\ }\textbf {\bibinfo {volume} {97}},\ \bibinfo
  {pages} {042337} (\bibinfo {year} {2018})}\BibitemShut {NoStop}%
\bibitem [{\citenamefont {Proctor}\ \emph {et~al.}(2018)\citenamefont
  {Proctor}, \citenamefont {Knott},\ and\ \citenamefont
  {Dunningham}}]{proctor2018}%
  \BibitemOpen
  \bibfield  {author} {\bibinfo {author} {\bibfnamefont {T.~J.}\ \bibnamefont
  {Proctor}}, \bibinfo {author} {\bibfnamefont {P.~A.}\ \bibnamefont {Knott}},\
  and\ \bibinfo {author} {\bibfnamefont {J.~A.}\ \bibnamefont {Dunningham}},\
  }\bibfield  {title} {\bibinfo {title} {Multiparameter {{Estimation}} in
  {{Networked Quantum Sensors}}},\ }\href
  {https://doi.org/10.1103/PhysRevLett.120.080501} {\bibfield  {journal}
  {\bibinfo  {journal} {Phys. Rev. Lett.}\ }\textbf {\bibinfo {volume} {120}},\
  \bibinfo {pages} {080501} (\bibinfo {year} {2018})}\BibitemShut {NoStop}%
\bibitem [{\citenamefont {K{\'o}m{\'a}r}\ \emph {et~al.}(2014)\citenamefont
  {K{\'o}m{\'a}r}, \citenamefont {Kessler}, \citenamefont {Bishof},
  \citenamefont {Jiang}, \citenamefont {S{\o}rensen}, \citenamefont {Ye},\ and\
  \citenamefont {Lukin}}]{komar2014}%
  \BibitemOpen
  \bibfield  {author} {\bibinfo {author} {\bibfnamefont {P.}~\bibnamefont
  {K{\'o}m{\'a}r}}, \bibinfo {author} {\bibfnamefont {E.~M.}\ \bibnamefont
  {Kessler}}, \bibinfo {author} {\bibfnamefont {M.}~\bibnamefont {Bishof}},
  \bibinfo {author} {\bibfnamefont {L.}~\bibnamefont {Jiang}}, \bibinfo
  {author} {\bibfnamefont {A.~S.}\ \bibnamefont {S{\o}rensen}}, \bibinfo
  {author} {\bibfnamefont {J.}~\bibnamefont {Ye}},\ and\ \bibinfo {author}
  {\bibfnamefont {M.~D.}\ \bibnamefont {Lukin}},\ }\bibfield  {title} {\bibinfo
  {title} {A quantum network of clocks},\ }\href
  {https://doi.org/10.1038/nphys3000} {\bibfield  {journal} {\bibinfo
  {journal} {Nature Physics}\ }\textbf {\bibinfo {volume} {10}},\ \bibinfo
  {pages} {582} (\bibinfo {year} {2014})}\BibitemShut {NoStop}%
\bibitem [{\citenamefont {Grover}(1997)}]{grover1997}%
  \BibitemOpen
  \bibfield  {author} {\bibinfo {author} {\bibfnamefont {L.~K.}\ \bibnamefont
  {Grover}},\ }\href {https://doi.org/10.48550/arXiv.quant-ph/9704012}
  {\bibinfo {title} {Quantum {{Telecomputation}}}} (\bibinfo {year} {1997}),\
  \Eprint {https://arxiv.org/abs/quant-ph/9704012} {arXiv:quant-ph/9704012}
  \BibitemShut {NoStop}%
\bibitem [{\citenamefont {Nickerson}\ \emph {et~al.}(2014)\citenamefont
  {Nickerson}, \citenamefont {Fitzsimons},\ and\ \citenamefont
  {Benjamin}}]{nickerson2014}%
  \BibitemOpen
  \bibfield  {author} {\bibinfo {author} {\bibfnamefont {N.~H.}\ \bibnamefont
  {Nickerson}}, \bibinfo {author} {\bibfnamefont {J.~F.}\ \bibnamefont
  {Fitzsimons}},\ and\ \bibinfo {author} {\bibfnamefont {S.~C.}\ \bibnamefont
  {Benjamin}},\ }\bibfield  {title} {\bibinfo {title} {Freely {{Scalable
  Quantum Technologies}} using {{Cells}} of 5-to-50 {{Qubits}} with {{Very
  Lossy}} and {{Noisy Photonic Links}}},\ }\href
  {https://doi.org/10.1103/PhysRevX.4.041041} {\bibfield  {journal} {\bibinfo
  {journal} {Phys. Rev. X}\ }\textbf {\bibinfo {volume} {4}},\ \bibinfo {pages}
  {041041} (\bibinfo {year} {2014})}\BibitemShut {NoStop}%
\bibitem [{\citenamefont {Cirac}\ \emph {et~al.}(1999)\citenamefont {Cirac},
  \citenamefont {Ekert}, \citenamefont {Huelga},\ and\ \citenamefont
  {Macchiavello}}]{cirac1999}%
  \BibitemOpen
  \bibfield  {author} {\bibinfo {author} {\bibfnamefont {J.~I.}\ \bibnamefont
  {Cirac}}, \bibinfo {author} {\bibfnamefont {A.~K.}\ \bibnamefont {Ekert}},
  \bibinfo {author} {\bibfnamefont {S.~F.}\ \bibnamefont {Huelga}},\ and\
  \bibinfo {author} {\bibfnamefont {C.}~\bibnamefont {Macchiavello}},\
  }\bibfield  {title} {\bibinfo {title} {Distributed quantum computation over
  noisy channels},\ }\href {https://doi.org/10.1103/PhysRevA.59.4249}
  {\bibfield  {journal} {\bibinfo  {journal} {Phys. Rev. A}\ }\textbf {\bibinfo
  {volume} {59}},\ \bibinfo {pages} {4249} (\bibinfo {year}
  {1999})}\BibitemShut {NoStop}%
\bibitem [{\citenamefont {Wehner}\ \emph {et~al.}(2018)\citenamefont {Wehner},
  \citenamefont {Elkouss},\ and\ \citenamefont {Hanson}}]{wehner2018}%
  \BibitemOpen
  \bibfield  {author} {\bibinfo {author} {\bibfnamefont {S.}~\bibnamefont
  {Wehner}}, \bibinfo {author} {\bibfnamefont {D.}~\bibnamefont {Elkouss}},\
  and\ \bibinfo {author} {\bibfnamefont {R.}~\bibnamefont {Hanson}},\
  }\bibfield  {title} {\bibinfo {title} {Quantum internet: {{A}} vision for the
  road ahead},\ }\href {https://doi.org/10.1126/science.aam9288} {\bibfield
  {journal} {\bibinfo  {journal} {Science}\ }\textbf {\bibinfo {volume}
  {362}},\ \bibinfo {pages} {eaam9288} (\bibinfo {year} {2018})}\BibitemShut
  {NoStop}%
\bibitem [{\citenamefont {Azuma}\ \emph {et~al.}(2015)\citenamefont {Azuma},
  \citenamefont {Tamaki},\ and\ \citenamefont {Lo}}]{azuma2015}%
  \BibitemOpen
  \bibfield  {author} {\bibinfo {author} {\bibfnamefont {K.}~\bibnamefont
  {Azuma}}, \bibinfo {author} {\bibfnamefont {K.}~\bibnamefont {Tamaki}},\ and\
  \bibinfo {author} {\bibfnamefont {H.-K.}\ \bibnamefont {Lo}},\ }\bibfield
  {title} {\bibinfo {title} {All-photonic quantum repeaters},\ }\href
  {https://doi.org/10.1038/ncomms7787} {\bibfield  {journal} {\bibinfo
  {journal} {Nature Communications}\ }\textbf {\bibinfo {volume} {6}},\
  \bibinfo {pages} {1} (\bibinfo {year} {2015})}\BibitemShut {NoStop}%
\bibitem [{\citenamefont {Briegel}\ \emph {et~al.}(1998)\citenamefont
  {Briegel}, \citenamefont {D{\"u}r}, \citenamefont {Cirac},\ and\
  \citenamefont {Zoller}}]{briegel1998}%
  \BibitemOpen
  \bibfield  {author} {\bibinfo {author} {\bibfnamefont {H.-J.}\ \bibnamefont
  {Briegel}}, \bibinfo {author} {\bibfnamefont {W.}~\bibnamefont {D{\"u}r}},
  \bibinfo {author} {\bibfnamefont {J.~I.}\ \bibnamefont {Cirac}},\ and\
  \bibinfo {author} {\bibfnamefont {P.}~\bibnamefont {Zoller}},\ }\bibfield
  {title} {\bibinfo {title} {Quantum {{Repeaters}}: {{The Role}} of {{Imperfect
  Local Operations}} in {{Quantum Communication}}},\ }\href
  {https://doi.org/10.1103/PhysRevLett.81.5932} {\bibfield  {journal} {\bibinfo
   {journal} {Phys. Rev. Lett.}\ }\textbf {\bibinfo {volume} {81}},\ \bibinfo
  {pages} {5932} (\bibinfo {year} {1998})}\BibitemShut {NoStop}%
\bibitem [{\citenamefont {Duan}\ \emph {et~al.}(2001)\citenamefont {Duan},
  \citenamefont {Lukin}, \citenamefont {Cirac},\ and\ \citenamefont
  {Zoller}}]{duan2001}%
  \BibitemOpen
  \bibfield  {author} {\bibinfo {author} {\bibfnamefont {L.-M.}\ \bibnamefont
  {Duan}}, \bibinfo {author} {\bibfnamefont {M.}~\bibnamefont {Lukin}},
  \bibinfo {author} {\bibfnamefont {I.}~\bibnamefont {Cirac}},\ and\ \bibinfo
  {author} {\bibfnamefont {P.}~\bibnamefont {Zoller}},\ }\bibfield  {title}
  {\bibinfo {title} {Long-distance quantum communication with atomic ensembles
  and linear optics},\ }\href {https://doi.org/10.1038/35106500} {\bibfield
  {journal} {\bibinfo  {journal} {Nature}\ }\textbf {\bibinfo {volume} {414}},\
  \bibinfo {pages} {413} (\bibinfo {year} {2001})}\BibitemShut {NoStop}%
\bibitem [{\citenamefont {D{\"u}r}\ \emph {et~al.}(1999)\citenamefont
  {D{\"u}r}, \citenamefont {Briegel}, \citenamefont {Cirac},\ and\
  \citenamefont {Zoller}}]{dur1999}%
  \BibitemOpen
  \bibfield  {author} {\bibinfo {author} {\bibfnamefont {W.}~\bibnamefont
  {D{\"u}r}}, \bibinfo {author} {\bibfnamefont {H.-J.}\ \bibnamefont
  {Briegel}}, \bibinfo {author} {\bibfnamefont {J.~I.}\ \bibnamefont {Cirac}},\
  and\ \bibinfo {author} {\bibfnamefont {P.}~\bibnamefont {Zoller}},\
  }\bibfield  {title} {\bibinfo {title} {Quantum repeaters based on
  entanglement purification},\ }\href {https://doi.org/10.1103/PhysRevA.59.169}
  {\bibfield  {journal} {\bibinfo  {journal} {Phys. Rev. A}\ }\textbf {\bibinfo
  {volume} {59}},\ \bibinfo {pages} {169} (\bibinfo {year} {1999})}\BibitemShut
  {NoStop}%
\bibitem [{\citenamefont {Jiang}\ \emph {et~al.}(2009)\citenamefont {Jiang},
  \citenamefont {Taylor}, \citenamefont {Nemoto}, \citenamefont {Munro},
  \citenamefont {Van~Meter},\ and\ \citenamefont {Lukin}}]{jiang2009}%
  \BibitemOpen
  \bibfield  {author} {\bibinfo {author} {\bibfnamefont {L.}~\bibnamefont
  {Jiang}}, \bibinfo {author} {\bibfnamefont {J.~M.}\ \bibnamefont {Taylor}},
  \bibinfo {author} {\bibfnamefont {K.}~\bibnamefont {Nemoto}}, \bibinfo
  {author} {\bibfnamefont {W.~J.}\ \bibnamefont {Munro}}, \bibinfo {author}
  {\bibfnamefont {R.}~\bibnamefont {Van~Meter}},\ and\ \bibinfo {author}
  {\bibfnamefont {M.~D.}\ \bibnamefont {Lukin}},\ }\bibfield  {title} {\bibinfo
  {title} {Quantum repeater with encoding},\ }\href
  {https://doi.org/10.1103/PhysRevA.79.032325} {\bibfield  {journal} {\bibinfo
  {journal} {Phys. Rev. A}\ }\textbf {\bibinfo {volume} {79}},\ \bibinfo
  {pages} {032325} (\bibinfo {year} {2009})}\BibitemShut {NoStop}%
\bibitem [{\citenamefont {Pompili}\ \emph {et~al.}(2021)\citenamefont
  {Pompili}, \citenamefont {Hermans}, \citenamefont {Baier}, \citenamefont
  {Beukers}, \citenamefont {Humphreys}, \citenamefont {Schouten}, \citenamefont
  {Vermeulen}, \citenamefont {Tiggelman}, \citenamefont {{dos Santos Martins}},
  \citenamefont {Dirkse}, \citenamefont {Wehner},\ and\ \citenamefont
  {Hanson}}]{pompili2021}%
  \BibitemOpen
  \bibfield  {author} {\bibinfo {author} {\bibfnamefont {M.}~\bibnamefont
  {Pompili}}, \bibinfo {author} {\bibfnamefont {S.~L.~N.}\ \bibnamefont
  {Hermans}}, \bibinfo {author} {\bibfnamefont {S.}~\bibnamefont {Baier}},
  \bibinfo {author} {\bibfnamefont {H.~K.~C.}\ \bibnamefont {Beukers}},
  \bibinfo {author} {\bibfnamefont {P.~C.}\ \bibnamefont {Humphreys}}, \bibinfo
  {author} {\bibfnamefont {R.~N.}\ \bibnamefont {Schouten}}, \bibinfo {author}
  {\bibfnamefont {R.~F.~L.}\ \bibnamefont {Vermeulen}}, \bibinfo {author}
  {\bibfnamefont {M.~J.}\ \bibnamefont {Tiggelman}}, \bibinfo {author}
  {\bibfnamefont {L.}~\bibnamefont {{dos Santos Martins}}}, \bibinfo {author}
  {\bibfnamefont {B.}~\bibnamefont {Dirkse}}, \bibinfo {author} {\bibfnamefont
  {S.}~\bibnamefont {Wehner}},\ and\ \bibinfo {author} {\bibfnamefont
  {R.}~\bibnamefont {Hanson}},\ }\bibfield  {title} {\bibinfo {title}
  {Realization of a multinode quantum network of remote solid-state qubits},\
  }\href {https://doi.org/10.1126/science.abg1919} {\bibfield  {journal}
  {\bibinfo  {journal} {Science}\ }\textbf {\bibinfo {volume} {372}},\ \bibinfo
  {pages} {259} (\bibinfo {year} {2021})}\BibitemShut {NoStop}%
\bibitem [{\citenamefont {Knaut}\ \emph {et~al.}(2024)\citenamefont {Knaut},
  \citenamefont {Suleymanzade}, \citenamefont {Wei}, \citenamefont {Assumpcao},
  \citenamefont {Stas}, \citenamefont {Huan}, \citenamefont {Machielse},
  \citenamefont {Knall}, \citenamefont {Sutula}, \citenamefont {Baranes},
  \citenamefont {Sinclair}, \citenamefont {{De-Eknamkul}}, \citenamefont
  {Levonian}, \citenamefont {Bhaskar}, \citenamefont {Park}, \citenamefont
  {Lon{\v c}ar},\ and\ \citenamefont {Lukin}}]{knaut2024}%
  \BibitemOpen
  \bibfield  {author} {\bibinfo {author} {\bibfnamefont {C.~M.}\ \bibnamefont
  {Knaut}}, \bibinfo {author} {\bibfnamefont {A.}~\bibnamefont {Suleymanzade}},
  \bibinfo {author} {\bibfnamefont {Y.-C.}\ \bibnamefont {Wei}}, \bibinfo
  {author} {\bibfnamefont {D.~R.}\ \bibnamefont {Assumpcao}}, \bibinfo {author}
  {\bibfnamefont {P.-J.}\ \bibnamefont {Stas}}, \bibinfo {author}
  {\bibfnamefont {Y.~Q.}\ \bibnamefont {Huan}}, \bibinfo {author}
  {\bibfnamefont {B.}~\bibnamefont {Machielse}}, \bibinfo {author}
  {\bibfnamefont {E.~N.}\ \bibnamefont {Knall}}, \bibinfo {author}
  {\bibfnamefont {M.}~\bibnamefont {Sutula}}, \bibinfo {author} {\bibfnamefont
  {G.}~\bibnamefont {Baranes}}, \bibinfo {author} {\bibfnamefont
  {N.}~\bibnamefont {Sinclair}}, \bibinfo {author} {\bibfnamefont
  {C.}~\bibnamefont {{De-Eknamkul}}}, \bibinfo {author} {\bibfnamefont {D.~S.}\
  \bibnamefont {Levonian}}, \bibinfo {author} {\bibfnamefont {M.~K.}\
  \bibnamefont {Bhaskar}}, \bibinfo {author} {\bibfnamefont {H.}~\bibnamefont
  {Park}}, \bibinfo {author} {\bibfnamefont {M.}~\bibnamefont {Lon{\v c}ar}},\
  and\ \bibinfo {author} {\bibfnamefont {M.~D.}\ \bibnamefont {Lukin}},\
  }\bibfield  {title} {\bibinfo {title} {Entanglement of nanophotonic quantum
  memory nodes in a telecom network},\ }\href
  {https://doi.org/10.1038/s41586-024-07252-z} {\bibfield  {journal} {\bibinfo
  {journal} {Nature}\ }\textbf {\bibinfo {volume} {629}},\ \bibinfo {pages}
  {573} (\bibinfo {year} {2024})}\BibitemShut {NoStop}%
\bibitem [{\citenamefont {Stolk}\ \emph {et~al.}(2024)\citenamefont {Stolk},
  \citenamefont {{van der Enden}}, \citenamefont {Slater}, \citenamefont {{te
  Raa-Derckx}}, \citenamefont {Botma}, \citenamefont {{van Rantwijk}},
  \citenamefont {Biemond}, \citenamefont {Hagen}, \citenamefont {Herfst},
  \citenamefont {Koek}, \citenamefont {Meskers}, \citenamefont {Vollmer},
  \citenamefont {{van Zwet}}, \citenamefont {Markham}, \citenamefont {Edmonds},
  \citenamefont {Geus}, \citenamefont {Elsen}, \citenamefont {Jungbluth},
  \citenamefont {Haefner}, \citenamefont {Tresp}, \citenamefont {Stuhler},
  \citenamefont {Ritter},\ and\ \citenamefont {Hanson}}]{stolk2024}%
  \BibitemOpen
  \bibfield  {author} {\bibinfo {author} {\bibfnamefont {A.~J.}\ \bibnamefont
  {Stolk}}, \bibinfo {author} {\bibfnamefont {K.~L.}\ \bibnamefont {{van der
  Enden}}}, \bibinfo {author} {\bibfnamefont {M.-C.}\ \bibnamefont {Slater}},
  \bibinfo {author} {\bibfnamefont {I.}~\bibnamefont {{te Raa-Derckx}}},
  \bibinfo {author} {\bibfnamefont {P.}~\bibnamefont {Botma}}, \bibinfo
  {author} {\bibfnamefont {J.}~\bibnamefont {{van Rantwijk}}}, \bibinfo
  {author} {\bibfnamefont {J.~J.~B.}\ \bibnamefont {Biemond}}, \bibinfo
  {author} {\bibfnamefont {R.~A.~J.}\ \bibnamefont {Hagen}}, \bibinfo {author}
  {\bibfnamefont {R.~W.}\ \bibnamefont {Herfst}}, \bibinfo {author}
  {\bibfnamefont {W.~D.}\ \bibnamefont {Koek}}, \bibinfo {author}
  {\bibfnamefont {A.~J.~H.}\ \bibnamefont {Meskers}}, \bibinfo {author}
  {\bibfnamefont {R.}~\bibnamefont {Vollmer}}, \bibinfo {author} {\bibfnamefont
  {E.~J.}\ \bibnamefont {{van Zwet}}}, \bibinfo {author} {\bibfnamefont
  {M.}~\bibnamefont {Markham}}, \bibinfo {author} {\bibfnamefont {A.~M.}\
  \bibnamefont {Edmonds}}, \bibinfo {author} {\bibfnamefont {J.~F.}\
  \bibnamefont {Geus}}, \bibinfo {author} {\bibfnamefont {F.}~\bibnamefont
  {Elsen}}, \bibinfo {author} {\bibfnamefont {B.}~\bibnamefont {Jungbluth}},
  \bibinfo {author} {\bibfnamefont {C.}~\bibnamefont {Haefner}}, \bibinfo
  {author} {\bibfnamefont {C.}~\bibnamefont {Tresp}}, \bibinfo {author}
  {\bibfnamefont {J.}~\bibnamefont {Stuhler}}, \bibinfo {author} {\bibfnamefont
  {S.}~\bibnamefont {Ritter}},\ and\ \bibinfo {author} {\bibfnamefont
  {R.}~\bibnamefont {Hanson}},\ }\bibfield  {title} {\bibinfo {title}
  {Metropolitan-scale heralded entanglement of solid-state qubits},\ }\href
  {https://doi.org/10.1126/sciadv.adp6442} {\bibfield  {journal} {\bibinfo
  {journal} {Science Advances}\ }\textbf {\bibinfo {volume} {10}},\ \bibinfo
  {pages} {eadp6442} (\bibinfo {year} {2024})}\BibitemShut {NoStop}%
\bibitem [{\citenamefont {Bhaskar}\ \emph {et~al.}(2020)\citenamefont
  {Bhaskar}, \citenamefont {Riedinger}, \citenamefont {Machielse},
  \citenamefont {Levonian}, \citenamefont {Nguyen}, \citenamefont {Knall},
  \citenamefont {Park}, \citenamefont {Englund}, \citenamefont {Lon{\v c}ar},
  \citenamefont {Sukachev},\ and\ \citenamefont {Lukin}}]{bhaskar2020}%
  \BibitemOpen
  \bibfield  {author} {\bibinfo {author} {\bibfnamefont {M.~K.}\ \bibnamefont
  {Bhaskar}}, \bibinfo {author} {\bibfnamefont {R.}~\bibnamefont {Riedinger}},
  \bibinfo {author} {\bibfnamefont {B.}~\bibnamefont {Machielse}}, \bibinfo
  {author} {\bibfnamefont {D.~S.}\ \bibnamefont {Levonian}}, \bibinfo {author}
  {\bibfnamefont {C.~T.}\ \bibnamefont {Nguyen}}, \bibinfo {author}
  {\bibfnamefont {E.~N.}\ \bibnamefont {Knall}}, \bibinfo {author}
  {\bibfnamefont {H.}~\bibnamefont {Park}}, \bibinfo {author} {\bibfnamefont
  {D.}~\bibnamefont {Englund}}, \bibinfo {author} {\bibfnamefont
  {M.}~\bibnamefont {Lon{\v c}ar}}, \bibinfo {author} {\bibfnamefont {D.~D.}\
  \bibnamefont {Sukachev}},\ and\ \bibinfo {author} {\bibfnamefont {M.~D.}\
  \bibnamefont {Lukin}},\ }\bibfield  {title} {\bibinfo {title} {Experimental
  demonstration of memory-enhanced quantum communication},\ }\href
  {https://doi.org/10.1038/s41586-020-2103-5} {\bibfield  {journal} {\bibinfo
  {journal} {Nature}\ }\textbf {\bibinfo {volume} {580}},\ \bibinfo {pages}
  {60} (\bibinfo {year} {2020})}\BibitemShut {NoStop}%
\bibitem [{\citenamefont {Langenfeld}\ \emph {et~al.}(2021)\citenamefont
  {Langenfeld}, \citenamefont {Thomas}, \citenamefont {Morin},\ and\
  \citenamefont {Rempe}}]{langenfeld2021}%
  \BibitemOpen
  \bibfield  {author} {\bibinfo {author} {\bibfnamefont {S.}~\bibnamefont
  {Langenfeld}}, \bibinfo {author} {\bibfnamefont {P.}~\bibnamefont {Thomas}},
  \bibinfo {author} {\bibfnamefont {O.}~\bibnamefont {Morin}},\ and\ \bibinfo
  {author} {\bibfnamefont {G.}~\bibnamefont {Rempe}},\ }\bibfield  {title}
  {\bibinfo {title} {Quantum {{Repeater Node Demonstrating Unconditionally
  Secure Key Distribution}}},\ }\href
  {https://doi.org/10.1103/PhysRevLett.126.230506} {\bibfield  {journal}
  {\bibinfo  {journal} {Phys. Rev. Lett.}\ }\textbf {\bibinfo {volume} {126}},\
  \bibinfo {pages} {230506} (\bibinfo {year} {2021})}\BibitemShut {NoStop}%
\bibitem [{\citenamefont {Goodenough}\ \emph {et~al.}(2021)\citenamefont
  {Goodenough}, \citenamefont {Elkouss},\ and\ \citenamefont
  {Wehner}}]{goodenough2021}%
  \BibitemOpen
  \bibfield  {author} {\bibinfo {author} {\bibfnamefont {K.}~\bibnamefont
  {Goodenough}}, \bibinfo {author} {\bibfnamefont {D.}~\bibnamefont
  {Elkouss}},\ and\ \bibinfo {author} {\bibfnamefont {S.}~\bibnamefont
  {Wehner}},\ }\bibfield  {title} {\bibinfo {title} {Optimizing repeater
  schemes for the quantum internet},\ }\href
  {https://doi.org/10.1103/PhysRevA.103.032610} {\bibfield  {journal} {\bibinfo
   {journal} {Phys. Rev. A}\ }\textbf {\bibinfo {volume} {103}},\ \bibinfo
  {pages} {032610} (\bibinfo {year} {2021})}\BibitemShut {NoStop}%
\bibitem [{\citenamefont {Vardoyan}\ and\ \citenamefont
  {Wehner}(2023)}]{vardoyan2023}%
  \BibitemOpen
  \bibfield  {author} {\bibinfo {author} {\bibfnamefont {G.}~\bibnamefont
  {Vardoyan}}\ and\ \bibinfo {author} {\bibfnamefont {S.}~\bibnamefont
  {Wehner}},\ }\bibfield  {title} {\bibinfo {title} {Quantum {{Network Utility
  Maximization}}},\ }in\ \href {https://doi.org/10.1109/QCE57702.2023.00140}
  {\emph {\bibinfo {booktitle} {2023 {{IEEE International Conference}} on
  {{Quantum Computing}} and {{Engineering}} ({{QCE}})}}},\ Vol.~\bibinfo
  {volume} {01}\ (\bibinfo {year} {2023})\ pp.\ \bibinfo {pages}
  {1238--1248}\BibitemShut {NoStop}%
\bibitem [{\citenamefont {Krastanov}\ \emph {et~al.}(2019)\citenamefont
  {Krastanov}, \citenamefont {Albert},\ and\ \citenamefont
  {Jiang}}]{krastanov2019}%
  \BibitemOpen
  \bibfield  {author} {\bibinfo {author} {\bibfnamefont {S.}~\bibnamefont
  {Krastanov}}, \bibinfo {author} {\bibfnamefont {V.~V.}\ \bibnamefont
  {Albert}},\ and\ \bibinfo {author} {\bibfnamefont {L.}~\bibnamefont
  {Jiang}},\ }\bibfield  {title} {\bibinfo {title} {Optimized {{Entanglement
  Purification}}},\ }\href {https://doi.org/10.22331/q-2019-02-18-123}
  {\bibfield  {journal} {\bibinfo  {journal} {Quantum}\ }\textbf {\bibinfo
  {volume} {3}},\ \bibinfo {pages} {123} (\bibinfo {year} {2019})}\BibitemShut
  {NoStop}%
\bibitem [{\citenamefont {Rozp{\k e}dek}\ \emph
  {et~al.}(2018{\natexlab{a}})\citenamefont {Rozp{\k e}dek}, \citenamefont
  {Schiet}, \citenamefont {Thinh}, \citenamefont {Elkouss}, \citenamefont
  {Doherty},\ and\ \citenamefont {Wehner}}]{rozpedek2018}%
  \BibitemOpen
  \bibfield  {author} {\bibinfo {author} {\bibfnamefont {F.}~\bibnamefont
  {Rozp{\k e}dek}}, \bibinfo {author} {\bibfnamefont {T.}~\bibnamefont
  {Schiet}}, \bibinfo {author} {\bibfnamefont {L.~P.}\ \bibnamefont {Thinh}},
  \bibinfo {author} {\bibfnamefont {D.}~\bibnamefont {Elkouss}}, \bibinfo
  {author} {\bibfnamefont {A.~C.}\ \bibnamefont {Doherty}},\ and\ \bibinfo
  {author} {\bibfnamefont {S.}~\bibnamefont {Wehner}},\ }\bibfield  {title}
  {\bibinfo {title} {Optimizing practical entanglement distillation},\ }\href
  {https://doi.org/10.1103/PhysRevA.97.062333} {\bibfield  {journal} {\bibinfo
  {journal} {Phys. Rev. A}\ }\textbf {\bibinfo {volume} {97}},\ \bibinfo
  {pages} {062333} (\bibinfo {year} {2018}{\natexlab{a}})}\BibitemShut
  {NoStop}%
\bibitem [{\citenamefont {Jiang}\ \emph {et~al.}(2007)\citenamefont {Jiang},
  \citenamefont {Taylor}, \citenamefont {Khaneja},\ and\ \citenamefont
  {Lukin}}]{jiang2007a}%
  \BibitemOpen
  \bibfield  {author} {\bibinfo {author} {\bibfnamefont {L.}~\bibnamefont
  {Jiang}}, \bibinfo {author} {\bibfnamefont {J.~M.}\ \bibnamefont {Taylor}},
  \bibinfo {author} {\bibfnamefont {N.}~\bibnamefont {Khaneja}},\ and\ \bibinfo
  {author} {\bibfnamefont {M.~D.}\ \bibnamefont {Lukin}},\ }\bibfield  {title}
  {\bibinfo {title} {Optimal approach to quantum communication using dynamic
  programming},\ }\href {https://doi.org/10.1073/pnas.0703284104} {\bibfield
  {journal} {\bibinfo  {journal} {Proceedings of the National Academy of
  Sciences}\ }\textbf {\bibinfo {volume} {104}},\ \bibinfo {pages} {17291}
  (\bibinfo {year} {2007})}\BibitemShut {NoStop}%
\bibitem [{\citenamefont {Haldar}\ \emph {et~al.}(2025)\citenamefont {Haldar},
  \citenamefont {Barge}, \citenamefont {Cheng}, \citenamefont {Chang},
  \citenamefont {Kirby}, \citenamefont {Khatri}, \citenamefont {Wong},\ and\
  \citenamefont {Lee}}]{haldar2025}%
  \BibitemOpen
  \bibfield  {author} {\bibinfo {author} {\bibfnamefont {S.}~\bibnamefont
  {Haldar}}, \bibinfo {author} {\bibfnamefont {P.~J.}\ \bibnamefont {Barge}},
  \bibinfo {author} {\bibfnamefont {X.}~\bibnamefont {Cheng}}, \bibinfo
  {author} {\bibfnamefont {K.-C.}\ \bibnamefont {Chang}}, \bibinfo {author}
  {\bibfnamefont {B.~T.}\ \bibnamefont {Kirby}}, \bibinfo {author}
  {\bibfnamefont {S.}~\bibnamefont {Khatri}}, \bibinfo {author} {\bibfnamefont
  {C.~W.}\ \bibnamefont {Wong}},\ and\ \bibinfo {author} {\bibfnamefont
  {H.}~\bibnamefont {Lee}},\ }\bibfield  {title} {\bibinfo {title} {Reducing
  classical communication costs in multiplexed quantum repeaters using
  hardware-aware quasi-local policies},\ }\href
  {https://doi.org/10.1038/s42005-025-02029-w} {\bibfield  {journal} {\bibinfo
  {journal} {Commun Phys}\ }\textbf {\bibinfo {volume} {8}},\ \bibinfo {pages}
  {1} (\bibinfo {year} {2025})}\BibitemShut {NoStop}%
\bibitem [{\citenamefont {Haldar}\ \emph {et~al.}(2024)\citenamefont {Haldar},
  \citenamefont {Barge}, \citenamefont {Khatri},\ and\ \citenamefont
  {Lee}}]{haldar2024a}%
  \BibitemOpen
  \bibfield  {author} {\bibinfo {author} {\bibfnamefont {S.}~\bibnamefont
  {Haldar}}, \bibinfo {author} {\bibfnamefont {P.~J.}\ \bibnamefont {Barge}},
  \bibinfo {author} {\bibfnamefont {S.}~\bibnamefont {Khatri}},\ and\ \bibinfo
  {author} {\bibfnamefont {H.}~\bibnamefont {Lee}},\ }\bibfield  {title}
  {\bibinfo {title} {Fast and reliable entanglement distribution with quantum
  repeaters: {{Principles}} for improving protocols using reinforcement
  learning},\ }\href {https://doi.org/10.1103/PhysRevApplied.21.024041}
  {\bibfield  {journal} {\bibinfo  {journal} {Phys. Rev. Appl.}\ }\textbf
  {\bibinfo {volume} {21}},\ \bibinfo {pages} {024041} (\bibinfo {year}
  {2024})}\BibitemShut {NoStop}%
\bibitem [{\citenamefont {Li}\ \emph {et~al.}(2021)\citenamefont {Li},
  \citenamefont {Coopmans},\ and\ \citenamefont {Elkouss}}]{li2021a}%
  \BibitemOpen
  \bibfield  {author} {\bibinfo {author} {\bibfnamefont {B.}~\bibnamefont
  {Li}}, \bibinfo {author} {\bibfnamefont {T.}~\bibnamefont {Coopmans}},\ and\
  \bibinfo {author} {\bibfnamefont {D.}~\bibnamefont {Elkouss}},\ }\bibfield
  {title} {\bibinfo {title} {Efficient {{Optimization}} of {{Cutoffs}} in
  {{Quantum Repeater Chains}}},\ }\href
  {https://doi.org/10.1109/TQE.2021.3099003} {\bibfield  {journal} {\bibinfo
  {journal} {IEEE Transactions on Quantum Engineering}\ }\textbf {\bibinfo
  {volume} {2}},\ \bibinfo {pages} {1} (\bibinfo {year} {2021})}\BibitemShut
  {NoStop}%
\bibitem [{\citenamefont {Pouryousef}\ \emph {et~al.}(2023)\citenamefont
  {Pouryousef}, \citenamefont {Panigrahy}, \citenamefont {Purkayastha},
  \citenamefont {Mukhopadhyay}, \citenamefont {Grammel}, \citenamefont
  {Di~Mola},\ and\ \citenamefont {Towsley}}]{pouryousef2023}%
  \BibitemOpen
  \bibfield  {author} {\bibinfo {author} {\bibfnamefont {S.}~\bibnamefont
  {Pouryousef}}, \bibinfo {author} {\bibfnamefont {N.~K.}\ \bibnamefont
  {Panigrahy}}, \bibinfo {author} {\bibfnamefont {M.~D.}\ \bibnamefont
  {Purkayastha}}, \bibinfo {author} {\bibfnamefont {S.}~\bibnamefont
  {Mukhopadhyay}}, \bibinfo {author} {\bibfnamefont {G.}~\bibnamefont
  {Grammel}}, \bibinfo {author} {\bibfnamefont {D.}~\bibnamefont {Di~Mola}},\
  and\ \bibinfo {author} {\bibfnamefont {D.}~\bibnamefont {Towsley}},\
  }\bibfield  {title} {\bibinfo {title} {Resource {{Management}} in {{Quantum
  Virtual Private Networks}}},\ }in\ \href
  {https://doi.org/10.1109/QCE57702.2023.10298} {\emph {\bibinfo {booktitle}
  {2023 {{IEEE International Conference}} on {{Quantum Computing}} and
  {{Engineering}} ({{QCE}})}}},\ Vol.~\bibinfo {volume} {02}\ (\bibinfo {year}
  {2023})\ pp.\ \bibinfo {pages} {393--394}\BibitemShut {NoStop}%
\bibitem [{\citenamefont {Pathumsoot}\ \emph {et~al.}(2021)\citenamefont
  {Pathumsoot}, \citenamefont {Benchasattabuse}, \citenamefont {Satoh},
  \citenamefont {Hajdu{\v s}ek}, \citenamefont {Van~Meter},\ and\ \citenamefont
  {Suwanna}}]{pathumsoot2021}%
  \BibitemOpen
  \bibfield  {author} {\bibinfo {author} {\bibfnamefont {P.}~\bibnamefont
  {Pathumsoot}}, \bibinfo {author} {\bibfnamefont {N.}~\bibnamefont
  {Benchasattabuse}}, \bibinfo {author} {\bibfnamefont {R.}~\bibnamefont
  {Satoh}}, \bibinfo {author} {\bibfnamefont {M.}~\bibnamefont {Hajdu{\v
  s}ek}}, \bibinfo {author} {\bibfnamefont {R.}~\bibnamefont {Van~Meter}},\
  and\ \bibinfo {author} {\bibfnamefont {S.}~\bibnamefont {Suwanna}},\
  }\bibfield  {title} {\bibinfo {title} {Optimizing {{Link-Level Entanglement
  Generation}} in {{Quantum Networks}} with {{Unequal Link Lengths}}},\ }in\
  \href {https://doi.org/10.1109/ICSEC53205.2021.9684634} {\emph {\bibinfo
  {booktitle} {2021 25th {{International Computer Science}} and {{Engineering
  Conference}} ({{ICSEC}})}}}\ (\bibinfo  {publisher} {IEEE},\ \bibinfo
  {address} {Chiang Rai, Thailand},\ \bibinfo {year} {2021})\ pp.\ \bibinfo
  {pages} {179--184}\BibitemShut {NoStop}%
\bibitem [{\citenamefont {Avis}\ \emph {et~al.}(2023)\citenamefont {Avis},
  \citenamefont {{Ferreira da Silva}}, \citenamefont {Coopmans}, \citenamefont
  {Dahlberg}, \citenamefont {Jirovsk{\'a}}, \citenamefont {Maier},
  \citenamefont {Rabbie}, \citenamefont {{Torres-Knoop}},\ and\ \citenamefont
  {Wehner}}]{avis2022a}%
  \BibitemOpen
  \bibfield  {author} {\bibinfo {author} {\bibfnamefont {G.}~\bibnamefont
  {Avis}}, \bibinfo {author} {\bibfnamefont {F.}~\bibnamefont {{Ferreira da
  Silva}}}, \bibinfo {author} {\bibfnamefont {T.}~\bibnamefont {Coopmans}},
  \bibinfo {author} {\bibfnamefont {A.}~\bibnamefont {Dahlberg}}, \bibinfo
  {author} {\bibfnamefont {H.}~\bibnamefont {Jirovsk{\'a}}}, \bibinfo {author}
  {\bibfnamefont {D.}~\bibnamefont {Maier}}, \bibinfo {author} {\bibfnamefont
  {J.}~\bibnamefont {Rabbie}}, \bibinfo {author} {\bibfnamefont
  {A.}~\bibnamefont {{Torres-Knoop}}},\ and\ \bibinfo {author} {\bibfnamefont
  {S.}~\bibnamefont {Wehner}},\ }\bibfield  {title} {\bibinfo {title}
  {Requirements for a processing-node quantum repeater on a real-world fiber
  grid},\ }\href {https://doi.org/10.1038/s41534-023-00765-x} {\bibfield
  {journal} {\bibinfo  {journal} {npj Quantum Inf}\ }\textbf {\bibinfo {volume}
  {9}},\ \bibinfo {pages} {1} (\bibinfo {year} {2023})}\BibitemShut {NoStop}%
\bibitem [{\citenamefont {da~Silva}\ \emph {et~al.}(2024)\citenamefont
  {da~Silva}, \citenamefont {Avis}, \citenamefont {Slater},\ and\ \citenamefont
  {Wehner}}]{silva2024}%
  \BibitemOpen
  \bibfield  {author} {\bibinfo {author} {\bibfnamefont {F.~F.}\ \bibnamefont
  {da~Silva}}, \bibinfo {author} {\bibfnamefont {G.}~\bibnamefont {Avis}},
  \bibinfo {author} {\bibfnamefont {J.~A.}\ \bibnamefont {Slater}},\ and\
  \bibinfo {author} {\bibfnamefont {S.}~\bibnamefont {Wehner}},\ }\bibfield
  {title} {\bibinfo {title} {Requirements for upgrading trusted nodes to a
  repeater chain over 900 km of optical fiber},\ }\href
  {https://doi.org/10.1088/2058-9565/ad7499} {\bibfield  {journal} {\bibinfo
  {journal} {Quantum Sci. Technol.}\ }\textbf {\bibinfo {volume} {9}},\
  \bibinfo {pages} {045041} (\bibinfo {year} {2024})}\BibitemShut {NoStop}%
\bibitem [{\citenamefont {da~Silva}\ \emph {et~al.}(2021)\citenamefont
  {da~Silva}, \citenamefont {{Torres-Knoop}}, \citenamefont {Coopmans},
  \citenamefont {Maier},\ and\ \citenamefont {Wehner}}]{silva2021}%
  \BibitemOpen
  \bibfield  {author} {\bibinfo {author} {\bibfnamefont {F.~F.}\ \bibnamefont
  {da~Silva}}, \bibinfo {author} {\bibfnamefont {A.}~\bibnamefont
  {{Torres-Knoop}}}, \bibinfo {author} {\bibfnamefont {T.}~\bibnamefont
  {Coopmans}}, \bibinfo {author} {\bibfnamefont {D.}~\bibnamefont {Maier}},\
  and\ \bibinfo {author} {\bibfnamefont {S.}~\bibnamefont {Wehner}},\
  }\bibfield  {title} {\bibinfo {title} {Optimizing entanglement generation and
  distribution using genetic algorithms},\ }\href
  {https://doi.org/10.1088/2058-9565/abfc93} {\bibfield  {journal} {\bibinfo
  {journal} {Quantum Sci. Technol.}\ }\textbf {\bibinfo {volume} {6}},\
  \bibinfo {pages} {035007} (\bibinfo {year} {2021})}\BibitemShut {NoStop}%
\bibitem [{\citenamefont {{van Dam}}\ \emph {et~al.}(2024)\citenamefont {{van
  Dam}}, \citenamefont {Avis}, \citenamefont {Propp}, \citenamefont {{Ferreira
  da Silva}}, \citenamefont {Slater}, \citenamefont {Northup},\ and\
  \citenamefont {Wehner}}]{vandam2024}%
  \BibitemOpen
  \bibfield  {author} {\bibinfo {author} {\bibfnamefont {J.}~\bibnamefont {{van
  Dam}}}, \bibinfo {author} {\bibfnamefont {G.}~\bibnamefont {Avis}}, \bibinfo
  {author} {\bibfnamefont {T.~B.}\ \bibnamefont {Propp}}, \bibinfo {author}
  {\bibfnamefont {F.}~\bibnamefont {{Ferreira da Silva}}}, \bibinfo {author}
  {\bibfnamefont {J.~A.}\ \bibnamefont {Slater}}, \bibinfo {author}
  {\bibfnamefont {T.~E.}\ \bibnamefont {Northup}},\ and\ \bibinfo {author}
  {\bibfnamefont {S.}~\bibnamefont {Wehner}},\ }\bibfield  {title} {\bibinfo
  {title} {Hardware requirements for trapped-ion-based verifiable blind quantum
  computing with a measurement-only client},\ }\href
  {https://doi.org/10.1088/2058-9565/ad6eb2} {\bibfield  {journal} {\bibinfo
  {journal} {Quantum Sci. Technol.}\ }\textbf {\bibinfo {volume} {9}},\
  \bibinfo {pages} {045031} (\bibinfo {year} {2024})}\BibitemShut {NoStop}%
\bibitem [{\citenamefont {Mora}\ \emph {et~al.}(2024)\citenamefont {Mora},
  \citenamefont {da~Silva},\ and\ \citenamefont {Wehner}}]{mora2024}%
  \BibitemOpen
  \bibfield  {author} {\bibinfo {author} {\bibfnamefont {A.~L.}\ \bibnamefont
  {Mora}}, \bibinfo {author} {\bibfnamefont {F.~F.}\ \bibnamefont {da~Silva}},\
  and\ \bibinfo {author} {\bibfnamefont {S.}~\bibnamefont {Wehner}},\
  }\bibfield  {title} {\bibinfo {title} {Reducing hardware requirements for
  entanglement distribution via joint hardware-protocol optimization},\ }\href
  {https://doi.org/10.1088/2058-9565/ad57e9} {\bibfield  {journal} {\bibinfo
  {journal} {Quantum Sci. Technol.}\ }\textbf {\bibinfo {volume} {9}},\
  \bibinfo {pages} {045001} (\bibinfo {year} {2024})}\BibitemShut {NoStop}%
\bibitem [{\citenamefont {Rabbie}\ \emph {et~al.}(2022)\citenamefont {Rabbie},
  \citenamefont {Chakraborty}, \citenamefont {Avis},\ and\ \citenamefont
  {Wehner}}]{rabbie2022}%
  \BibitemOpen
  \bibfield  {author} {\bibinfo {author} {\bibfnamefont {J.}~\bibnamefont
  {Rabbie}}, \bibinfo {author} {\bibfnamefont {K.}~\bibnamefont {Chakraborty}},
  \bibinfo {author} {\bibfnamefont {G.}~\bibnamefont {Avis}},\ and\ \bibinfo
  {author} {\bibfnamefont {S.}~\bibnamefont {Wehner}},\ }\bibfield  {title}
  {\bibinfo {title} {Designing {{Quantum Networks Using Preexisting
  Infrastructure}}},\ }\href {https://doi.org/10.1038/s41534-021-00501-3}
  {\bibfield  {journal} {\bibinfo  {journal} {npj Quantum Inf}\ }\textbf
  {\bibinfo {volume} {8}},\ \bibinfo {pages} {5} (\bibinfo {year}
  {2022})}\BibitemShut {NoStop}%
\bibitem [{\citenamefont {Islam}\ and\ \citenamefont
  {Arslan}(2023)}]{islam2023}%
  \BibitemOpen
  \bibfield  {author} {\bibinfo {author} {\bibfnamefont {T.}~\bibnamefont
  {Islam}}\ and\ \bibinfo {author} {\bibfnamefont {E.}~\bibnamefont {Arslan}},\
  }\bibfield  {title} {\bibinfo {title} {A {{Heuristic Approach}} for
  {{Scalable Quantum Repeater Deployment Modeling}}},\ }in\ \href
  {https://doi.org/10.1109/LCN58197.2023.10223375} {\emph {\bibinfo {booktitle}
  {2023 {{IEEE}} 48th {{Conference}} on {{Local Computer Networks}}
  ({{LCN}})}}}\ (\bibinfo {year} {2023})\ pp.\ \bibinfo {pages}
  {1--9}\BibitemShut {NoStop}%
\bibitem [{\citenamefont {Pouryousef}\ \emph {et~al.}(2024)\citenamefont
  {Pouryousef}, \citenamefont {Shapourian}, \citenamefont {Shabani},
  \citenamefont {Kompella},\ and\ \citenamefont {Towsley}}]{pouryousef2024a}%
  \BibitemOpen
  \bibfield  {author} {\bibinfo {author} {\bibfnamefont {S.}~\bibnamefont
  {Pouryousef}}, \bibinfo {author} {\bibfnamefont {H.}~\bibnamefont
  {Shapourian}}, \bibinfo {author} {\bibfnamefont {A.}~\bibnamefont {Shabani}},
  \bibinfo {author} {\bibfnamefont {R.}~\bibnamefont {Kompella}},\ and\
  \bibinfo {author} {\bibfnamefont {D.}~\bibnamefont {Towsley}},\ }\bibfield
  {title} {\bibinfo {title} {Resource {{Placement}} for {{Rate}} and {{Fidelity
  Maximization}} in {{Quantum Networks}}},\ }\href
  {https://doi.org/10.1109/TQE.2024.3432390} {\bibfield  {journal} {\bibinfo
  {journal} {IEEE Transactions on Quantum Engineering}\ }\textbf {\bibinfo
  {volume} {5}},\ \bibinfo {pages} {1} (\bibinfo {year} {2024})}\BibitemShut
  {NoStop}%
\bibitem [{\citenamefont {Sripotchanart}\ \emph {et~al.}(2024)\citenamefont
  {Sripotchanart}, \citenamefont {Si}, \citenamefont {Calheiros}, \citenamefont
  {Cao},\ and\ \citenamefont {Qiu}}]{sripotchanart2024}%
  \BibitemOpen
  \bibfield  {author} {\bibinfo {author} {\bibfnamefont {R.}~\bibnamefont
  {Sripotchanart}}, \bibinfo {author} {\bibfnamefont {W.}~\bibnamefont {Si}},
  \bibinfo {author} {\bibfnamefont {R.~N.}\ \bibnamefont {Calheiros}}, \bibinfo
  {author} {\bibfnamefont {Q.}~\bibnamefont {Cao}},\ and\ \bibinfo {author}
  {\bibfnamefont {T.}~\bibnamefont {Qiu}},\ }\bibfield  {title} {\bibinfo
  {title} {A two-step linear programming approach for repeater placement in
  large-scale quantum networks},\ }\href
  {https://doi.org/10.1016/j.comnet.2024.110795} {\bibfield  {journal}
  {\bibinfo  {journal} {Computer Networks}\ }\textbf {\bibinfo {volume}
  {254}},\ \bibinfo {pages} {110795} (\bibinfo {year} {2024})}\BibitemShut
  {NoStop}%
\bibitem [{\citenamefont {Avis}\ \emph {et~al.}(2024)\citenamefont {Avis},
  \citenamefont {Knegjens}, \citenamefont {S{\o}rensen},\ and\ \citenamefont
  {Wehner}}]{avis2024}%
  \BibitemOpen
  \bibfield  {author} {\bibinfo {author} {\bibfnamefont {G.}~\bibnamefont
  {Avis}}, \bibinfo {author} {\bibfnamefont {R.}~\bibnamefont {Knegjens}},
  \bibinfo {author} {\bibfnamefont {A.~S.}\ \bibnamefont {S{\o}rensen}},\ and\
  \bibinfo {author} {\bibfnamefont {S.}~\bibnamefont {Wehner}},\ }\bibfield
  {title} {\bibinfo {title} {Asymmetric node placement in fiber-based quantum
  networks},\ }\href {https://doi.org/10.1103/PhysRevA.109.052627} {\bibfield
  {journal} {\bibinfo  {journal} {Phys. Rev. A}\ }\textbf {\bibinfo {volume}
  {109}},\ \bibinfo {pages} {052627} (\bibinfo {year} {2024})}\BibitemShut
  {NoStop}%
\bibitem [{\citenamefont {Chehimi}\ \emph {et~al.}(2023)\citenamefont
  {Chehimi}, \citenamefont {Pouryousef}, \citenamefont {Panigrahy},
  \citenamefont {Towsley},\ and\ \citenamefont {Saad}}]{chehimi2023}%
  \BibitemOpen
  \bibfield  {author} {\bibinfo {author} {\bibfnamefont {M.}~\bibnamefont
  {Chehimi}}, \bibinfo {author} {\bibfnamefont {S.}~\bibnamefont {Pouryousef}},
  \bibinfo {author} {\bibfnamefont {N.~K.}\ \bibnamefont {Panigrahy}}, \bibinfo
  {author} {\bibfnamefont {D.}~\bibnamefont {Towsley}},\ and\ \bibinfo {author}
  {\bibfnamefont {W.}~\bibnamefont {Saad}},\ }\bibfield  {title} {\bibinfo
  {title} {Scaling {{Limits}} of {{Quantum Repeater Networks}}},\ }in\ \href
  {https://doi.org/10.1109/QCE57702.2023.00136} {\emph {\bibinfo {booktitle}
  {2023 {{IEEE International Conference}} on {{Quantum Computing}} and
  {{Engineering}} ({{QCE}})}}},\ Vol.~\bibinfo {volume} {01}\ (\bibinfo {year}
  {2023})\ pp.\ \bibinfo {pages} {1205--1210}\BibitemShut {NoStop}%
\bibitem [{\citenamefont {Dhara}\ \emph {et~al.}(2021)\citenamefont {Dhara},
  \citenamefont {Patil}, \citenamefont {Krovi},\ and\ \citenamefont
  {Guha}}]{dhara2021a}%
  \BibitemOpen
  \bibfield  {author} {\bibinfo {author} {\bibfnamefont {P.}~\bibnamefont
  {Dhara}}, \bibinfo {author} {\bibfnamefont {A.}~\bibnamefont {Patil}},
  \bibinfo {author} {\bibfnamefont {H.}~\bibnamefont {Krovi}},\ and\ \bibinfo
  {author} {\bibfnamefont {S.}~\bibnamefont {Guha}},\ }\bibfield  {title}
  {\bibinfo {title} {Subexponential rate versus distance with time-multiplexed
  quantum repeaters},\ }\href {https://doi.org/10.1103/PhysRevA.104.052612}
  {\bibfield  {journal} {\bibinfo  {journal} {Phys. Rev. A}\ }\textbf {\bibinfo
  {volume} {104}},\ \bibinfo {pages} {052612} (\bibinfo {year}
  {2021})}\BibitemShut {NoStop}%
\bibitem [{\citenamefont {Luong}\ \emph {et~al.}(2016)\citenamefont {Luong},
  \citenamefont {Jiang}, \citenamefont {Kim},\ and\ \citenamefont
  {L{\"u}tkenhaus}}]{luong2016}%
  \BibitemOpen
  \bibfield  {author} {\bibinfo {author} {\bibfnamefont {D.}~\bibnamefont
  {Luong}}, \bibinfo {author} {\bibfnamefont {L.}~\bibnamefont {Jiang}},
  \bibinfo {author} {\bibfnamefont {J.}~\bibnamefont {Kim}},\ and\ \bibinfo
  {author} {\bibfnamefont {N.}~\bibnamefont {L{\"u}tkenhaus}},\ }\bibfield
  {title} {\bibinfo {title} {Overcoming lossy channel bounds using a single
  quantum repeater node},\ }\href {https://doi.org/10.1007/s00340-016-6373-4}
  {\bibfield  {journal} {\bibinfo  {journal} {Appl. Phys. B}\ }\textbf
  {\bibinfo {volume} {122}},\ \bibinfo {pages} {96} (\bibinfo {year}
  {2016})}\BibitemShut {NoStop}%
\bibitem [{\citenamefont {Rozp{\k e}dek}\ \emph
  {et~al.}(2018{\natexlab{b}})\citenamefont {Rozp{\k e}dek}, \citenamefont
  {Goodenough}, \citenamefont {Ribeiro}, \citenamefont {Kalb}, \citenamefont
  {Vivoli}, \citenamefont {Reiserer}, \citenamefont {Hanson}, \citenamefont
  {Wehner},\ and\ \citenamefont {Elkouss}}]{rozpedek2018a}%
  \BibitemOpen
  \bibfield  {author} {\bibinfo {author} {\bibfnamefont {F.}~\bibnamefont
  {Rozp{\k e}dek}}, \bibinfo {author} {\bibfnamefont {K.}~\bibnamefont
  {Goodenough}}, \bibinfo {author} {\bibfnamefont {J.}~\bibnamefont {Ribeiro}},
  \bibinfo {author} {\bibfnamefont {N.}~\bibnamefont {Kalb}}, \bibinfo {author}
  {\bibfnamefont {V.~C.}\ \bibnamefont {Vivoli}}, \bibinfo {author}
  {\bibfnamefont {A.}~\bibnamefont {Reiserer}}, \bibinfo {author}
  {\bibfnamefont {R.}~\bibnamefont {Hanson}}, \bibinfo {author} {\bibfnamefont
  {S.}~\bibnamefont {Wehner}},\ and\ \bibinfo {author} {\bibfnamefont
  {D.}~\bibnamefont {Elkouss}},\ }\bibfield  {title} {\bibinfo {title}
  {Parameter regimes for a single sequential quantum repeater},\ }\href
  {https://doi.org/10.1088/2058-9565/aab31b} {\bibfield  {journal} {\bibinfo
  {journal} {Quantum Sci. Technol.}\ }\textbf {\bibinfo {volume} {3}},\
  \bibinfo {pages} {034002} (\bibinfo {year} {2018}{\natexlab{b}})}\BibitemShut
  {NoStop}%
\bibitem [{\citenamefont {Rozp{\k e}dek}\ \emph {et~al.}(2021)\citenamefont
  {Rozp{\k e}dek}, \citenamefont {Noh}, \citenamefont {Xu}, \citenamefont
  {Guha},\ and\ \citenamefont {Jiang}}]{rozpedek2021}%
  \BibitemOpen
  \bibfield  {author} {\bibinfo {author} {\bibfnamefont {F.}~\bibnamefont
  {Rozp{\k e}dek}}, \bibinfo {author} {\bibfnamefont {K.}~\bibnamefont {Noh}},
  \bibinfo {author} {\bibfnamefont {Q.}~\bibnamefont {Xu}}, \bibinfo {author}
  {\bibfnamefont {S.}~\bibnamefont {Guha}},\ and\ \bibinfo {author}
  {\bibfnamefont {L.}~\bibnamefont {Jiang}},\ }\bibfield  {title} {\bibinfo
  {title} {Quantum repeaters based on concatenated bosonic and
  discrete-variable quantum codes},\ }\href
  {https://doi.org/10.1038/s41534-021-00438-7} {\bibfield  {journal} {\bibinfo
  {journal} {npj Quantum Inf}\ }\textbf {\bibinfo {volume} {7}},\ \bibinfo
  {pages} {1} (\bibinfo {year} {2021})}\BibitemShut {NoStop}%
\bibitem [{\citenamefont {Wo}\ \emph {et~al.}(2023)\citenamefont {Wo},
  \citenamefont {Avis}, \citenamefont {Rozp{\k e}dek}, \citenamefont
  {{Mor-Ruiz}}, \citenamefont {Pieplow}, \citenamefont {Schr{\"o}der},
  \citenamefont {Jiang}, \citenamefont {S{\o}rensen},\ and\ \citenamefont
  {Borregaard}}]{wo2023}%
  \BibitemOpen
  \bibfield  {author} {\bibinfo {author} {\bibfnamefont {K.~J.}\ \bibnamefont
  {Wo}}, \bibinfo {author} {\bibfnamefont {G.}~\bibnamefont {Avis}}, \bibinfo
  {author} {\bibfnamefont {F.}~\bibnamefont {Rozp{\k e}dek}}, \bibinfo {author}
  {\bibfnamefont {M.~F.}\ \bibnamefont {{Mor-Ruiz}}}, \bibinfo {author}
  {\bibfnamefont {G.}~\bibnamefont {Pieplow}}, \bibinfo {author} {\bibfnamefont
  {T.}~\bibnamefont {Schr{\"o}der}}, \bibinfo {author} {\bibfnamefont
  {L.}~\bibnamefont {Jiang}}, \bibinfo {author} {\bibfnamefont {A.~S.}\
  \bibnamefont {S{\o}rensen}},\ and\ \bibinfo {author} {\bibfnamefont
  {J.}~\bibnamefont {Borregaard}},\ }\bibfield  {title} {\bibinfo {title}
  {Resource-efficient fault-tolerant one-way quantum repeater with code
  concatenation},\ }\href {https://doi.org/10.1038/s41534-023-00792-8}
  {\bibfield  {journal} {\bibinfo  {journal} {npj Quantum Inf}\ }\textbf
  {\bibinfo {volume} {9}},\ \bibinfo {pages} {1} (\bibinfo {year}
  {2023})}\BibitemShut {NoStop}%
\bibitem [{\citenamefont {Yao}\ \emph {et~al.}(2021)\citenamefont {Yao},
  \citenamefont {Zou}, \citenamefont {Li},\ and\ \citenamefont
  {Jiang}}]{yao2021}%
  \BibitemOpen
  \bibfield  {author} {\bibinfo {author} {\bibfnamefont {J.}~\bibnamefont
  {Yao}}, \bibinfo {author} {\bibfnamefont {K.}~\bibnamefont {Zou}}, \bibinfo
  {author} {\bibfnamefont {D.}~\bibnamefont {Li}},\ and\ \bibinfo {author}
  {\bibfnamefont {Z.}~\bibnamefont {Jiang}},\ }\bibfield  {title} {\bibinfo
  {title} {Optimal deployment design of repeaters and memories in quantum
  networks},\ }in\ \href
  {https://doi.org/10.1109/HPCC-DSS-SmartCity-DependSys53884.2021.00072} {\emph
  {\bibinfo {booktitle} {2021 {{IEEE}} 23rd {{Int Conf}} on {{High Performance
  Computing}} \& {{Communications}}; 7th {{Int Conf}} on {{Data Science}} \&
  {{Systems}}; 19th {{Int Conf}} on {{Smart City}}; 7th {{Int Conf}} on
  {{Dependability}} in {{Sensor}}, {{Cloud}} \& {{Big Data Systems}} \&
  {{Application}} ({{HPCC}}/{{DSS}}/{{SmartCity}}/{{DependSys}})}}}\ (\bibinfo
  {year} {2021})\ pp.\ \bibinfo {pages} {361--368}\BibitemShut {NoStop}%
\bibitem [{\citenamefont {Dai}\ and\ \citenamefont {Towsley}(2021)}]{dai2021a}%
  \BibitemOpen
  \bibfield  {author} {\bibinfo {author} {\bibfnamefont {W.}~\bibnamefont
  {Dai}}\ and\ \bibinfo {author} {\bibfnamefont {D.}~\bibnamefont {Towsley}},\
  }\bibfield  {title} {\bibinfo {title} {Entanglement {{Swapping}} for
  {{Repeater Chains}} with {{Finite Memory Sizes}}},\ }\href
  {http://arxiv.org/abs/2111.10994} {\bibfield  {journal} {\bibinfo  {journal}
  {arXiv:2111.10994 [quant-ph]}\ } (\bibinfo {year} {2021})}\BibitemShut
  {NoStop}%
\bibitem [{\citenamefont {Kamin}\ \emph {et~al.}(2023)\citenamefont {Kamin},
  \citenamefont {Shchukin}, \citenamefont {Schmidt},\ and\ \citenamefont {{van
  Loock}}}]{kamin2023}%
  \BibitemOpen
  \bibfield  {author} {\bibinfo {author} {\bibfnamefont {L.}~\bibnamefont
  {Kamin}}, \bibinfo {author} {\bibfnamefont {E.}~\bibnamefont {Shchukin}},
  \bibinfo {author} {\bibfnamefont {F.}~\bibnamefont {Schmidt}},\ and\ \bibinfo
  {author} {\bibfnamefont {P.}~\bibnamefont {{van Loock}}},\ }\bibfield
  {title} {\bibinfo {title} {Exact rate analysis for quantum repeaters with
  imperfect memories and entanglement swapping as soon as possible},\ }\href
  {https://doi.org/10.1103/PhysRevResearch.5.023086} {\bibfield  {journal}
  {\bibinfo  {journal} {Phys. Rev. Res.}\ }\textbf {\bibinfo {volume} {5}},\
  \bibinfo {pages} {023086} (\bibinfo {year} {2023})}\BibitemShut {NoStop}%
\bibitem [{\citenamefont {Goodenough}\ \emph {et~al.}(2025)\citenamefont
  {Goodenough}, \citenamefont {Coopmans},\ and\ \citenamefont
  {Towsley}}]{goodenough2025}%
  \BibitemOpen
  \bibfield  {author} {\bibinfo {author} {\bibfnamefont {K.}~\bibnamefont
  {Goodenough}}, \bibinfo {author} {\bibfnamefont {T.}~\bibnamefont
  {Coopmans}},\ and\ \bibinfo {author} {\bibfnamefont {D.}~\bibnamefont
  {Towsley}},\ }\bibfield  {title} {\bibinfo {title} {On noise in swap {{ASAP}}
  repeater chains: Exact analytics, distributions and tight approximations},\
  }\href {https://doi.org/10.22331/q-2025-05-15-1744} {\bibfield  {journal}
  {\bibinfo  {journal} {Quantum}\ }\textbf {\bibinfo {volume} {9}},\ \bibinfo
  {pages} {1744} (\bibinfo {year} {2025})}\BibitemShut {NoStop}%
\bibitem [{\citenamefont {Coopmans}\ \emph {et~al.}(2022)\citenamefont
  {Coopmans}, \citenamefont {Brand},\ and\ \citenamefont
  {Elkouss}}]{coopmans2022}%
  \BibitemOpen
  \bibfield  {author} {\bibinfo {author} {\bibfnamefont {T.}~\bibnamefont
  {Coopmans}}, \bibinfo {author} {\bibfnamefont {S.}~\bibnamefont {Brand}},\
  and\ \bibinfo {author} {\bibfnamefont {D.}~\bibnamefont {Elkouss}},\
  }\bibfield  {title} {\bibinfo {title} {Improved analytical bounds on delivery
  times of long-distance entanglement},\ }\href
  {https://doi.org/10.1103/PhysRevA.105.012608} {\bibfield  {journal} {\bibinfo
   {journal} {Phys. Rev. A}\ }\textbf {\bibinfo {volume} {105}},\ \bibinfo
  {pages} {012608} (\bibinfo {year} {2022})}\BibitemShut {NoStop}%
\bibitem [{\citenamefont {Sangouard}\ \emph {et~al.}(2007)\citenamefont
  {Sangouard}, \citenamefont {Simon}, \citenamefont {Min{\'a}{\v r}},
  \citenamefont {Zbinden}, \citenamefont {{de Riedmatten}},\ and\ \citenamefont
  {Gisin}}]{sangouard2007}%
  \BibitemOpen
  \bibfield  {author} {\bibinfo {author} {\bibfnamefont {N.}~\bibnamefont
  {Sangouard}}, \bibinfo {author} {\bibfnamefont {C.}~\bibnamefont {Simon}},
  \bibinfo {author} {\bibfnamefont {J.}~\bibnamefont {Min{\'a}{\v r}}},
  \bibinfo {author} {\bibfnamefont {H.}~\bibnamefont {Zbinden}}, \bibinfo
  {author} {\bibfnamefont {H.}~\bibnamefont {{de Riedmatten}}},\ and\ \bibinfo
  {author} {\bibfnamefont {N.}~\bibnamefont {Gisin}},\ }\bibfield  {title}
  {\bibinfo {title} {Long-distance entanglement distribution with single-photon
  sources},\ }\href {https://doi.org/10.1103/PhysRevA.76.050301} {\bibfield
  {journal} {\bibinfo  {journal} {Phys. Rev. A}\ }\textbf {\bibinfo {volume}
  {76}},\ \bibinfo {pages} {050301} (\bibinfo {year} {2007})}\BibitemShut
  {NoStop}%
\bibitem [{\citenamefont {Guha}\ \emph {et~al.}(2015)\citenamefont {Guha},
  \citenamefont {Krovi}, \citenamefont {Fuchs}, \citenamefont {Dutton},
  \citenamefont {Slater}, \citenamefont {Simon},\ and\ \citenamefont
  {Tittel}}]{guha2015}%
  \BibitemOpen
  \bibfield  {author} {\bibinfo {author} {\bibfnamefont {S.}~\bibnamefont
  {Guha}}, \bibinfo {author} {\bibfnamefont {H.}~\bibnamefont {Krovi}},
  \bibinfo {author} {\bibfnamefont {C.~A.}\ \bibnamefont {Fuchs}}, \bibinfo
  {author} {\bibfnamefont {Z.}~\bibnamefont {Dutton}}, \bibinfo {author}
  {\bibfnamefont {J.~A.}\ \bibnamefont {Slater}}, \bibinfo {author}
  {\bibfnamefont {C.}~\bibnamefont {Simon}},\ and\ \bibinfo {author}
  {\bibfnamefont {W.}~\bibnamefont {Tittel}},\ }\bibfield  {title} {\bibinfo
  {title} {Rate-loss analysis of an efficient quantum repeater architecture},\
  }\href {https://doi.org/10.1103/PhysRevA.92.022357} {\bibfield  {journal}
  {\bibinfo  {journal} {Phys. Rev. A}\ }\textbf {\bibinfo {volume} {92}},\
  \bibinfo {pages} {022357} (\bibinfo {year} {2015})}\BibitemShut {NoStop}%
\bibitem [{\citenamefont {Shchukin}\ \emph {et~al.}(2019)\citenamefont
  {Shchukin}, \citenamefont {Schmidt},\ and\ \citenamefont {{van
  Loock}}}]{shchukin2019}%
  \BibitemOpen
  \bibfield  {author} {\bibinfo {author} {\bibfnamefont {E.}~\bibnamefont
  {Shchukin}}, \bibinfo {author} {\bibfnamefont {F.}~\bibnamefont {Schmidt}},\
  and\ \bibinfo {author} {\bibfnamefont {P.}~\bibnamefont {{van Loock}}},\
  }\bibfield  {title} {\bibinfo {title} {Waiting time in quantum repeaters with
  probabilistic entanglement swapping},\ }\href
  {https://doi.org/10.1103/PhysRevA.100.032322} {\bibfield  {journal} {\bibinfo
   {journal} {Phys. Rev. A}\ }\textbf {\bibinfo {volume} {100}},\ \bibinfo
  {pages} {032322} (\bibinfo {year} {2019})}\BibitemShut {NoStop}%
\bibitem [{\citenamefont {{de Andrade}}\ \emph {et~al.}(2024)\citenamefont {{de
  Andrade}}, \citenamefont {Van~Milligen}, \citenamefont {Bacciottini},
  \citenamefont {Chandra}, \citenamefont {Pouryousef}, \citenamefont
  {Panigrahy}, \citenamefont {Vardoyan},\ and\ \citenamefont
  {Towsley}}]{deandrade2024b}%
  \BibitemOpen
  \bibfield  {author} {\bibinfo {author} {\bibfnamefont {M.~G.}\ \bibnamefont
  {{de Andrade}}}, \bibinfo {author} {\bibfnamefont {E.~A.}\ \bibnamefont
  {Van~Milligen}}, \bibinfo {author} {\bibfnamefont {L.}~\bibnamefont
  {Bacciottini}}, \bibinfo {author} {\bibfnamefont {A.}~\bibnamefont
  {Chandra}}, \bibinfo {author} {\bibfnamefont {S.}~\bibnamefont {Pouryousef}},
  \bibinfo {author} {\bibfnamefont {N.~K.}\ \bibnamefont {Panigrahy}}, \bibinfo
  {author} {\bibfnamefont {G.}~\bibnamefont {Vardoyan}},\ and\ \bibinfo
  {author} {\bibfnamefont {D.}~\bibnamefont {Towsley}},\ }\href
  {https://doi.org/10.48550/arXiv.2405.18252} {\bibinfo {title} {On the
  {{Analysis}} of {{Quantum Repeater Chains}} with {{Sequential Swaps}}}}
  (\bibinfo {year} {2024}),\ \Eprint {https://arxiv.org/abs/2405.18252}
  {arXiv:2405.18252 [quant-ph]} \BibitemShut {NoStop}%
\bibitem [{\citenamefont {Coopmans}\ \emph {et~al.}(2021)\citenamefont
  {Coopmans}, \citenamefont {Knegjens}, \citenamefont {Dahlberg}, \citenamefont
  {Maier}, \citenamefont {Nijsten}, \citenamefont {{de Oliveira Filho}},
  \citenamefont {Papendrecht}, \citenamefont {Rabbie}, \citenamefont {Rozp{\k
  e}dek}, \citenamefont {Skrzypczyk}, \citenamefont {Wubben}, \citenamefont
  {{de Jong}}, \citenamefont {Podareanu}, \citenamefont {{Torres-Knoop}},
  \citenamefont {Elkouss},\ and\ \citenamefont {Wehner}}]{coopmans2021}%
  \BibitemOpen
  \bibfield  {author} {\bibinfo {author} {\bibfnamefont {T.}~\bibnamefont
  {Coopmans}}, \bibinfo {author} {\bibfnamefont {R.}~\bibnamefont {Knegjens}},
  \bibinfo {author} {\bibfnamefont {A.}~\bibnamefont {Dahlberg}}, \bibinfo
  {author} {\bibfnamefont {D.}~\bibnamefont {Maier}}, \bibinfo {author}
  {\bibfnamefont {L.}~\bibnamefont {Nijsten}}, \bibinfo {author} {\bibfnamefont
  {J.}~\bibnamefont {{de Oliveira Filho}}}, \bibinfo {author} {\bibfnamefont
  {M.}~\bibnamefont {Papendrecht}}, \bibinfo {author} {\bibfnamefont
  {J.}~\bibnamefont {Rabbie}}, \bibinfo {author} {\bibfnamefont
  {F.}~\bibnamefont {Rozp{\k e}dek}}, \bibinfo {author} {\bibfnamefont
  {M.}~\bibnamefont {Skrzypczyk}}, \bibinfo {author} {\bibfnamefont
  {L.}~\bibnamefont {Wubben}}, \bibinfo {author} {\bibfnamefont
  {W.}~\bibnamefont {{de Jong}}}, \bibinfo {author} {\bibfnamefont
  {D.}~\bibnamefont {Podareanu}}, \bibinfo {author} {\bibfnamefont
  {A.}~\bibnamefont {{Torres-Knoop}}}, \bibinfo {author} {\bibfnamefont
  {D.}~\bibnamefont {Elkouss}},\ and\ \bibinfo {author} {\bibfnamefont
  {S.}~\bibnamefont {Wehner}},\ }\bibfield  {title} {\bibinfo {title}
  {{{NetSquid}}, a {{NETwork Simulator}} for {{QUantum Information}} using
  {{Discrete}} events},\ }\href {https://doi.org/10.1038/s42005-021-00647-8}
  {\bibfield  {journal} {\bibinfo  {journal} {Commun Phys}\ }\textbf {\bibinfo
  {volume} {4}},\ \bibinfo {pages} {1} (\bibinfo {year} {2021})}\BibitemShut
  {NoStop}%
\bibitem [{\citenamefont {Wu}\ \emph {et~al.}(2021)\citenamefont {Wu},
  \citenamefont {Kolar}, \citenamefont {Chung}, \citenamefont {Jin},
  \citenamefont {Zhong}, \citenamefont {Kettimuthu},\ and\ \citenamefont
  {Suchara}}]{wu2021b}%
  \BibitemOpen
  \bibfield  {author} {\bibinfo {author} {\bibfnamefont {X.}~\bibnamefont
  {Wu}}, \bibinfo {author} {\bibfnamefont {A.}~\bibnamefont {Kolar}}, \bibinfo
  {author} {\bibfnamefont {J.}~\bibnamefont {Chung}}, \bibinfo {author}
  {\bibfnamefont {D.}~\bibnamefont {Jin}}, \bibinfo {author} {\bibfnamefont
  {T.}~\bibnamefont {Zhong}}, \bibinfo {author} {\bibfnamefont
  {R.}~\bibnamefont {Kettimuthu}},\ and\ \bibinfo {author} {\bibfnamefont
  {M.}~\bibnamefont {Suchara}},\ }\bibfield  {title} {\bibinfo {title}
  {{{SeQUeNCe}}: A customizable discrete-event simulator of quantum networks},\
  }\href {https://doi.org/10.1088/2058-9565/ac22f6} {\bibfield  {journal}
  {\bibinfo  {journal} {Quantum Sci. Technol.}\ }\textbf {\bibinfo {volume}
  {6}},\ \bibinfo {pages} {045027} (\bibinfo {year} {2021})}\BibitemShut
  {NoStop}%
\bibitem [{\citenamefont {Walln{\"o}fer}\ \emph {et~al.}(2024)\citenamefont
  {Walln{\"o}fer}, \citenamefont {Hahn}, \citenamefont {Wiesner}, \citenamefont
  {Walk},\ and\ \citenamefont {Eisert}}]{wallnofer2024}%
  \BibitemOpen
  \bibfield  {author} {\bibinfo {author} {\bibfnamefont {J.}~\bibnamefont
  {Walln{\"o}fer}}, \bibinfo {author} {\bibfnamefont {F.}~\bibnamefont {Hahn}},
  \bibinfo {author} {\bibfnamefont {F.}~\bibnamefont {Wiesner}}, \bibinfo
  {author} {\bibfnamefont {N.}~\bibnamefont {Walk}},\ and\ \bibinfo {author}
  {\bibfnamefont {J.}~\bibnamefont {Eisert}},\ }\bibfield  {title} {\bibinfo
  {title} {Faithfully {{Simulating Near-Term Quantum Repeaters}}},\ }\href
  {https://doi.org/10.1103/PRXQuantum.5.010351} {\bibfield  {journal} {\bibinfo
   {journal} {PRX Quantum}\ }\textbf {\bibinfo {volume} {5}},\ \bibinfo {pages}
  {010351} (\bibinfo {year} {2024})}\BibitemShut {NoStop}%
\bibitem [{\citenamefont {Chen}\ \emph {et~al.}(2023)\citenamefont {Chen},
  \citenamefont {Xue}, \citenamefont {Li}, \citenamefont {Yu}, \citenamefont
  {Li}, \citenamefont {Sun},\ and\ \citenamefont {Lu}}]{chen2023}%
  \BibitemOpen
  \bibfield  {author} {\bibinfo {author} {\bibfnamefont {L.}~\bibnamefont
  {Chen}}, \bibinfo {author} {\bibfnamefont {K.}~\bibnamefont {Xue}}, \bibinfo
  {author} {\bibfnamefont {J.}~\bibnamefont {Li}}, \bibinfo {author}
  {\bibfnamefont {N.}~\bibnamefont {Yu}}, \bibinfo {author} {\bibfnamefont
  {R.}~\bibnamefont {Li}}, \bibinfo {author} {\bibfnamefont {Q.}~\bibnamefont
  {Sun}},\ and\ \bibinfo {author} {\bibfnamefont {J.}~\bibnamefont {Lu}},\
  }\bibfield  {title} {\bibinfo {title} {{{SimQN}}: A {{Network-layer
  Simulator}} for the {{Quantum Network Investigation}}},\ }\href
  {https://doi.org/10.1109/MNET.130.2200481} {\bibfield  {journal} {\bibinfo
  {journal} {IEEE Network}\ ,\ \bibinfo {pages} {1}} (\bibinfo {year}
  {2023})}\BibitemShut {NoStop}%
\bibitem [{\citenamefont {Satoh}\ \emph {et~al.}(2022)\citenamefont {Satoh},
  \citenamefont {Hajdu{\v s}ek}, \citenamefont {Benchasattabuse}, \citenamefont
  {Nagayama}, \citenamefont {Teramoto}, \citenamefont {Matsuo}, \citenamefont
  {Metwalli}, \citenamefont {Pathumsoot}, \citenamefont {Satoh}, \citenamefont
  {Suzuki},\ and\ \citenamefont {Meter}}]{satoh2022}%
  \BibitemOpen
  \bibfield  {author} {\bibinfo {author} {\bibfnamefont {R.}~\bibnamefont
  {Satoh}}, \bibinfo {author} {\bibfnamefont {M.}~\bibnamefont {Hajdu{\v
  s}ek}}, \bibinfo {author} {\bibfnamefont {N.}~\bibnamefont
  {Benchasattabuse}}, \bibinfo {author} {\bibfnamefont {S.}~\bibnamefont
  {Nagayama}}, \bibinfo {author} {\bibfnamefont {K.}~\bibnamefont {Teramoto}},
  \bibinfo {author} {\bibfnamefont {T.}~\bibnamefont {Matsuo}}, \bibinfo
  {author} {\bibfnamefont {S.~A.}\ \bibnamefont {Metwalli}}, \bibinfo {author}
  {\bibfnamefont {P.}~\bibnamefont {Pathumsoot}}, \bibinfo {author}
  {\bibfnamefont {T.}~\bibnamefont {Satoh}}, \bibinfo {author} {\bibfnamefont
  {S.}~\bibnamefont {Suzuki}},\ and\ \bibinfo {author} {\bibfnamefont {R.~V.}\
  \bibnamefont {Meter}},\ }\bibfield  {title} {\bibinfo {title} {{{QuISP}}: A
  {{Quantum Internet Simulation Package}}},\ }in\ \href
  {https://doi.org/10.1109/QCE53715.2022.00056} {\emph {\bibinfo {booktitle}
  {2022 {{IEEE International Conference}} on {{Quantum Computing}} and
  {{Engineering}} ({{QCE}})}}}\ (\bibinfo {year} {2022})\ pp.\ \bibinfo {pages}
  {353--364}\BibitemShut {NoStop}%
\bibitem [{\citenamefont {LeVeque}(2007)}]{leveque2007}%
  \BibitemOpen
  \bibfield  {author} {\bibinfo {author} {\bibfnamefont {R.~J.}\ \bibnamefont
  {LeVeque}},\ }\href {https://doi.org/10.1137/1.9780898717839} {\emph
  {\bibinfo {title} {Finite {{Difference Methods}} for {{Ordinary}} and
  {{Partial Differential Equations}}}}},\ Other {{Titles}} in {{Applied
  Mathematics}}\ (\bibinfo  {publisher} {{Society for Industrial and Applied
  Mathematics}},\ \bibinfo {year} {2007})\BibitemShut {NoStop}%
\bibitem [{\citenamefont {Glynn}(1990)}]{glynn1990}%
  \BibitemOpen
  \bibfield  {author} {\bibinfo {author} {\bibfnamefont {P.~W.}\ \bibnamefont
  {Glynn}},\ }\bibfield  {title} {\bibinfo {title} {Likelihood ratio gradient
  estimation for stochastic systems},\ }\href
  {https://doi.org/10.1145/84537.84552} {\bibfield  {journal} {\bibinfo
  {journal} {Commun. ACM}\ }\textbf {\bibinfo {volume} {33}},\ \bibinfo {pages}
  {75} (\bibinfo {year} {1990})}\BibitemShut {NoStop}%
\bibitem [{\citenamefont {Grathwohl}\ \emph {et~al.}(2018)\citenamefont
  {Grathwohl}, \citenamefont {Choi}, \citenamefont {Wu}, \citenamefont
  {Roeder},\ and\ \citenamefont {Duvenaud}}]{grathwohl2018}%
  \BibitemOpen
  \bibfield  {author} {\bibinfo {author} {\bibfnamefont {W.}~\bibnamefont
  {Grathwohl}}, \bibinfo {author} {\bibfnamefont {D.}~\bibnamefont {Choi}},
  \bibinfo {author} {\bibfnamefont {Y.}~\bibnamefont {Wu}}, \bibinfo {author}
  {\bibfnamefont {G.}~\bibnamefont {Roeder}},\ and\ \bibinfo {author}
  {\bibfnamefont {D.}~\bibnamefont {Duvenaud}},\ }\href
  {https://doi.org/10.48550/arXiv.1711.00123} {\bibinfo {title}
  {Backpropagation through the {{Void}}: {{Optimizing}} control variates for
  black-box gradient estimation}} (\bibinfo {year} {2018}),\ \Eprint
  {https://arxiv.org/abs/1711.00123} {arXiv:1711.00123 [cs]} \BibitemShut
  {NoStop}%
\bibitem [{\citenamefont {Heidergott}\ and\ \citenamefont
  {{V{\'a}zquez-Abad}}(2008)}]{heidergott2008}%
  \BibitemOpen
  \bibfield  {author} {\bibinfo {author} {\bibfnamefont {B.}~\bibnamefont
  {Heidergott}}\ and\ \bibinfo {author} {\bibfnamefont {F.~J.}\ \bibnamefont
  {{V{\'a}zquez-Abad}}},\ }\bibfield  {title} {\bibinfo {title}
  {Measure-{{Valued Differentiation}} for {{Markov Chains}}},\ }\href
  {https://doi.org/10.1007/s10957-007-9297-7} {\bibfield  {journal} {\bibinfo
  {journal} {J Optim Theory Appl}\ }\textbf {\bibinfo {volume} {136}},\
  \bibinfo {pages} {187} (\bibinfo {year} {2008})}\BibitemShut {NoStop}%
\bibitem [{\citenamefont {Jang}\ \emph {et~al.}(2017)\citenamefont {Jang},
  \citenamefont {Gu},\ and\ \citenamefont {Poole}}]{jang2017}%
  \BibitemOpen
  \bibfield  {author} {\bibinfo {author} {\bibfnamefont {E.}~\bibnamefont
  {Jang}}, \bibinfo {author} {\bibfnamefont {S.}~\bibnamefont {Gu}},\ and\
  \bibinfo {author} {\bibfnamefont {B.}~\bibnamefont {Poole}},\ }\bibfield
  {title} {\bibinfo {title} {Categorical {{Reparameterization}} with
  {{Gumbel-Softmax}}},\ }in\ \href {https://openreview.net/forum?id=rkE3y85ee}
  {\emph {\bibinfo {booktitle} {International {{Conference}} on {{Learning
  Representations}}}}}\ (\bibinfo {address} {Toulon, France},\ \bibinfo {year}
  {2017})\BibitemShut {NoStop}%
\bibitem [{\citenamefont {Kingma}\ and\ \citenamefont
  {Welling}(2022)}]{kingma2022a}%
  \BibitemOpen
  \bibfield  {author} {\bibinfo {author} {\bibfnamefont {D.~P.}\ \bibnamefont
  {Kingma}}\ and\ \bibinfo {author} {\bibfnamefont {M.}~\bibnamefont
  {Welling}},\ }\href {https://doi.org/10.48550/arXiv.1312.6114} {\bibinfo
  {title} {Auto-{{Encoding Variational Bayes}}}} (\bibinfo {year} {2022}),\
  \Eprint {https://arxiv.org/abs/1312.6114} {arXiv:1312.6114 [stat]}
  \BibitemShut {NoStop}%
\bibitem [{\citenamefont {Kleijnen}\ and\ \citenamefont
  {Rubinstein}(1996)}]{kleijnen1996}%
  \BibitemOpen
  \bibfield  {author} {\bibinfo {author} {\bibfnamefont {J.~P.~C.}\
  \bibnamefont {Kleijnen}}\ and\ \bibinfo {author} {\bibfnamefont {R.~Y.}\
  \bibnamefont {Rubinstein}},\ }\bibfield  {title} {\bibinfo {title}
  {Optimization and sensitivity analysis of computer simulation models by the
  score function method},\ }\href
  {https://doi.org/10.1016/0377-2217(95)00107-7} {\bibfield  {journal}
  {\bibinfo  {journal} {European Journal of Operational Research}\ }\textbf
  {\bibinfo {volume} {88}},\ \bibinfo {pages} {413} (\bibinfo {year}
  {1996})}\BibitemShut {NoStop}%
\bibitem [{\citenamefont {Mohamed}\ \emph {et~al.}(2020)\citenamefont
  {Mohamed}, \citenamefont {Rosca}, \citenamefont {Figurnov},\ and\
  \citenamefont {Mnih}}]{mohamed2020}%
  \BibitemOpen
  \bibfield  {author} {\bibinfo {author} {\bibfnamefont {S.}~\bibnamefont
  {Mohamed}}, \bibinfo {author} {\bibfnamefont {M.}~\bibnamefont {Rosca}},
  \bibinfo {author} {\bibfnamefont {M.}~\bibnamefont {Figurnov}},\ and\
  \bibinfo {author} {\bibfnamefont {A.}~\bibnamefont {Mnih}},\ }\bibfield
  {title} {\bibinfo {title} {Monte carlo gradient estimation in machine
  learning},\ }\href {http://jmlr.org/papers/v21/19-346.html} {\bibfield
  {journal} {\bibinfo  {journal} {Journal of Machine Learning Research}\
  }\textbf {\bibinfo {volume} {21}},\ \bibinfo {pages} {1} (\bibinfo {year}
  {2020})}\BibitemShut {NoStop}%
\bibitem [{\citenamefont {Pflug}(1989)}]{pflug1989}%
  \BibitemOpen
  \bibfield  {author} {\bibinfo {author} {\bibfnamefont {G.~{\relax Ch}.}\
  \bibnamefont {Pflug}},\ }\bibfield  {title} {\bibinfo {title} {Sampling
  derivatives of probabilities},\ }\href {https://doi.org/10.1007/BF02243227}
  {\bibfield  {journal} {\bibinfo  {journal} {Computing}\ }\textbf {\bibinfo
  {volume} {42}},\ \bibinfo {pages} {315} (\bibinfo {year} {1989})}\BibitemShut
  {NoStop}%
\bibitem [{\citenamefont {Tucker}\ \emph {et~al.}(2017)\citenamefont {Tucker},
  \citenamefont {Mnih}, \citenamefont {Maddison}, \citenamefont {Lawson},\ and\
  \citenamefont {{Sohl-Dickstein}}}]{tucker2017}%
  \BibitemOpen
  \bibfield  {author} {\bibinfo {author} {\bibfnamefont {G.}~\bibnamefont
  {Tucker}}, \bibinfo {author} {\bibfnamefont {A.}~\bibnamefont {Mnih}},
  \bibinfo {author} {\bibfnamefont {C.~J.}\ \bibnamefont {Maddison}}, \bibinfo
  {author} {\bibfnamefont {D.}~\bibnamefont {Lawson}},\ and\ \bibinfo {author}
  {\bibfnamefont {J.}~\bibnamefont {{Sohl-Dickstein}}},\ }\bibfield  {title}
  {\bibinfo {title} {{{REBAR}}: Low-variance, unbiased gradient estimates for
  discrete latent variable models},\ }in\ \href@noop {} {\emph {\bibinfo
  {booktitle} {Proceedings of the 31st {{International Conference}} on {{Neural
  Information Processing Systems}}}}},\ \bibinfo {series and number}
  {{{NIPS}}'17}\ (\bibinfo  {publisher} {Curran Associates Inc.},\ \bibinfo
  {address} {Red Hook, NY, USA},\ \bibinfo {year} {2017})\ pp.\ \bibinfo
  {pages} {2624--2633}\BibitemShut {NoStop}%
\bibitem [{\citenamefont {Williams}(1992)}]{williams1992}%
  \BibitemOpen
  \bibfield  {author} {\bibinfo {author} {\bibfnamefont {R.~J.}\ \bibnamefont
  {Williams}},\ }\bibfield  {title} {\bibinfo {title} {Simple statistical
  gradient-following algorithms for connectionist reinforcement learning},\
  }\href {https://doi.org/10.1007/BF00992696} {\bibfield  {journal} {\bibinfo
  {journal} {Mach Learn}\ }\textbf {\bibinfo {volume} {8}},\ \bibinfo {pages}
  {229} (\bibinfo {year} {1992})}\BibitemShut {NoStop}%
\bibitem [{\citenamefont {Glasserman}(1991)}]{glasserman1991}%
  \BibitemOpen
  \bibfield  {author} {\bibinfo {author} {\bibfnamefont {P.}~\bibnamefont
  {Glasserman}},\ }\href@noop {} {\emph {\bibinfo {title} {Gradient
  {{Estimation Via Perturbation Analysis}}}}}\ (\bibinfo  {publisher} {Springer
  Science \& Business Media},\ \bibinfo {year} {1991})\BibitemShut {NoStop}%
\bibitem [{\citenamefont {Arya}\ \emph {et~al.}(2022)\citenamefont {Arya},
  \citenamefont {Schauer}, \citenamefont {Sch{\"a}fer},\ and\ \citenamefont
  {Rackauckas}}]{arya2023}%
  \BibitemOpen
  \bibfield  {author} {\bibinfo {author} {\bibfnamefont {G.}~\bibnamefont
  {Arya}}, \bibinfo {author} {\bibfnamefont {M.}~\bibnamefont {Schauer}},
  \bibinfo {author} {\bibfnamefont {F.}~\bibnamefont {Sch{\"a}fer}},\ and\
  \bibinfo {author} {\bibfnamefont {C.}~\bibnamefont {Rackauckas}},\ }\bibfield
   {title} {\bibinfo {title} {Automatic {{Differentiation}} of {{Programs}}
  with {{Discrete Randomness}}},\ }\href
  {https://papers.nips.cc/paper_files/paper/2022/hash/43d8e5fc816c692f342493331d5e98fc-Abstract-Conference.html}
  {\bibfield  {journal} {\bibinfo  {journal} {Advances in Neural Information
  Processing Systems}\ }\textbf {\bibinfo {volume} {35}},\ \bibinfo {pages}
  {10435} (\bibinfo {year} {2022})}\BibitemShut {NoStop}%
\bibitem [{\citenamefont {Porotti}\ \emph {et~al.}(2023)\citenamefont
  {Porotti}, \citenamefont {Peano},\ and\ \citenamefont
  {Marquardt}}]{porotti2023}%
  \BibitemOpen
  \bibfield  {author} {\bibinfo {author} {\bibfnamefont {R.}~\bibnamefont
  {Porotti}}, \bibinfo {author} {\bibfnamefont {V.}~\bibnamefont {Peano}},\
  and\ \bibinfo {author} {\bibfnamefont {F.}~\bibnamefont {Marquardt}},\
  }\bibfield  {title} {\bibinfo {title} {Gradient-{{Ascent Pulse Engineering}}
  with {{Feedback}}},\ }\href {https://doi.org/10.1103/PRXQuantum.4.030305}
  {\bibfield  {journal} {\bibinfo  {journal} {PRX Quantum}\ }\textbf {\bibinfo
  {volume} {4}},\ \bibinfo {pages} {030305} (\bibinfo {year}
  {2023})}\BibitemShut {NoStop}%
\bibitem [{\citenamefont {Belliardo}\ \emph
  {et~al.}(2024{\natexlab{a}})\citenamefont {Belliardo}, \citenamefont
  {Zoratti}, \citenamefont {Marquardt},\ and\ \citenamefont
  {Giovannetti}}]{belliardo2024}%
  \BibitemOpen
  \bibfield  {author} {\bibinfo {author} {\bibfnamefont {F.}~\bibnamefont
  {Belliardo}}, \bibinfo {author} {\bibfnamefont {F.}~\bibnamefont {Zoratti}},
  \bibinfo {author} {\bibfnamefont {F.}~\bibnamefont {Marquardt}},\ and\
  \bibinfo {author} {\bibfnamefont {V.}~\bibnamefont {Giovannetti}},\
  }\bibfield  {title} {\bibinfo {title} {Model-aware reinforcement learning for
  high-performance {{Bayesian}} experimental design in quantum metrology},\
  }\href {https://doi.org/10.22331/q-2024-12-10-1555} {\bibfield  {journal}
  {\bibinfo  {journal} {Quantum}\ }\textbf {\bibinfo {volume} {8}},\ \bibinfo
  {pages} {1555} (\bibinfo {year} {2024}{\natexlab{a}})}\BibitemShut {NoStop}%
\bibitem [{\citenamefont {Belliardo}\ \emph
  {et~al.}(2024{\natexlab{b}})\citenamefont {Belliardo}, \citenamefont
  {Zoratti},\ and\ \citenamefont {Giovannetti}}]{belliardo2024a}%
  \BibitemOpen
  \bibfield  {author} {\bibinfo {author} {\bibfnamefont {F.}~\bibnamefont
  {Belliardo}}, \bibinfo {author} {\bibfnamefont {F.}~\bibnamefont {Zoratti}},\
  and\ \bibinfo {author} {\bibfnamefont {V.}~\bibnamefont {Giovannetti}},\
  }\bibfield  {title} {\bibinfo {title} {Applications of model-aware
  reinforcement learning in {{Bayesian}} quantum metrology},\ }\href
  {https://doi.org/10.1103/PhysRevA.109.062609} {\bibfield  {journal} {\bibinfo
   {journal} {Phys. Rev. A}\ }\textbf {\bibinfo {volume} {109}},\ \bibinfo
  {pages} {062609} (\bibinfo {year} {2024}{\natexlab{b}})}\BibitemShut
  {NoStop}%
\bibitem [{\citenamefont {Nolan}(1953)}]{nolan1953}%
  \BibitemOpen
  \bibfield  {author} {\bibinfo {author} {\bibfnamefont {J.~F.}\ \bibnamefont
  {Nolan}},\ }\emph {\bibinfo {title} {Analytical Differentiation on a Digital
  Computer}},\ \href {https://dspace.mit.edu/handle/1721.1/12297} {Master's
  thesis},\ \bibinfo  {school} {Massachusetts Institute of Technology}
  (\bibinfo {year} {1953})\BibitemShut {NoStop}%
\bibitem [{\citenamefont {Wengert}(1964)}]{wengert1964}%
  \BibitemOpen
  \bibfield  {author} {\bibinfo {author} {\bibfnamefont {R.~E.}\ \bibnamefont
  {Wengert}},\ }\bibfield  {title} {\bibinfo {title} {A simple automatic
  derivative evaluation program},\ }\href
  {https://doi.org/10.1145/355586.364791} {\bibfield  {journal} {\bibinfo
  {journal} {Commun. ACM}\ }\textbf {\bibinfo {volume} {7}},\ \bibinfo {pages}
  {463} (\bibinfo {year} {1964})}\BibitemShut {NoStop}%
\bibitem [{\citenamefont {Baydin}\ \emph {et~al.}(2018)\citenamefont {Baydin},
  \citenamefont {Pearlmutter}, \citenamefont {Radul},\ and\ \citenamefont
  {Siskind}}]{baydin2018}%
  \BibitemOpen
  \bibfield  {author} {\bibinfo {author} {\bibfnamefont {A.~G.}\ \bibnamefont
  {Baydin}}, \bibinfo {author} {\bibfnamefont {B.~A.}\ \bibnamefont
  {Pearlmutter}}, \bibinfo {author} {\bibfnamefont {A.~A.}\ \bibnamefont
  {Radul}},\ and\ \bibinfo {author} {\bibfnamefont {J.~M.}\ \bibnamefont
  {Siskind}},\ }\bibfield  {title} {\bibinfo {title} {Automatic differentiation
  in machine learning: A survey},\ }\href
  {http://jmlr.org/papers/v18/17-468.html} {\bibfield  {journal} {\bibinfo
  {journal} {Journal of Machine Learning Research}\ }\textbf {\bibinfo {volume}
  {18}},\ \bibinfo {pages} {1} (\bibinfo {year} {2018})}\BibitemShut {NoStop}%
\bibitem [{\citenamefont {Arya}(2024)}]{arya2024}%
  \BibitemOpen
  \bibfield  {author} {\bibinfo {author} {\bibfnamefont {G.}~\bibnamefont
  {Arya}},\ }\href {https://github.com/gaurav-arya/StochasticAD.jl} {\bibinfo
  {title} {gaurav-arya/stochasticad.jl}} (\bibinfo {year} {2024})\BibitemShut
  {NoStop}%
\bibitem [{\citenamefont {Werner}(1989)}]{werner1989}%
  \BibitemOpen
  \bibfield  {author} {\bibinfo {author} {\bibfnamefont {R.~F.}\ \bibnamefont
  {Werner}},\ }\bibfield  {title} {\bibinfo {title} {Quantum states with
  {{Einstein-Podolsky-Rosen}} correlations admitting a hidden-variable model},\
  }\href {https://doi.org/10.1103/PhysRevA.40.4277} {\bibfield  {journal}
  {\bibinfo  {journal} {Phys. Rev. A}\ }\textbf {\bibinfo {volume} {40}},\
  \bibinfo {pages} {4277} (\bibinfo {year} {1989})}\BibitemShut {NoStop}%
\bibitem [{\citenamefont {I{\~n}esta}\ \emph {et~al.}(2023)\citenamefont
  {I{\~n}esta}, \citenamefont {Vardoyan}, \citenamefont {Scavuzzo},\ and\
  \citenamefont {Wehner}}]{inesta2023}%
  \BibitemOpen
  \bibfield  {author} {\bibinfo {author} {\bibfnamefont {{\'A}.~G.}\
  \bibnamefont {I{\~n}esta}}, \bibinfo {author} {\bibfnamefont
  {G.}~\bibnamefont {Vardoyan}}, \bibinfo {author} {\bibfnamefont
  {L.}~\bibnamefont {Scavuzzo}},\ and\ \bibinfo {author} {\bibfnamefont
  {S.}~\bibnamefont {Wehner}},\ }\bibfield  {title} {\bibinfo {title} {Optimal
  entanglement distribution policies in homogeneous repeater chains with
  cutoffs},\ }\href {https://doi.org/10.1038/s41534-023-00713-9} {\bibfield
  {journal} {\bibinfo  {journal} {npj Quantum Inf}\ }\textbf {\bibinfo {volume}
  {9}},\ \bibinfo {pages} {46} (\bibinfo {year} {2023})}\BibitemShut {NoStop}%
\bibitem [{\citenamefont {Bennett}\ \emph {et~al.}(1993)\citenamefont
  {Bennett}, \citenamefont {Brassard}, \citenamefont {Cr{\'e}peau},
  \citenamefont {Jozsa}, \citenamefont {Peres},\ and\ \citenamefont
  {Wootters}}]{bennett1993a}%
  \BibitemOpen
  \bibfield  {author} {\bibinfo {author} {\bibfnamefont {C.~H.}\ \bibnamefont
  {Bennett}}, \bibinfo {author} {\bibfnamefont {G.}~\bibnamefont {Brassard}},
  \bibinfo {author} {\bibfnamefont {C.}~\bibnamefont {Cr{\'e}peau}}, \bibinfo
  {author} {\bibfnamefont {R.}~\bibnamefont {Jozsa}}, \bibinfo {author}
  {\bibfnamefont {A.}~\bibnamefont {Peres}},\ and\ \bibinfo {author}
  {\bibfnamefont {W.~K.}\ \bibnamefont {Wootters}},\ }\bibfield  {title}
  {\bibinfo {title} {Teleporting an unknown quantum state via dual classical
  and {{Einstein-Podolsky-Rosen}} channels},\ }\href
  {https://doi.org/10.1103/PhysRevLett.70.1895} {\bibfield  {journal} {\bibinfo
   {journal} {Phys. Rev. Lett.}\ }\textbf {\bibinfo {volume} {70}},\ \bibinfo
  {pages} {1895} (\bibinfo {year} {1993})}\BibitemShut {NoStop}%
\bibitem [{\citenamefont {Santra}\ \emph {et~al.}(2019)\citenamefont {Santra},
  \citenamefont {Jiang},\ and\ \citenamefont {Malinovsky}}]{santra2019}%
  \BibitemOpen
  \bibfield  {author} {\bibinfo {author} {\bibfnamefont {S.}~\bibnamefont
  {Santra}}, \bibinfo {author} {\bibfnamefont {L.}~\bibnamefont {Jiang}},\ and\
  \bibinfo {author} {\bibfnamefont {V.~S.}\ \bibnamefont {Malinovsky}},\
  }\bibfield  {title} {\bibinfo {title} {Quantum repeater architecture with
  hierarchically optimized memory buffer times},\ }\href
  {https://doi.org/10.1088/2058-9565/ab0bc2} {\bibfield  {journal} {\bibinfo
  {journal} {Quantum Sci. Technol.}\ }\textbf {\bibinfo {volume} {4}},\
  \bibinfo {pages} {025010} (\bibinfo {year} {2019})}\BibitemShut {NoStop}%
\bibitem [{\citenamefont {Shor}\ and\ \citenamefont
  {Preskill}(2000)}]{shor2000}%
  \BibitemOpen
  \bibfield  {author} {\bibinfo {author} {\bibfnamefont {P.~W.}\ \bibnamefont
  {Shor}}\ and\ \bibinfo {author} {\bibfnamefont {J.}~\bibnamefont
  {Preskill}},\ }\bibfield  {title} {\bibinfo {title} {Simple {{Proof}} of
  {{Security}} of the {{BB84 Quantum Key Distribution Protocol}}},\ }\href
  {https://doi.org/10.1103/PhysRevLett.85.441} {\bibfield  {journal} {\bibinfo
  {journal} {Phys. Rev. Lett.}\ }\textbf {\bibinfo {volume} {85}},\ \bibinfo
  {pages} {441} (\bibinfo {year} {2000})}\BibitemShut {NoStop}%
\bibitem [{\citenamefont
  {Avis}(2025{\natexlab{a}})}]{quantum_network_recipes.jl}%
  \BibitemOpen
  \bibfield  {author} {\bibinfo {author} {\bibfnamefont {G.}~\bibnamefont
  {Avis}},\ }\href {https://github.com/GuusAvis/QuantumNetworkRecipes.jl}
  {\bibinfo {title} {Guusavis/quantumnetworkrecipes.jl}} (\bibinfo {year}
  {2025}{\natexlab{a}})\BibitemShut {NoStop}%
\bibitem [{\citenamefont {Cabrillo}\ \emph {et~al.}(1999)\citenamefont
  {Cabrillo}, \citenamefont {Cirac}, \citenamefont
  {{Garc{\'i}a-Fern{\'a}ndez}},\ and\ \citenamefont {Zoller}}]{cabrillo1999}%
  \BibitemOpen
  \bibfield  {author} {\bibinfo {author} {\bibfnamefont {C.}~\bibnamefont
  {Cabrillo}}, \bibinfo {author} {\bibfnamefont {J.~I.}\ \bibnamefont {Cirac}},
  \bibinfo {author} {\bibfnamefont {P.}~\bibnamefont
  {{Garc{\'i}a-Fern{\'a}ndez}}},\ and\ \bibinfo {author} {\bibfnamefont
  {P.}~\bibnamefont {Zoller}},\ }\bibfield  {title} {\bibinfo {title} {Creation
  of entangled states of distant atoms by interference},\ }\href
  {https://doi.org/10.1103/PhysRevA.59.1025} {\bibfield  {journal} {\bibinfo
  {journal} {Phys. Rev. A}\ }\textbf {\bibinfo {volume} {59}},\ \bibinfo
  {pages} {1025} (\bibinfo {year} {1999})}\BibitemShut {NoStop}%
\bibitem [{\citenamefont {Bose}\ \emph {et~al.}(1999)\citenamefont {Bose},
  \citenamefont {Knight}, \citenamefont {Plenio},\ and\ \citenamefont
  {Vedral}}]{bose1999}%
  \BibitemOpen
  \bibfield  {author} {\bibinfo {author} {\bibfnamefont {S.}~\bibnamefont
  {Bose}}, \bibinfo {author} {\bibfnamefont {P.~L.}\ \bibnamefont {Knight}},
  \bibinfo {author} {\bibfnamefont {M.~B.}\ \bibnamefont {Plenio}},\ and\
  \bibinfo {author} {\bibfnamefont {V.}~\bibnamefont {Vedral}},\ }\bibfield
  {title} {\bibinfo {title} {Proposal for teleportation of an atomic state via
  cavity decay},\ }\href {https://doi.org/10.1103/PhysRevLett.83.5158}
  {\bibfield  {journal} {\bibinfo  {journal} {Phys. Rev. Lett.}\ }\textbf
  {\bibinfo {volume} {83}},\ \bibinfo {pages} {5158} (\bibinfo {year}
  {1999})}\BibitemShut {NoStop}%
\bibitem [{\citenamefont {Humphreys}\ \emph {et~al.}(2018)\citenamefont
  {Humphreys}, \citenamefont {Kalb}, \citenamefont {Morits}, \citenamefont
  {Schouten}, \citenamefont {Vermeulen}, \citenamefont {Twitchen},
  \citenamefont {Markham},\ and\ \citenamefont {Hanson}}]{humphreys2018}%
  \BibitemOpen
  \bibfield  {author} {\bibinfo {author} {\bibfnamefont {P.~C.}\ \bibnamefont
  {Humphreys}}, \bibinfo {author} {\bibfnamefont {N.}~\bibnamefont {Kalb}},
  \bibinfo {author} {\bibfnamefont {J.~P.~J.}\ \bibnamefont {Morits}}, \bibinfo
  {author} {\bibfnamefont {R.~N.}\ \bibnamefont {Schouten}}, \bibinfo {author}
  {\bibfnamefont {R.~F.~L.}\ \bibnamefont {Vermeulen}}, \bibinfo {author}
  {\bibfnamefont {D.~J.}\ \bibnamefont {Twitchen}}, \bibinfo {author}
  {\bibfnamefont {M.}~\bibnamefont {Markham}},\ and\ \bibinfo {author}
  {\bibfnamefont {R.}~\bibnamefont {Hanson}},\ }\bibfield  {title} {\bibinfo
  {title} {Deterministic delivery of remote entanglement on a quantum
  network},\ }\href {https://doi.org/10.1038/s41586-018-0200-5} {\bibfield
  {journal} {\bibinfo  {journal} {Nature}\ }\textbf {\bibinfo {volume} {558}},\
  \bibinfo {pages} {268} (\bibinfo {year} {2018})}\BibitemShut {NoStop}%
\bibitem [{\citenamefont {Schupp}\ \emph {et~al.}(2021)\citenamefont {Schupp},
  \citenamefont {Krcmarsky}, \citenamefont {Krutyanskiy}, \citenamefont
  {Meraner}, \citenamefont {Northup},\ and\ \citenamefont
  {Lanyon}}]{schupp2021}%
  \BibitemOpen
  \bibfield  {author} {\bibinfo {author} {\bibfnamefont {J.}~\bibnamefont
  {Schupp}}, \bibinfo {author} {\bibfnamefont {V.}~\bibnamefont {Krcmarsky}},
  \bibinfo {author} {\bibfnamefont {V.}~\bibnamefont {Krutyanskiy}}, \bibinfo
  {author} {\bibfnamefont {M.}~\bibnamefont {Meraner}}, \bibinfo {author}
  {\bibfnamefont {T.}~\bibnamefont {Northup}},\ and\ \bibinfo {author}
  {\bibfnamefont {B.}~\bibnamefont {Lanyon}},\ }\bibfield  {title} {\bibinfo
  {title} {Interface between {{Trapped-Ion Qubits}} and {{Traveling Photons}}
  with {{Close-to-Optimal Efficiency}}},\ }\href
  {https://doi.org/10.1103/PRXQuantum.2.020331} {\bibfield  {journal} {\bibinfo
   {journal} {PRX Quantum}\ }\textbf {\bibinfo {volume} {2}},\ \bibinfo {pages}
  {020331} (\bibinfo {year} {2021})}\BibitemShut {NoStop}%
\bibitem [{\citenamefont {Krutyanskiy}\ \emph {et~al.}(2023)\citenamefont
  {Krutyanskiy}, \citenamefont {Galli}, \citenamefont {Krcmarsky},
  \citenamefont {Baier}, \citenamefont {Fioretto}, \citenamefont {Pu},
  \citenamefont {Mazloom}, \citenamefont {Sekatski}, \citenamefont {Canteri},
  \citenamefont {Teller}, \citenamefont {Schupp}, \citenamefont {Bate},
  \citenamefont {Meraner}, \citenamefont {Sangouard}, \citenamefont {Lanyon},\
  and\ \citenamefont {Northup}}]{krutyanskiy2023}%
  \BibitemOpen
  \bibfield  {author} {\bibinfo {author} {\bibfnamefont {V.}~\bibnamefont
  {Krutyanskiy}}, \bibinfo {author} {\bibfnamefont {M.}~\bibnamefont {Galli}},
  \bibinfo {author} {\bibfnamefont {V.}~\bibnamefont {Krcmarsky}}, \bibinfo
  {author} {\bibfnamefont {S.}~\bibnamefont {Baier}}, \bibinfo {author}
  {\bibfnamefont {D.~A.}\ \bibnamefont {Fioretto}}, \bibinfo {author}
  {\bibfnamefont {Y.}~\bibnamefont {Pu}}, \bibinfo {author} {\bibfnamefont
  {A.}~\bibnamefont {Mazloom}}, \bibinfo {author} {\bibfnamefont
  {P.}~\bibnamefont {Sekatski}}, \bibinfo {author} {\bibfnamefont
  {M.}~\bibnamefont {Canteri}}, \bibinfo {author} {\bibfnamefont
  {M.}~\bibnamefont {Teller}}, \bibinfo {author} {\bibfnamefont
  {J.}~\bibnamefont {Schupp}}, \bibinfo {author} {\bibfnamefont
  {J.}~\bibnamefont {Bate}}, \bibinfo {author} {\bibfnamefont {M.}~\bibnamefont
  {Meraner}}, \bibinfo {author} {\bibfnamefont {N.}~\bibnamefont {Sangouard}},
  \bibinfo {author} {\bibfnamefont {B.~P.}\ \bibnamefont {Lanyon}},\ and\
  \bibinfo {author} {\bibfnamefont {T.~E.}\ \bibnamefont {Northup}},\
  }\bibfield  {title} {\bibinfo {title} {Entanglement of trapped-ion qubits
  separated by 230 meters},\ }\href
  {https://doi.org/10.1103/PhysRevLett.130.050803} {\bibfield  {journal}
  {\bibinfo  {journal} {Phys. Rev. Lett.}\ }\textbf {\bibinfo {volume} {130}},\
  \bibinfo {pages} {050803} (\bibinfo {year} {2023})}\BibitemShut {NoStop}%
\bibitem [{\citenamefont {Hensen}\ \emph {et~al.}(2015)\citenamefont {Hensen},
  \citenamefont {Bernien}, \citenamefont {Dr{\'e}au}, \citenamefont {Reiserer},
  \citenamefont {Kalb}, \citenamefont {Blok}, \citenamefont {Ruitenberg},
  \citenamefont {Vermeulen}, \citenamefont {Schouten}, \citenamefont
  {Abell{\'a}n}, \citenamefont {Amaya}, \citenamefont {Pruneri}, \citenamefont
  {Mitchell}, \citenamefont {Markham}, \citenamefont {Twitchen}, \citenamefont
  {Elkouss}, \citenamefont {Wehner}, \citenamefont {Taminiau},\ and\
  \citenamefont {Hanson}}]{hensen2015}%
  \BibitemOpen
  \bibfield  {author} {\bibinfo {author} {\bibfnamefont {B.}~\bibnamefont
  {Hensen}}, \bibinfo {author} {\bibfnamefont {H.}~\bibnamefont {Bernien}},
  \bibinfo {author} {\bibfnamefont {A.~E.}\ \bibnamefont {Dr{\'e}au}}, \bibinfo
  {author} {\bibfnamefont {A.}~\bibnamefont {Reiserer}}, \bibinfo {author}
  {\bibfnamefont {N.}~\bibnamefont {Kalb}}, \bibinfo {author} {\bibfnamefont
  {M.~S.}\ \bibnamefont {Blok}}, \bibinfo {author} {\bibfnamefont
  {J.}~\bibnamefont {Ruitenberg}}, \bibinfo {author} {\bibfnamefont {R.~F.~L.}\
  \bibnamefont {Vermeulen}}, \bibinfo {author} {\bibfnamefont {R.~N.}\
  \bibnamefont {Schouten}}, \bibinfo {author} {\bibfnamefont {C.}~\bibnamefont
  {Abell{\'a}n}}, \bibinfo {author} {\bibfnamefont {W.}~\bibnamefont {Amaya}},
  \bibinfo {author} {\bibfnamefont {V.}~\bibnamefont {Pruneri}}, \bibinfo
  {author} {\bibfnamefont {M.~W.}\ \bibnamefont {Mitchell}}, \bibinfo {author}
  {\bibfnamefont {M.}~\bibnamefont {Markham}}, \bibinfo {author} {\bibfnamefont
  {D.~J.}\ \bibnamefont {Twitchen}}, \bibinfo {author} {\bibfnamefont
  {D.}~\bibnamefont {Elkouss}}, \bibinfo {author} {\bibfnamefont
  {S.}~\bibnamefont {Wehner}}, \bibinfo {author} {\bibfnamefont {T.~H.}\
  \bibnamefont {Taminiau}},\ and\ \bibinfo {author} {\bibfnamefont
  {R.}~\bibnamefont {Hanson}},\ }\bibfield  {title} {\bibinfo {title}
  {Loophole-free {{Bell}} inequality violation using electron spins separated
  by 1.3 kilometres},\ }\href {https://doi.org/10.1038/nature15759} {\bibfield
  {journal} {\bibinfo  {journal} {Nature}\ }\textbf {\bibinfo {volume} {526}},\
  \bibinfo {pages} {682} (\bibinfo {year} {2015})}\BibitemShut {NoStop}%
\bibitem [{\citenamefont {Krovi}\ \emph {et~al.}(2016)\citenamefont {Krovi},
  \citenamefont {Guha}, \citenamefont {Dutton}, \citenamefont {Slater},
  \citenamefont {Simon},\ and\ \citenamefont {Tittel}}]{krovi2016}%
  \BibitemOpen
  \bibfield  {author} {\bibinfo {author} {\bibfnamefont {H.}~\bibnamefont
  {Krovi}}, \bibinfo {author} {\bibfnamefont {S.}~\bibnamefont {Guha}},
  \bibinfo {author} {\bibfnamefont {Z.}~\bibnamefont {Dutton}}, \bibinfo
  {author} {\bibfnamefont {J.~A.}\ \bibnamefont {Slater}}, \bibinfo {author}
  {\bibfnamefont {C.}~\bibnamefont {Simon}},\ and\ \bibinfo {author}
  {\bibfnamefont {W.}~\bibnamefont {Tittel}},\ }\bibfield  {title} {\bibinfo
  {title} {Practical quantum repeaters with parametric down-conversion
  sources},\ }\href {https://doi.org/10.1007/s00340-015-6297-4} {\bibfield
  {journal} {\bibinfo  {journal} {Appl. Phys. B}\ }\textbf {\bibinfo {volume}
  {122}},\ \bibinfo {pages} {52} (\bibinfo {year} {2016})}\BibitemShut
  {NoStop}%
\bibitem [{\citenamefont {Kar}\ and\ \citenamefont {Wehner}(2025)}]{kar2025}%
  \BibitemOpen
  \bibfield  {author} {\bibinfo {author} {\bibfnamefont {S.}~\bibnamefont
  {Kar}}\ and\ \bibinfo {author} {\bibfnamefont {S.}~\bibnamefont {Wehner}},\
  }\bibfield  {title} {\bibinfo {title} {Convexification of the {{Quantum
  Network Utility Maximization Problem}}},\ }\href
  {https://doi.org/10.1109/TQE.2024.3523889} {\bibfield  {journal} {\bibinfo
  {journal} {IEEE Transactions on Quantum Engineering}\ }\textbf {\bibinfo
  {volume} {6}},\ \bibinfo {pages} {1} (\bibinfo {year} {2025})}\BibitemShut
  {NoStop}%
\bibitem [{\citenamefont {Kingma}\ and\ \citenamefont {Ba}(2017)}]{kingma2017}%
  \BibitemOpen
  \bibfield  {author} {\bibinfo {author} {\bibfnamefont {D.~P.}\ \bibnamefont
  {Kingma}}\ and\ \bibinfo {author} {\bibfnamefont {J.}~\bibnamefont {Ba}},\
  }\href {https://doi.org/10.48550/arXiv.1412.6980} {\bibinfo {title} {Adam:
  {{A Method}} for {{Stochastic Optimization}}}} (\bibinfo {year} {2017}),\
  \Eprint {https://arxiv.org/abs/1412.6980} {arXiv:1412.6980 [cs]} \BibitemShut
  {NoStop}%
\bibitem [{\citenamefont {Stephens}(2023)}]{stephens2022}%
  \BibitemOpen
  \bibfield  {author} {\bibinfo {author} {\bibfnamefont {E.}~\bibnamefont
  {Stephens}},\ }\emph {\bibinfo {title} {Influence of Coherence Time Drift on
  the Secret Key Rate}},\ \href
  {https://repository.tudelft.nl/islandora/object/uuid%3Adffed3f3-c92b-43a1-a639-dbaa1bd8ad4f}
  {Master's thesis} (\bibinfo {year} {2023})\BibitemShut {NoStop}%
\bibitem [{\citenamefont {Danageozian}(2023)}]{danageozian2023}%
  \BibitemOpen
  \bibfield  {author} {\bibinfo {author} {\bibfnamefont {A.}~\bibnamefont
  {Danageozian}},\ }\bibfield  {title} {\bibinfo {title} {Recovery {{With
  Incomplete Knowledge}}: {{Fundamental Bounds}} on {{Real-Time Quantum
  Memories}}},\ }\href {https://doi.org/10.22331/q-2023-12-04-1195} {\bibfield
  {journal} {\bibinfo  {journal} {Quantum}\ }\textbf {\bibinfo {volume} {7}},\
  \bibinfo {pages} {1195} (\bibinfo {year} {2023})}\BibitemShut {NoStop}%
\bibitem [{\citenamefont {Klimov}\ \emph {et~al.}(2018)\citenamefont {Klimov},
  \citenamefont {Kelly}, \citenamefont {Chen}, \citenamefont {Neeley},
  \citenamefont {Megrant}, \citenamefont {Burkett}, \citenamefont {Barends},
  \citenamefont {Arya}, \citenamefont {Chiaro}, \citenamefont {Chen},
  \citenamefont {Dunsworth}, \citenamefont {Fowler}, \citenamefont {Foxen},
  \citenamefont {Gidney}, \citenamefont {Giustina}, \citenamefont {Graff},
  \citenamefont {Huang}, \citenamefont {Jeffrey}, \citenamefont {Lucero},
  \citenamefont {Mutus}, \citenamefont {Naaman}, \citenamefont {Neill},
  \citenamefont {Quintana}, \citenamefont {Roushan}, \citenamefont {Sank},
  \citenamefont {Vainsencher}, \citenamefont {Wenner}, \citenamefont {White},
  \citenamefont {Boixo}, \citenamefont {Babbush}, \citenamefont {Smelyanskiy},
  \citenamefont {Neven},\ and\ \citenamefont {Martinis}}]{klimov2018}%
  \BibitemOpen
  \bibfield  {author} {\bibinfo {author} {\bibfnamefont {P.~V.}\ \bibnamefont
  {Klimov}}, \bibinfo {author} {\bibfnamefont {J.}~\bibnamefont {Kelly}},
  \bibinfo {author} {\bibfnamefont {Z.}~\bibnamefont {Chen}}, \bibinfo {author}
  {\bibfnamefont {M.}~\bibnamefont {Neeley}}, \bibinfo {author} {\bibfnamefont
  {A.}~\bibnamefont {Megrant}}, \bibinfo {author} {\bibfnamefont
  {B.}~\bibnamefont {Burkett}}, \bibinfo {author} {\bibfnamefont
  {R.}~\bibnamefont {Barends}}, \bibinfo {author} {\bibfnamefont
  {K.}~\bibnamefont {Arya}}, \bibinfo {author} {\bibfnamefont {B.}~\bibnamefont
  {Chiaro}}, \bibinfo {author} {\bibfnamefont {Y.}~\bibnamefont {Chen}},
  \bibinfo {author} {\bibfnamefont {A.}~\bibnamefont {Dunsworth}}, \bibinfo
  {author} {\bibfnamefont {A.}~\bibnamefont {Fowler}}, \bibinfo {author}
  {\bibfnamefont {B.}~\bibnamefont {Foxen}}, \bibinfo {author} {\bibfnamefont
  {C.}~\bibnamefont {Gidney}}, \bibinfo {author} {\bibfnamefont
  {M.}~\bibnamefont {Giustina}}, \bibinfo {author} {\bibfnamefont
  {R.}~\bibnamefont {Graff}}, \bibinfo {author} {\bibfnamefont
  {T.}~\bibnamefont {Huang}}, \bibinfo {author} {\bibfnamefont
  {E.}~\bibnamefont {Jeffrey}}, \bibinfo {author} {\bibfnamefont
  {E.}~\bibnamefont {Lucero}}, \bibinfo {author} {\bibfnamefont {J.~Y.}\
  \bibnamefont {Mutus}}, \bibinfo {author} {\bibfnamefont {O.}~\bibnamefont
  {Naaman}}, \bibinfo {author} {\bibfnamefont {C.}~\bibnamefont {Neill}},
  \bibinfo {author} {\bibfnamefont {C.}~\bibnamefont {Quintana}}, \bibinfo
  {author} {\bibfnamefont {P.}~\bibnamefont {Roushan}}, \bibinfo {author}
  {\bibfnamefont {D.}~\bibnamefont {Sank}}, \bibinfo {author} {\bibfnamefont
  {A.}~\bibnamefont {Vainsencher}}, \bibinfo {author} {\bibfnamefont
  {J.}~\bibnamefont {Wenner}}, \bibinfo {author} {\bibfnamefont {T.~C.}\
  \bibnamefont {White}}, \bibinfo {author} {\bibfnamefont {S.}~\bibnamefont
  {Boixo}}, \bibinfo {author} {\bibfnamefont {R.}~\bibnamefont {Babbush}},
  \bibinfo {author} {\bibfnamefont {V.~N.}\ \bibnamefont {Smelyanskiy}},
  \bibinfo {author} {\bibfnamefont {H.}~\bibnamefont {Neven}},\ and\ \bibinfo
  {author} {\bibfnamefont {J.~M.}\ \bibnamefont {Martinis}},\ }\bibfield
  {title} {\bibinfo {title} {Fluctuations of {{Energy-Relaxation Times}} in
  {{Superconducting Qubits}}},\ }\href
  {https://doi.org/10.1103/PhysRevLett.121.090502} {\bibfield  {journal}
  {\bibinfo  {journal} {Phys. Rev. Lett.}\ }\textbf {\bibinfo {volume} {121}},\
  \bibinfo {pages} {090502} (\bibinfo {year} {2018})}\BibitemShut {NoStop}%
\bibitem [{\citenamefont {Matityahu}\ \emph {et~al.}(2021)\citenamefont
  {Matityahu}, \citenamefont {Shnirman},\ and\ \citenamefont
  {Schechter}}]{matityahu2021}%
  \BibitemOpen
  \bibfield  {author} {\bibinfo {author} {\bibfnamefont {S.}~\bibnamefont
  {Matityahu}}, \bibinfo {author} {\bibfnamefont {A.}~\bibnamefont
  {Shnirman}},\ and\ \bibinfo {author} {\bibfnamefont {M.}~\bibnamefont
  {Schechter}},\ }\bibfield  {title} {\bibinfo {title} {Stabilization of
  {{Qubit Relaxation Rates}} by {{Frequency Modulation}}},\ }\href
  {https://doi.org/10.1103/PhysRevApplied.16.044036} {\bibfield  {journal}
  {\bibinfo  {journal} {Phys. Rev. Appl.}\ }\textbf {\bibinfo {volume} {16}},\
  \bibinfo {pages} {044036} (\bibinfo {year} {2021})}\BibitemShut {NoStop}%
\bibitem [{\citenamefont {Gottesman}\ \emph {et~al.}(2012)\citenamefont
  {Gottesman}, \citenamefont {Jennewein},\ and\ \citenamefont
  {Croke}}]{gottesman2012}%
  \BibitemOpen
  \bibfield  {author} {\bibinfo {author} {\bibfnamefont {D.}~\bibnamefont
  {Gottesman}}, \bibinfo {author} {\bibfnamefont {T.}~\bibnamefont
  {Jennewein}},\ and\ \bibinfo {author} {\bibfnamefont {S.}~\bibnamefont
  {Croke}},\ }\bibfield  {title} {\bibinfo {title} {Longer-{{Baseline
  Telescopes Using Quantum Repeaters}}},\ }\href
  {https://doi.org/10.1103/PhysRevLett.109.070503} {\bibfield  {journal}
  {\bibinfo  {journal} {Phys. Rev. Lett.}\ }\textbf {\bibinfo {volume} {109}},\
  \bibinfo {pages} {070503} (\bibinfo {year} {2012})}\BibitemShut {NoStop}%
\bibitem [{\citenamefont {Khabiboulline}\ \emph {et~al.}(2019)\citenamefont
  {Khabiboulline}, \citenamefont {Borregaard}, \citenamefont {De~Greve},\ and\
  \citenamefont {Lukin}}]{khabiboulline2019a}%
  \BibitemOpen
  \bibfield  {author} {\bibinfo {author} {\bibfnamefont {E.~T.}\ \bibnamefont
  {Khabiboulline}}, \bibinfo {author} {\bibfnamefont {J.}~\bibnamefont
  {Borregaard}}, \bibinfo {author} {\bibfnamefont {K.}~\bibnamefont
  {De~Greve}},\ and\ \bibinfo {author} {\bibfnamefont {M.~D.}\ \bibnamefont
  {Lukin}},\ }\bibfield  {title} {\bibinfo {title} {Quantum-assisted telescope
  arrays},\ }\href {https://doi.org/10.1103/PhysRevA.100.022316} {\bibfield
  {journal} {\bibinfo  {journal} {Phys. Rev. A}\ }\textbf {\bibinfo {volume}
  {100}},\ \bibinfo {pages} {022316} (\bibinfo {year} {2019})}\BibitemShut
  {NoStop}%
\bibitem [{\citenamefont {Brady}\ \emph {et~al.}(2022)\citenamefont {Brady},
  \citenamefont {Gao}, \citenamefont {Harnik}, \citenamefont {Liu},
  \citenamefont {Zhang},\ and\ \citenamefont {Zhuang}}]{brady2022}%
  \BibitemOpen
  \bibfield  {author} {\bibinfo {author} {\bibfnamefont {A.~J.}\ \bibnamefont
  {Brady}}, \bibinfo {author} {\bibfnamefont {C.}~\bibnamefont {Gao}}, \bibinfo
  {author} {\bibfnamefont {R.}~\bibnamefont {Harnik}}, \bibinfo {author}
  {\bibfnamefont {Z.}~\bibnamefont {Liu}}, \bibinfo {author} {\bibfnamefont
  {Z.}~\bibnamefont {Zhang}},\ and\ \bibinfo {author} {\bibfnamefont
  {Q.}~\bibnamefont {Zhuang}},\ }\bibfield  {title} {\bibinfo {title}
  {Entangled {{Sensor-Networks}} for {{Dark-Matter Searches}}},\ }\href
  {https://doi.org/10.1103/PRXQuantum.3.030333} {\bibfield  {journal} {\bibinfo
   {journal} {PRX Quantum}\ }\textbf {\bibinfo {volume} {3}},\ \bibinfo {pages}
  {030333} (\bibinfo {year} {2022})}\BibitemShut {NoStop}%
\bibitem [{\citenamefont {Guo}\ \emph {et~al.}(2020)\citenamefont {Guo},
  \citenamefont {Breum}, \citenamefont {Borregaard}, \citenamefont {Izumi},
  \citenamefont {Larsen}, \citenamefont {Gehring}, \citenamefont {Christandl},
  \citenamefont {{Neergaard-Nielsen}},\ and\ \citenamefont
  {Andersen}}]{guo2020a}%
  \BibitemOpen
  \bibfield  {author} {\bibinfo {author} {\bibfnamefont {X.}~\bibnamefont
  {Guo}}, \bibinfo {author} {\bibfnamefont {C.~R.}\ \bibnamefont {Breum}},
  \bibinfo {author} {\bibfnamefont {J.}~\bibnamefont {Borregaard}}, \bibinfo
  {author} {\bibfnamefont {S.}~\bibnamefont {Izumi}}, \bibinfo {author}
  {\bibfnamefont {M.~V.}\ \bibnamefont {Larsen}}, \bibinfo {author}
  {\bibfnamefont {T.}~\bibnamefont {Gehring}}, \bibinfo {author} {\bibfnamefont
  {M.}~\bibnamefont {Christandl}}, \bibinfo {author} {\bibfnamefont {J.~S.}\
  \bibnamefont {{Neergaard-Nielsen}}},\ and\ \bibinfo {author} {\bibfnamefont
  {U.~L.}\ \bibnamefont {Andersen}},\ }\bibfield  {title} {\bibinfo {title}
  {Distributed quantum sensing in a continuous-variable entangled network},\
  }\href {https://doi.org/10.1038/s41567-019-0743-x} {\bibfield  {journal}
  {\bibinfo  {journal} {Nat. Phys.}\ }\textbf {\bibinfo {volume} {16}},\
  \bibinfo {pages} {281} (\bibinfo {year} {2020})}\BibitemShut {NoStop}%
\bibitem [{\citenamefont {Sekatski}\ \emph {et~al.}(2020)\citenamefont
  {Sekatski}, \citenamefont {W{\"o}lk},\ and\ \citenamefont
  {D{\"u}r}}]{sekatski2020}%
  \BibitemOpen
  \bibfield  {author} {\bibinfo {author} {\bibfnamefont {P.}~\bibnamefont
  {Sekatski}}, \bibinfo {author} {\bibfnamefont {S.}~\bibnamefont {W{\"o}lk}},\
  and\ \bibinfo {author} {\bibfnamefont {W.}~\bibnamefont {D{\"u}r}},\
  }\bibfield  {title} {\bibinfo {title} {Optimal distributed sensing in noisy
  environments},\ }\href {https://doi.org/10.1103/PhysRevResearch.2.023052}
  {\bibfield  {journal} {\bibinfo  {journal} {Phys. Rev. Res.}\ }\textbf
  {\bibinfo {volume} {2}},\ \bibinfo {pages} {023052} (\bibinfo {year}
  {2020})}\BibitemShut {NoStop}%
\bibitem [{\citenamefont {Xia}\ \emph {et~al.}(2019)\citenamefont {Xia},
  \citenamefont {Zhuang}, \citenamefont {Clark},\ and\ \citenamefont
  {Zhang}}]{xia2019}%
  \BibitemOpen
  \bibfield  {author} {\bibinfo {author} {\bibfnamefont {Y.}~\bibnamefont
  {Xia}}, \bibinfo {author} {\bibfnamefont {Q.}~\bibnamefont {Zhuang}},
  \bibinfo {author} {\bibfnamefont {W.}~\bibnamefont {Clark}},\ and\ \bibinfo
  {author} {\bibfnamefont {Z.}~\bibnamefont {Zhang}},\ }\bibfield  {title}
  {\bibinfo {title} {Repeater-enhanced distributed quantum sensing based on
  continuous-variable multipartite entanglement},\ }\href
  {https://doi.org/10.1103/PhysRevA.99.012328} {\bibfield  {journal} {\bibinfo
  {journal} {Phys. Rev. A}\ }\textbf {\bibinfo {volume} {99}},\ \bibinfo
  {pages} {012328} (\bibinfo {year} {2019})}\BibitemShut {NoStop}%
\bibitem [{\citenamefont {Zang}\ \emph {et~al.}(2024)\citenamefont {Zang},
  \citenamefont {Kolar}, \citenamefont {Gonzales}, \citenamefont {Chung},
  \citenamefont {Gray}, \citenamefont {Kettimuthu}, \citenamefont {Zhong},\
  and\ \citenamefont {Saleem}}]{zang2024}%
  \BibitemOpen
  \bibfield  {author} {\bibinfo {author} {\bibfnamefont {A.}~\bibnamefont
  {Zang}}, \bibinfo {author} {\bibfnamefont {A.}~\bibnamefont {Kolar}},
  \bibinfo {author} {\bibfnamefont {A.}~\bibnamefont {Gonzales}}, \bibinfo
  {author} {\bibfnamefont {J.}~\bibnamefont {Chung}}, \bibinfo {author}
  {\bibfnamefont {S.~K.}\ \bibnamefont {Gray}}, \bibinfo {author}
  {\bibfnamefont {R.}~\bibnamefont {Kettimuthu}}, \bibinfo {author}
  {\bibfnamefont {T.}~\bibnamefont {Zhong}},\ and\ \bibinfo {author}
  {\bibfnamefont {Z.~H.}\ \bibnamefont {Saleem}},\ }\href
  {https://doi.org/10.48550/arXiv.2409.17089} {\bibinfo {title} {Quantum
  {{Advantage}} in {{Distributed Sensing}} with {{Noisy Quantum Networks}}}}
  (\bibinfo {year} {2024}),\ \Eprint {https://arxiv.org/abs/2409.17089}
  {arXiv:2409.17089 [quant-ph]} \BibitemShut {NoStop}%
\bibitem [{\citenamefont {Zhang}\ and\ \citenamefont
  {Zhuang}(2021)}]{zhang2021}%
  \BibitemOpen
  \bibfield  {author} {\bibinfo {author} {\bibfnamefont {Z.}~\bibnamefont
  {Zhang}}\ and\ \bibinfo {author} {\bibfnamefont {Q.}~\bibnamefont {Zhuang}},\
  }\bibfield  {title} {\bibinfo {title} {Distributed quantum sensing},\ }\href
  {https://doi.org/10.1088/2058-9565/abd4c3} {\bibfield  {journal} {\bibinfo
  {journal} {Quantum Sci. Technol.}\ }\textbf {\bibinfo {volume} {6}},\
  \bibinfo {pages} {043001} (\bibinfo {year} {2021})}\BibitemShut {NoStop}%
\bibitem [{\citenamefont {Kumar}\ \emph {et~al.}(2022)\citenamefont {Kumar},
  \citenamefont {{Augusto de Jesus Pacheco}}, \citenamefont {Kaushik},\ and\
  \citenamefont {Rodrigues}}]{kumar2022}%
  \BibitemOpen
  \bibfield  {author} {\bibinfo {author} {\bibfnamefont {A.}~\bibnamefont
  {Kumar}}, \bibinfo {author} {\bibfnamefont {D.}~\bibnamefont {{Augusto de
  Jesus Pacheco}}}, \bibinfo {author} {\bibfnamefont {K.}~\bibnamefont
  {Kaushik}},\ and\ \bibinfo {author} {\bibfnamefont {J.~J. P.~C.}\
  \bibnamefont {Rodrigues}},\ }\bibfield  {title} {\bibinfo {title} {Futuristic
  view of the {{Internet}} of {{Quantum Drones}}: {{Review}}, challenges and
  research agenda},\ }\href {https://doi.org/10.1016/j.vehcom.2022.100487}
  {\bibfield  {journal} {\bibinfo  {journal} {Vehicular Communications}\
  }\textbf {\bibinfo {volume} {36}},\ \bibinfo {pages} {100487} (\bibinfo
  {year} {2022})}\BibitemShut {NoStop}%
\bibitem [{\citenamefont {Liu}\ \emph {et~al.}(2020)\citenamefont {Liu},
  \citenamefont {Tian}, \citenamefont {Gu}, \citenamefont {Fan}, \citenamefont
  {Ni}, \citenamefont {Yang}, \citenamefont {Zhang}, \citenamefont {Hu},
  \citenamefont {Guo}, \citenamefont {Cao}, \citenamefont {Hu}, \citenamefont
  {Zhao}, \citenamefont {Lu}, \citenamefont {Gong}, \citenamefont {Xie},\ and\
  \citenamefont {Zhu}}]{liu2020b}%
  \BibitemOpen
  \bibfield  {author} {\bibinfo {author} {\bibfnamefont {H.-Y.}\ \bibnamefont
  {Liu}}, \bibinfo {author} {\bibfnamefont {X.-H.}\ \bibnamefont {Tian}},
  \bibinfo {author} {\bibfnamefont {C.}~\bibnamefont {Gu}}, \bibinfo {author}
  {\bibfnamefont {P.}~\bibnamefont {Fan}}, \bibinfo {author} {\bibfnamefont
  {X.}~\bibnamefont {Ni}}, \bibinfo {author} {\bibfnamefont {R.}~\bibnamefont
  {Yang}}, \bibinfo {author} {\bibfnamefont {J.-N.}\ \bibnamefont {Zhang}},
  \bibinfo {author} {\bibfnamefont {M.}~\bibnamefont {Hu}}, \bibinfo {author}
  {\bibfnamefont {J.}~\bibnamefont {Guo}}, \bibinfo {author} {\bibfnamefont
  {X.}~\bibnamefont {Cao}}, \bibinfo {author} {\bibfnamefont {X.}~\bibnamefont
  {Hu}}, \bibinfo {author} {\bibfnamefont {G.}~\bibnamefont {Zhao}}, \bibinfo
  {author} {\bibfnamefont {Y.-Q.}\ \bibnamefont {Lu}}, \bibinfo {author}
  {\bibfnamefont {Y.-X.}\ \bibnamefont {Gong}}, \bibinfo {author}
  {\bibfnamefont {Z.}~\bibnamefont {Xie}},\ and\ \bibinfo {author}
  {\bibfnamefont {S.-N.}\ \bibnamefont {Zhu}},\ }\bibfield  {title} {\bibinfo
  {title} {Drone-based entanglement distribution towards mobile quantum
  networks},\ }\href {https://doi.org/10.1093/nsr/nwz227} {\bibfield  {journal}
  {\bibinfo  {journal} {National Science Review}\ }\textbf {\bibinfo {volume}
  {7}},\ \bibinfo {pages} {921} (\bibinfo {year} {2020})}\BibitemShut {NoStop}%
\bibitem [{\citenamefont {Conrad}\ \emph {et~al.}(2023)\citenamefont {Conrad},
  \citenamefont {Isaac}, \citenamefont {Cochran}, \citenamefont
  {{Sanchez-Rosales}}, \citenamefont {Rezaei}, \citenamefont {Javid},
  \citenamefont {Schroeder}, \citenamefont {Golba}, \citenamefont {Gauthier},\
  and\ \citenamefont {Kwiat}}]{conrad2023}%
  \BibitemOpen
  \bibfield  {author} {\bibinfo {author} {\bibfnamefont {A.}~\bibnamefont
  {Conrad}}, \bibinfo {author} {\bibfnamefont {S.}~\bibnamefont {Isaac}},
  \bibinfo {author} {\bibfnamefont {R.}~\bibnamefont {Cochran}}, \bibinfo
  {author} {\bibfnamefont {D.}~\bibnamefont {{Sanchez-Rosales}}}, \bibinfo
  {author} {\bibfnamefont {T.}~\bibnamefont {Rezaei}}, \bibinfo {author}
  {\bibfnamefont {T.}~\bibnamefont {Javid}}, \bibinfo {author} {\bibfnamefont
  {A.~J.}\ \bibnamefont {Schroeder}}, \bibinfo {author} {\bibfnamefont
  {G.}~\bibnamefont {Golba}}, \bibinfo {author} {\bibfnamefont
  {D.}~\bibnamefont {Gauthier}},\ and\ \bibinfo {author} {\bibfnamefont
  {P.}~\bibnamefont {Kwiat}},\ }\bibfield  {title} {\bibinfo {title}
  {Drone-based quantum communication links},\ }in\ \href
  {https://doi.org/10.1117/12.2647923} {\emph {\bibinfo {booktitle} {Quantum
  {{Computing}}, {{Communication}}, and {{Simulation III}}}}},\ Vol.\ \bibinfo
  {volume} {12446}\ (\bibinfo  {publisher} {SPIE},\ \bibinfo {year} {2023})\
  pp.\ \bibinfo {pages} {99--106}\BibitemShut {NoStop}%
\bibitem [{\citenamefont {{Karakosta-Amarantidou}}\ \emph
  {et~al.}(2025)\citenamefont {{Karakosta-Amarantidou}}, \citenamefont
  {Yehia},\ and\ \citenamefont {Schiavon}}]{karakosta-amarantidou2025}%
  \BibitemOpen
  \bibfield  {author} {\bibinfo {author} {\bibfnamefont {I.}~\bibnamefont
  {{Karakosta-Amarantidou}}}, \bibinfo {author} {\bibfnamefont
  {R.}~\bibnamefont {Yehia}},\ and\ \bibinfo {author} {\bibfnamefont
  {M.}~\bibnamefont {Schiavon}},\ }\bibfield  {title} {\bibinfo {title}
  {Free-space model for a balloon-based quantum network},\ }\href
  {https://doi.org/10.1103/PhysRevResearch.7.023199} {\bibfield  {journal}
  {\bibinfo  {journal} {Phys. Rev. Res.}\ }\textbf {\bibinfo {volume} {7}},\
  \bibinfo {pages} {023199} (\bibinfo {year} {2025})}\BibitemShut {NoStop}%
\bibitem [{\citenamefont {Hill}\ \emph {et~al.}(2016)\citenamefont {Hill},
  \citenamefont {Christensen},\ and\ \citenamefont {Kwiat}}]{hill2016}%
  \BibitemOpen
  \bibfield  {author} {\bibinfo {author} {\bibfnamefont {A.~D.}\ \bibnamefont
  {Hill}}, \bibinfo {author} {\bibfnamefont {B.}~\bibnamefont {Christensen}},\
  and\ \bibinfo {author} {\bibfnamefont {P.~G.}\ \bibnamefont {Kwiat}},\
  }\bibfield  {title} {\bibinfo {title} {Advanced techniques for free-space
  optical quantum cryptography over water},\ }in\ \href
  {https://doi.org/10.1117/12.2218270} {\emph {\bibinfo {booktitle}
  {Free-{{Space Laser Communication}} and {{Atmospheric Propagation
  XXVIII}}}}},\ Vol.\ \bibinfo {volume} {9739}\ (\bibinfo  {publisher} {SPIE},\
  \bibinfo {year} {2016})\ pp.\ \bibinfo {pages} {306--311}\BibitemShut
  {NoStop}%
\bibitem [{\citenamefont {Qin}\ \emph {et~al.}(2021)\citenamefont {Qin},
  \citenamefont {Li},\ and\ \citenamefont {Wang}}]{qin2021}%
  \BibitemOpen
  \bibfield  {author} {\bibinfo {author} {\bibfnamefont {L.}~\bibnamefont
  {Qin}}, \bibinfo {author} {\bibfnamefont {Y.}~\bibnamefont {Li}},\ and\
  \bibinfo {author} {\bibfnamefont {S.}~\bibnamefont {Wang}},\ }\bibfield
  {title} {\bibinfo {title} {A key distribution protocol based on quantum
  entanglement swapping for unmanned surface vehicle},\ }\href
  {https://doi.org/10.1088/1742-6596/1976/1/012034} {\bibfield  {journal}
  {\bibinfo  {journal} {J. Phys.: Conf. Ser.}\ }\textbf {\bibinfo {volume}
  {1976}},\ \bibinfo {pages} {012034} (\bibinfo {year} {2021})}\BibitemShut
  {NoStop}%
\bibitem [{\citenamefont {Gariano}\ and\ \citenamefont
  {Djordjevic}(2019)}]{gariano2019}%
  \BibitemOpen
  \bibfield  {author} {\bibinfo {author} {\bibfnamefont {J.}~\bibnamefont
  {Gariano}}\ and\ \bibinfo {author} {\bibfnamefont {I.~B.}\ \bibnamefont
  {Djordjevic}},\ }\bibfield  {title} {\bibinfo {title} {Theoretical study of a
  submarine to submarine quantum key distribution systems},\ }\href
  {https://doi.org/10.1364/OE.27.003055} {\bibfield  {journal} {\bibinfo
  {journal} {Opt. Express, OE}\ }\textbf {\bibinfo {volume} {27}},\ \bibinfo
  {pages} {3055} (\bibinfo {year} {2019})}\BibitemShut {NoStop}%
\bibitem [{\citenamefont {Yin}\ \emph {et~al.}(2020)\citenamefont {Yin},
  \citenamefont {Li}, \citenamefont {Liao}, \citenamefont {Yang}, \citenamefont
  {Cao}, \citenamefont {Zhang}, \citenamefont {Ren}, \citenamefont {Cai},
  \citenamefont {Liu}, \citenamefont {Li}, \citenamefont {Shu}, \citenamefont
  {Huang}, \citenamefont {Deng}, \citenamefont {Li}, \citenamefont {Zhang},
  \citenamefont {Liu}, \citenamefont {Chen}, \citenamefont {Lu}, \citenamefont
  {Wang}, \citenamefont {Xu}, \citenamefont {Wang}, \citenamefont {Peng},
  \citenamefont {Ekert},\ and\ \citenamefont {Pan}}]{yin2020}%
  \BibitemOpen
  \bibfield  {author} {\bibinfo {author} {\bibfnamefont {J.}~\bibnamefont
  {Yin}}, \bibinfo {author} {\bibfnamefont {Y.-H.}\ \bibnamefont {Li}},
  \bibinfo {author} {\bibfnamefont {S.-K.}\ \bibnamefont {Liao}}, \bibinfo
  {author} {\bibfnamefont {M.}~\bibnamefont {Yang}}, \bibinfo {author}
  {\bibfnamefont {Y.}~\bibnamefont {Cao}}, \bibinfo {author} {\bibfnamefont
  {L.}~\bibnamefont {Zhang}}, \bibinfo {author} {\bibfnamefont {J.-G.}\
  \bibnamefont {Ren}}, \bibinfo {author} {\bibfnamefont {W.-Q.}\ \bibnamefont
  {Cai}}, \bibinfo {author} {\bibfnamefont {W.-Y.}\ \bibnamefont {Liu}},
  \bibinfo {author} {\bibfnamefont {S.-L.}\ \bibnamefont {Li}}, \bibinfo
  {author} {\bibfnamefont {R.}~\bibnamefont {Shu}}, \bibinfo {author}
  {\bibfnamefont {Y.-M.}\ \bibnamefont {Huang}}, \bibinfo {author}
  {\bibfnamefont {L.}~\bibnamefont {Deng}}, \bibinfo {author} {\bibfnamefont
  {L.}~\bibnamefont {Li}}, \bibinfo {author} {\bibfnamefont {Q.}~\bibnamefont
  {Zhang}}, \bibinfo {author} {\bibfnamefont {N.-L.}\ \bibnamefont {Liu}},
  \bibinfo {author} {\bibfnamefont {Y.-A.}\ \bibnamefont {Chen}}, \bibinfo
  {author} {\bibfnamefont {C.-Y.}\ \bibnamefont {Lu}}, \bibinfo {author}
  {\bibfnamefont {X.-B.}\ \bibnamefont {Wang}}, \bibinfo {author}
  {\bibfnamefont {F.}~\bibnamefont {Xu}}, \bibinfo {author} {\bibfnamefont
  {J.-Y.}\ \bibnamefont {Wang}}, \bibinfo {author} {\bibfnamefont {C.-Z.}\
  \bibnamefont {Peng}}, \bibinfo {author} {\bibfnamefont {A.~K.}\ \bibnamefont
  {Ekert}},\ and\ \bibinfo {author} {\bibfnamefont {J.-W.}\ \bibnamefont
  {Pan}},\ }\bibfield  {title} {\bibinfo {title} {Entanglement-based secure
  quantum cryptography over 1,120 kilometres},\ }\href
  {https://doi.org/10.1038/s41586-020-2401-y} {\bibfield  {journal} {\bibinfo
  {journal} {Nature}\ ,\ \bibinfo {pages} {1}} (\bibinfo {year}
  {2020})}\BibitemShut {NoStop}%
\bibitem [{\citenamefont {Vergoossen}\ \emph {et~al.}(2020)\citenamefont
  {Vergoossen}, \citenamefont {Loarte}, \citenamefont {Bedington},
  \citenamefont {Kuiper},\ and\ \citenamefont {Ling}}]{vergoossen2020}%
  \BibitemOpen
  \bibfield  {author} {\bibinfo {author} {\bibfnamefont {T.}~\bibnamefont
  {Vergoossen}}, \bibinfo {author} {\bibfnamefont {S.}~\bibnamefont {Loarte}},
  \bibinfo {author} {\bibfnamefont {R.}~\bibnamefont {Bedington}}, \bibinfo
  {author} {\bibfnamefont {H.}~\bibnamefont {Kuiper}},\ and\ \bibinfo {author}
  {\bibfnamefont {A.}~\bibnamefont {Ling}},\ }\bibfield  {title} {\bibinfo
  {title} {Modelling of satellite constellations for trusted node {{QKD}}
  networks},\ }\href {https://doi.org/10.1016/j.actaastro.2020.02.010}
  {\bibfield  {journal} {\bibinfo  {journal} {Acta Astronautica}\ }\textbf
  {\bibinfo {volume} {173}},\ \bibinfo {pages} {164} (\bibinfo {year}
  {2020})}\BibitemShut {NoStop}%
\bibitem [{\citenamefont {Sidhu}\ \emph {et~al.}(2021)\citenamefont {Sidhu},
  \citenamefont {Joshi}, \citenamefont {G{\"u}ndo{\u g}an}, \citenamefont
  {Brougham}, \citenamefont {Lowndes}, \citenamefont {Mazzarella},
  \citenamefont {Krutzik}, \citenamefont {Mohapatra}, \citenamefont {Dequal},
  \citenamefont {Vallone}, \citenamefont {Villoresi}, \citenamefont {Ling},
  \citenamefont {Jennewein}, \citenamefont {Mohageg}, \citenamefont {Rarity},
  \citenamefont {Fuentes}, \citenamefont {Pirandola},\ and\ \citenamefont
  {Oi}}]{sidhu2021}%
  \BibitemOpen
  \bibfield  {author} {\bibinfo {author} {\bibfnamefont {J.~S.}\ \bibnamefont
  {Sidhu}}, \bibinfo {author} {\bibfnamefont {S.~K.}\ \bibnamefont {Joshi}},
  \bibinfo {author} {\bibfnamefont {M.}~\bibnamefont {G{\"u}ndo{\u g}an}},
  \bibinfo {author} {\bibfnamefont {T.}~\bibnamefont {Brougham}}, \bibinfo
  {author} {\bibfnamefont {D.}~\bibnamefont {Lowndes}}, \bibinfo {author}
  {\bibfnamefont {L.}~\bibnamefont {Mazzarella}}, \bibinfo {author}
  {\bibfnamefont {M.}~\bibnamefont {Krutzik}}, \bibinfo {author} {\bibfnamefont
  {S.}~\bibnamefont {Mohapatra}}, \bibinfo {author} {\bibfnamefont
  {D.}~\bibnamefont {Dequal}}, \bibinfo {author} {\bibfnamefont
  {G.}~\bibnamefont {Vallone}}, \bibinfo {author} {\bibfnamefont
  {P.}~\bibnamefont {Villoresi}}, \bibinfo {author} {\bibfnamefont
  {A.}~\bibnamefont {Ling}}, \bibinfo {author} {\bibfnamefont {T.}~\bibnamefont
  {Jennewein}}, \bibinfo {author} {\bibfnamefont {M.}~\bibnamefont {Mohageg}},
  \bibinfo {author} {\bibfnamefont {J.~G.}\ \bibnamefont {Rarity}}, \bibinfo
  {author} {\bibfnamefont {I.}~\bibnamefont {Fuentes}}, \bibinfo {author}
  {\bibfnamefont {S.}~\bibnamefont {Pirandola}},\ and\ \bibinfo {author}
  {\bibfnamefont {D.~K.~L.}\ \bibnamefont {Oi}},\ }\bibfield  {title} {\bibinfo
  {title} {Advances in space quantum communications},\ }\href
  {https://doi.org/10.1049/qtc2.12015} {\bibfield  {journal} {\bibinfo
  {journal} {IET Quantum Communication}\ }\textbf {\bibinfo {volume} {2}},\
  \bibinfo {pages} {182} (\bibinfo {year} {2021})}\BibitemShut {NoStop}%
\bibitem [{\citenamefont {Delahaye}\ \emph {et~al.}(2019)\citenamefont
  {Delahaye}, \citenamefont {Chaimatanan},\ and\ \citenamefont
  {Mongeau}}]{delahaye2019}%
  \BibitemOpen
  \bibfield  {author} {\bibinfo {author} {\bibfnamefont {D.}~\bibnamefont
  {Delahaye}}, \bibinfo {author} {\bibfnamefont {S.}~\bibnamefont
  {Chaimatanan}},\ and\ \bibinfo {author} {\bibfnamefont {M.}~\bibnamefont
  {Mongeau}},\ }\bibfield  {title} {\bibinfo {title} {Simulated {{Annealing}}:
  {{From Basics}} to {{Applications}}},\ }in\ \href
  {https://doi.org/10.1007/978-3-319-91086-4_1} {\emph {\bibinfo {booktitle}
  {Handbook of {{Metaheuristics}}}}},\ Vol.\ \bibinfo {volume} {272},\ \bibinfo
  {editor} {edited by\ \bibinfo {editor} {\bibfnamefont {M.}~\bibnamefont
  {Gendreau}}\ and\ \bibinfo {editor} {\bibfnamefont {J.-Y.}\ \bibnamefont
  {Potvin}}}\ (\bibinfo  {publisher} {Springer International Publishing},\
  \bibinfo {address} {Cham},\ \bibinfo {year} {2019})\ pp.\ \bibinfo {pages}
  {1--35}\BibitemShut {NoStop}%
\bibitem [{\citenamefont {Avis}(2025{\natexlab{b}})}]{repeater_placement.jl}%
  \BibitemOpen
  \bibfield  {author} {\bibinfo {author} {\bibfnamefont {G.}~\bibnamefont
  {Avis}},\ }\href {https://github.com/GuusAvis/RepeaterPlacement.jl} {\bibinfo
  {title} {Guusavis/repeaterplacement.jl}} (\bibinfo {year}
  {2025}{\natexlab{b}})\BibitemShut {NoStop}%
\bibitem [{\citenamefont {Epping}\ \emph {et~al.}(2017)\citenamefont {Epping},
  \citenamefont {Kampermann}, \citenamefont {Macchiavello},\ and\ \citenamefont
  {Bru{\ss}}}]{epping2017}%
  \BibitemOpen
  \bibfield  {author} {\bibinfo {author} {\bibfnamefont {M.}~\bibnamefont
  {Epping}}, \bibinfo {author} {\bibfnamefont {H.}~\bibnamefont {Kampermann}},
  \bibinfo {author} {\bibfnamefont {C.}~\bibnamefont {Macchiavello}},\ and\
  \bibinfo {author} {\bibfnamefont {D.}~\bibnamefont {Bru{\ss}}},\ }\bibfield
  {title} {\bibinfo {title} {Multi-partite entanglement speeds up quantum key
  distribution in networks},\ }\href {https://doi.org/10.1088/1367-2630/aa8487}
  {\bibfield  {journal} {\bibinfo  {journal} {New J. Phys.}\ }\textbf {\bibinfo
  {volume} {19}},\ \bibinfo {pages} {093012} (\bibinfo {year}
  {2017})}\BibitemShut {NoStop}%
\bibitem [{\citenamefont {Murta}\ \emph {et~al.}(2020)\citenamefont {Murta},
  \citenamefont {Grasselli}, \citenamefont {Kampermann},\ and\ \citenamefont
  {Bru{\ss}}}]{murta2020a}%
  \BibitemOpen
  \bibfield  {author} {\bibinfo {author} {\bibfnamefont {G.}~\bibnamefont
  {Murta}}, \bibinfo {author} {\bibfnamefont {F.}~\bibnamefont {Grasselli}},
  \bibinfo {author} {\bibfnamefont {H.}~\bibnamefont {Kampermann}},\ and\
  \bibinfo {author} {\bibfnamefont {D.}~\bibnamefont {Bru{\ss}}},\ }\bibfield
  {title} {\bibinfo {title} {Quantum {{Conference Key Agreement}}: {{A
  Review}}},\ }\href {https://doi.org/10.1002/qute.202000025} {\bibfield
  {journal} {\bibinfo  {journal} {Advanced Quantum Technologies}\ }\textbf
  {\bibinfo {volume} {3}},\ \bibinfo {pages} {2000025} (\bibinfo {year}
  {2020})}\BibitemShut {NoStop}%
\bibitem [{\citenamefont {Colson}\ \emph {et~al.}(2007)\citenamefont {Colson},
  \citenamefont {Marcotte},\ and\ \citenamefont {Savard}}]{colson2007}%
  \BibitemOpen
  \bibfield  {author} {\bibinfo {author} {\bibfnamefont {B.}~\bibnamefont
  {Colson}}, \bibinfo {author} {\bibfnamefont {P.}~\bibnamefont {Marcotte}},\
  and\ \bibinfo {author} {\bibfnamefont {G.}~\bibnamefont {Savard}},\
  }\bibfield  {title} {\bibinfo {title} {An overview of bilevel optimization},\
  }\href {https://doi.org/10.1007/s10479-007-0176-2} {\bibfield  {journal}
  {\bibinfo  {journal} {Ann Oper Res}\ }\textbf {\bibinfo {volume} {153}},\
  \bibinfo {pages} {235} (\bibinfo {year} {2007})}\BibitemShut {NoStop}%
\bibitem [{\citenamefont {Avis}(2025{\natexlab{c}})}]{code_for_autodiff_paper}%
  \BibitemOpen
  \bibfield  {author} {\bibinfo {author} {\bibfnamefont {G.}~\bibnamefont
  {Avis}},\ }\href
  {https://github.com/GuusAvis/code_for_Optimization_of_Quantum-Repeater_Networks_using_Stochastic_Automatic_Differentiation}
  {\bibinfo {title}
  {Guusavis/code\allowbreak\_for\allowbreak\_optimization\allowbreak\_of\allowbreak\_quantum-repeater\allowbreak\_networks\allowbreak\_using\allowbreak\_stochastic\allowbreak\_automatic\allowbreak\_differentiation}}
  (\bibinfo {year} {2025}{\natexlab{c}})\BibitemShut {NoStop}%
\bibitem [{\citenamefont {Avis}(2025{\natexlab{d}})}]{avis2025a}%
  \BibitemOpen
  \bibfield  {author} {\bibinfo {author} {\bibfnamefont {G.}~\bibnamefont
  {Avis}},\ }\href {https://doi.org/10.5281/zenodo.14595893} {\bibinfo {title}
  {{{GuusAvis}}/code\_for\_{{Optimization}}\_of\_{{Quantum-Repeater}}\_{{Networks}}\_using\_{{Stochastic}}\_{{Automatic}}\_{{Differentiation}}:
  V1}},\ \bibinfo {howpublished} {Zenodo} (\bibinfo {year}
  {2025}{\natexlab{d}})\BibitemShut {NoStop}%
\bibitem [{\citenamefont {Avis}(2025{\natexlab{e}})}]{avis2025b}%
  \BibitemOpen
  \bibfield  {author} {\bibinfo {author} {\bibfnamefont {G.}~\bibnamefont
  {Avis}},\ }\href {https://zenodo.org/records/14721267} {\bibinfo {title}
  {Data for {{Optimization}} of {{Quantum-Repeater Networks}} using
  {{Stochastic Automatic Differentiation}}}} (\bibinfo {year}
  {2025}{\natexlab{e}})\BibitemShut {NoStop}%
\bibitem [{\citenamefont {Sobrinho}(2005)}]{sobrinho2005}%
  \BibitemOpen
  \bibfield  {author} {\bibinfo {author} {\bibfnamefont {J.}~\bibnamefont
  {Sobrinho}},\ }\bibfield  {title} {\bibinfo {title} {An algebraic theory of
  dynamic network routing},\ }\href {https://doi.org/10.1109/TNET.2005.857111}
  {\bibfield  {journal} {\bibinfo  {journal} {IEEE/ACM Transactions on
  Networking}\ }\textbf {\bibinfo {volume} {13}},\ \bibinfo {pages} {1160}
  (\bibinfo {year} {2005})}\BibitemShut {NoStop}%
\bibitem [{\citenamefont {Caleffi}(2017)}]{caleffi2017}%
  \BibitemOpen
  \bibfield  {author} {\bibinfo {author} {\bibfnamefont {M.}~\bibnamefont
  {Caleffi}},\ }\bibfield  {title} {\bibinfo {title} {Optimal {{Routing}} for
  {{Quantum Networks}}},\ }\href {https://doi.org/10.1109/ACCESS.2017.2763325}
  {\bibfield  {journal} {\bibinfo  {journal} {IEEE Access}\ }\textbf {\bibinfo
  {volume} {5}},\ \bibinfo {pages} {22299} (\bibinfo {year}
  {2017})}\BibitemShut {NoStop}%
\end{thebibliography}%

\appendix

\section{Benchmarking} \label{app:benchmarking}

\begin{figure}
    \centering 
    \includegraphics[width=\columnwidth]{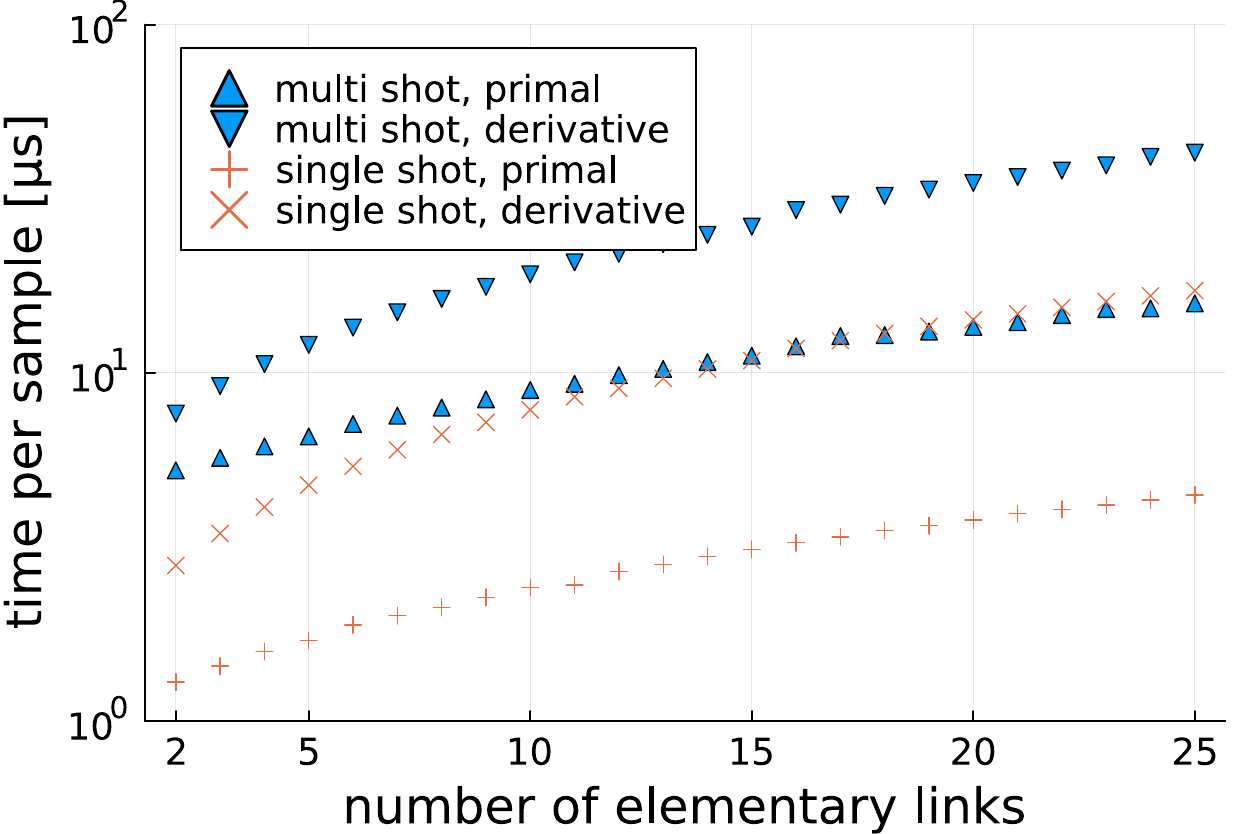}
    \caption{
    Duration of Monte Carlo simulation as a function of the number of elementary links in a repeater chain.
    Each elementary link is 65 km long.
    Includes both the time required to estimate the secret-key rate (``primal'') and its derivative (``derivative''), for both the single-shot and multi-shot protocols.
    For each of 1000 different values of the bright-state parameter (chosen uniformly at random between 0.001 and 0.101), a repeater chain using that bright-state parameter uniformly for all elementary links is evaluated using 1000 samples.
    The derivative is taken with respect to the bright-state parameter.
    The total duration is timed and divided by $1000^2$ to determine the time per sample.
    Error bars are smaller than the marker size and have been suppressed.
    }
    \label{fig:benchmarking}
\end{figure}

In this appendix, we discuss the performance of applying stochastic AD to our quantum-repeater-chain simulations.
In Figure~\ref{fig:benchmarking} the relative performance of evaluating the secret-key rate and its derivative are shown, both for the multi-shot and single-shot cases.
We have used the model described in Section~\ref{sec:model} in the case of the single-click protocol described in Section~\ref{sec:optimizing_bright_state_params}.
The performance gap between the primal and derivative evaluation can be seen to grow with the size of the repeater chain.
For the single-shot case, the gap starts at a factor of ~2.7 for two links and builds up to ~3.9 for 25 links.
In the multi-shot case, the performance gap varies from a factor of ~1.6 to ~2.7.

\section{Repeater chain: single shot and multi shot} \label{app:shots}

The optimization of the bright-state parameters, as reported in Section~\ref{sec:optimizing_bright_state_params}, was executed not only for the multi-shot protocol but also for the single-shot protocol (both described in Section~\ref{sec:model}).
The single-shot optimization was used to compare the coherence-time sensitivities between the two protocols in Figure~\ref{fig:repeater_chain} (c).
\begin{figure}[!tbp]
    \centering 
    \includegraphics[width=\columnwidth]{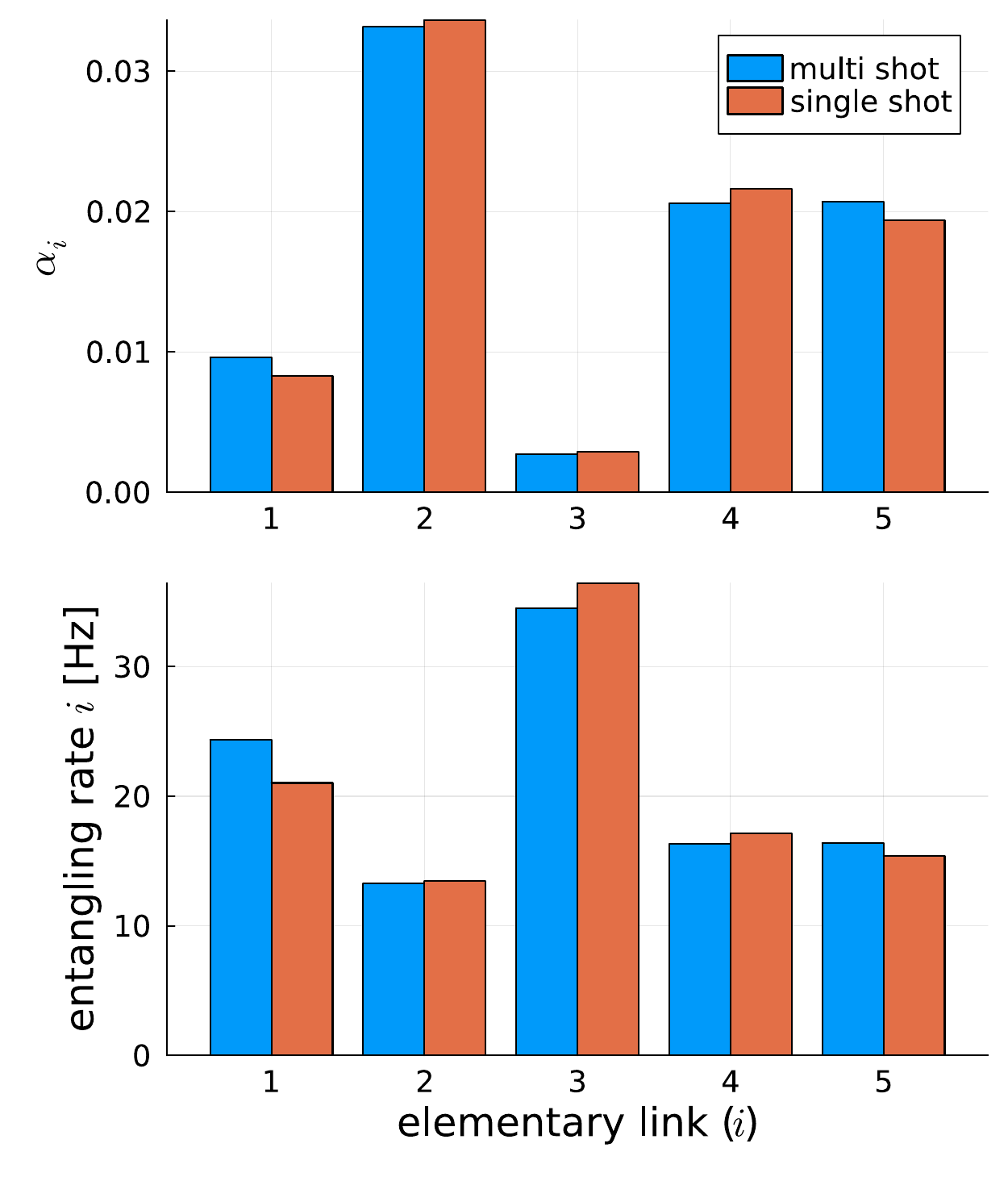}
    \caption{
    Optimized bright-state parameters ($\alpha$) and corresponding entangling rates for the repeater chain shown in Figure~\ref{fig:repeater_chain} (d), for both the single-shot and multi-shot protocols.
    }
\label{fig:single_multi_shot_compare}
\end{figure} 
Here, we further compare the results of these two optimizations.
First of all, we note that the optimized secret-key rate for the multi-shot protocol is $2.002 \pm 0.001$ Hz, while it is $1.966 \pm 0.001$ for the single-shot case.
Hence, the multi-shot protocol outperforms the single-shot protocol, but only by a very small margin.
Furthermore, we compare the optimal bright-state parameters for the two different protocols in Figure~\ref{fig:single_multi_shot_compare}.
In this figure, we also compare the corresponding entangling rates, defined as the success probability divided by the attempt duration, $p_i / \Delta t_i$.
It can be seen that the optimized bright-state parameters do not differ much between the two protocols, suggesting that choosing bright-state parameters by studying the simpler single-shot protocol could sometimes be a good approximation to optimize the multi-shot protocol.
It is remarkable that despite the similarity between the two protocols in secret-key rate and optimal bright-state parameters, the sensitivity of the secret-key rate with respect to the different coherence times is quite different between the two protocols, as shown in Figure~\ref{fig:repeater_chain} (c).
\\

\begin{figure}
    \centering 
    \includegraphics[width=\columnwidth]{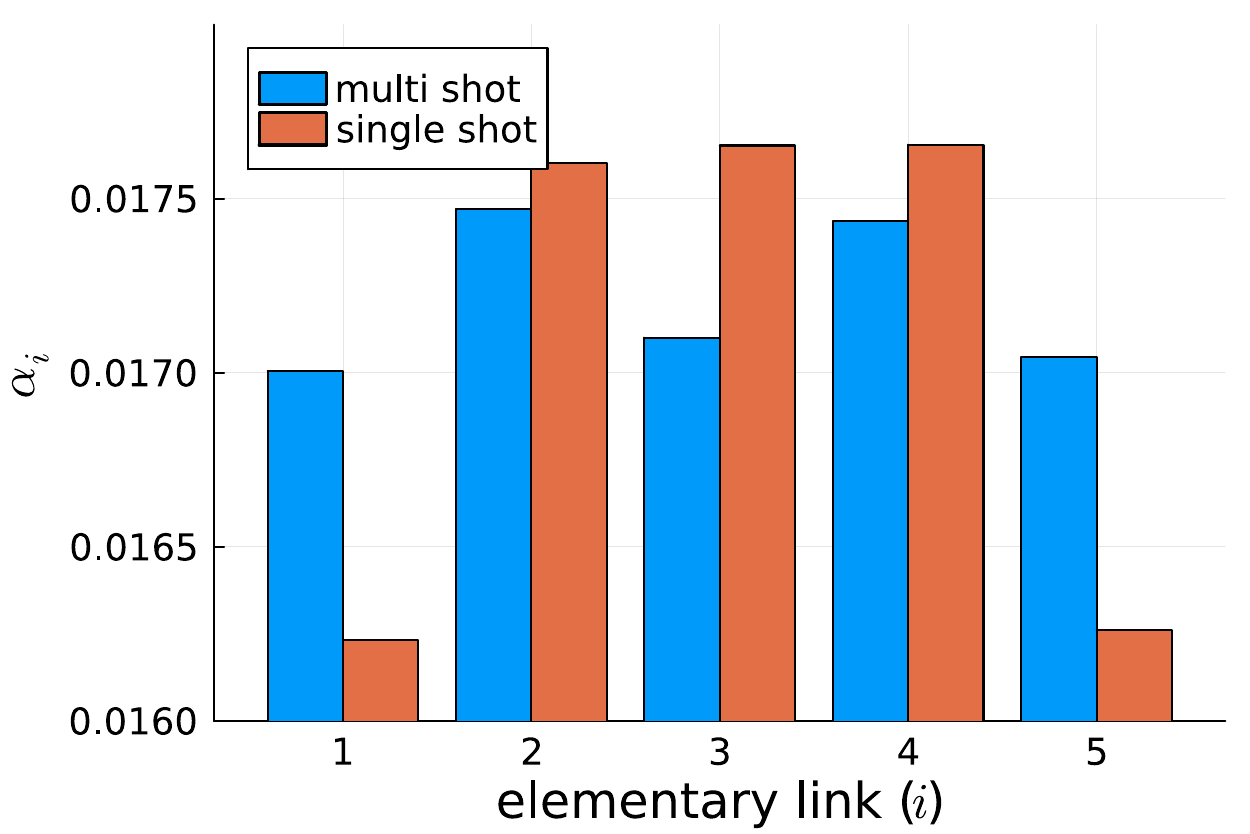}
    \caption{
    Optimized bright-state parameters ($\alpha$) for a repeater chain of five links that are each 65 km long.
    }
\label{fig:single_multi_shot_compare_symmetric}
\end{figure}

We note that in the chain that we study, links 4 and 5 have the same length (80 km).
A qualitative difference between the single-shot and multi-shot protocols is that for the multi-shot protocol elementary links 4 and 5 are assigned almost the same bright-state parameter, while in the single-shot protocol link 4 is assigned a larger bright-state parameter than elementary link 5.
To determine whether this points to a more general difference between the two protocols, we have also optimized the bright-state parameters for a homogeneous repeater chain of five elementary links that are each 65 km long.
The results can be found in Figure~\ref{fig:single_multi_shot_compare_symmetric}.
It appears that both protocols divide the elementary links into two groups with all in the same group being assigned approximately the same bright-state parameter.
Interestingly though, how they are divided differs between the two protocols.
For the single-shot protocol, the extremal elementary links (1 and 5) are assigned a smaller value than the links in the interior.
This can perhaps be understood from the fact that the extremal links are less important to hasten entanglement swapping (and hence reduce qubit storage times), as they are connected to only a single repeater as opposed to two.
For the multi-shot protocol, on the other hand, the elementary links are divided in an alternating fashion, with elementary links 1, 3, and 5 assigned a smaller value than elementary links 2 and 4.
Understanding why this is the case, or how these patterns depend on the number of repeaters, is beyond the scope of this paper.
We note that the difference between the bright-state parameters is smaller for the multi-shot protocol than for the single-shot protocol.
However, in both the single-shot protocol and the multi-shot protocol, using the same bright-state parameter everywhere appears to be a good approximation of the optimal bright-state parameters.
For the single-shot protocol, we found a secret-key rate of $3.039 \pm 0.002$ Hz when optimizing bright-state parameters separately and $3.047 \pm 0.002$ Hz when optimizing them all in unison.
For the multi-shot protocol, we found a secret-key rate of $3.147 \pm 0.002$ Hz vs $3.143 \pm 0.002$ Hz.

\section{Pathfinding} \label{app:pathfinding}

The runtime of optimizations of the repeater placement, as discussed in Section~\ref{sec:repeater_placement}, is typically dominated by our algorithm for finding the optimal path between each pair of end nodes as required to evaluate the utility function $\text{SKR}_\text{min}$ defined in Eq. \eqref{eq:skr_min}.
Traditional pathfinding algorithms like Dijkstra's cannot be applied to this problem, as the per-path utility function SKR as defined in Eq.~\eqref{eq:skr} is not isotonic~\cite{sobrinho2005, caleffi2017}.
This means that when there are paths $p_1$ and $p_2$, and paths $p_1'$ and $p_2'$ that are obtained by adding some edge $e$ to paths $p_1$ and $p_2$ respectively, it is possible that $\text{SKR}(p_1) > \text{SKR}(p_2)$ but $\text{SKR}(p_1') < \text{SKR}(p_2')$.
A naive solution to finding the best path between nodes $s$ and $t$ is to simply calculate SKR for all paths between those nodes.
However, this is inefficient as the number of paths scales superexponentially with the number of repeaters. 
Instead, we use an algorithm that gradually builds a list of paths by adding edges step-by-step to paths originating from $s$ until they reach $t$.
At each step, the utility function is evaluated for the partial path.
If the result is worse than the best solution for a path between $s$ and $t$ found so far, the partial path is removed from the list and not further expanded (this can be safely done because the SKR is monotonic, i.e., adding an extra edge to a path can never increase the SKR).
Moreover, when the optimization algorithm perturbs the node locations and restarts the pathfinding algorithm, it immediately adds the best path found at the previous iteration to the list of solutions, bypassing the gradual construction.
While this path may no longer be the best path as the edge lengths are now different, it will often still be a good solution, allowing some of the partial paths to be pruned early on.
A second technique we have applied to some of the optimizations is to only allow for elementary links of length $L \leq D$.
In this case, direct connections between end nodes are not allowed, and the graph that the pathfinding has to search through is sparser than the complete graph that corresponds to the unconstrained case.
We have found that combining these two strategies allows us to find solutions in a reasonable amount of time even when placing 40 repeaters.

\section{Simulated annealing} \label{app:annealing}

In order to optimize repeater placement as discussed in Section~\ref{sec:repeater_placement}, we have used simulated annealing.
There is a temperature parameter that starts at a high value and is decreased every epoch.
Each of the optimizations reported here consists of 25 epochs, with 15 iterations per epoch.
For each iteration in epoch $i = 1, 2, ..., 25$, the used temperature is $t = t_0 * 0.8^i$.
The value of $t_0$ at $D = 300$ km is either 5, 10, or 15, and for other values of $D$, these values are changed in a way that is directly proportional to $D$.
As reported in Section~\ref{sec:repeater_placement}, the temperature is used as the learning rate of the gradient-descent algorithm; this is the reason why we set $t_0$ proportional to $D$, as the step size with which the repeaters are displaced should scale with the distances between the end nodes, as otherwise the algorithm will place repeater nodes far outside of the convex hull of the end nodes for small values of $D$ while they will not be varied significantly enough for large values of $D$.
Apart from being used as the learning rate, the temperature affects the optimization in the following ways:
\begin{itemize}
\item
    The number of samples taken to estimate the secret-key rate of each path is set to $\left \lceil 30 / (1 + t) \right \rceil$.
    For high temperatures the number of samples is then smaller, resulting in cheaper evaluations and, more importantly,  more randomness and hence more exploration.

\item
    The per-path utility function that the network utility function ($\text{SKR}_\text{min}$) is actually based on is not the secret-key rate (SKR), but $\text{SKR} + 0.1 \cdot t / \mathbb E[T_\text{ent}]$, where $T_\text{ent}$ is the distribution time as in Eq.~\eqref{eq:skr}.
    Hence for low temperatures, the utility function is approximately equal to the secret-key rate.
    For higher temperatures this gives an incentive to increase entangling rates even when the error rates are too high to generate any key at all, often allowing the optimization to escape an otherwise barren plateau.
\end{itemize}

\section{More repeater-placement results} \label{app:more_solutions}

\begin{figure}[!tbp]
    \centering 
    \includegraphics[width=\columnwidth]{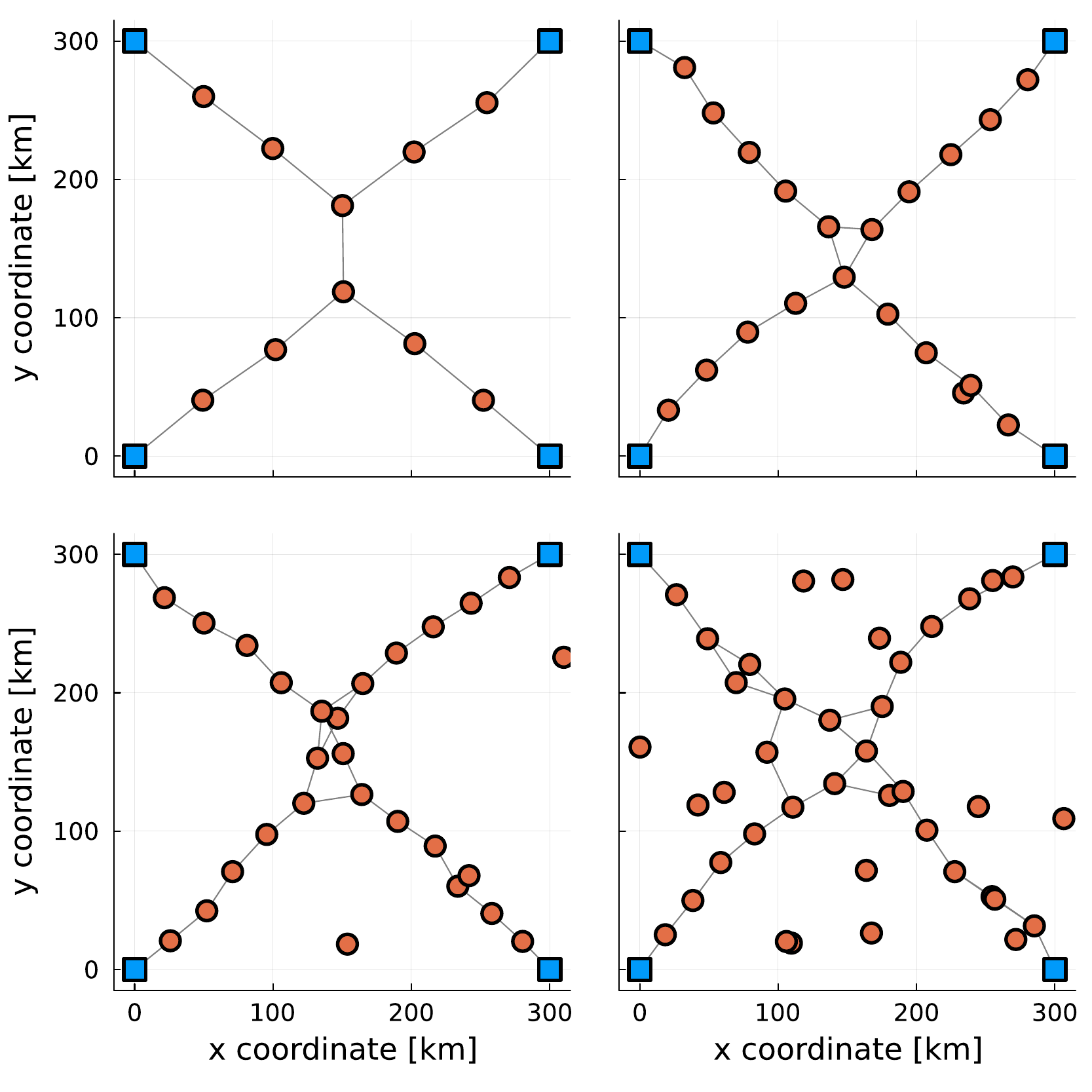}
    \caption{
    Examples of solutions for the optimization of repeater locations.
    End nodes (blue squares) form a square with edges of length $D = 300$ km.
    The total numbers of repeaters (orange circles) are $N = 10, 20, 30, 40$.
    For $N = 30$, three repeaters have been placed so far outside of the square that they are not visible in the figure.
    For $N = 40$, this is the case for one repeater.
    }
\label{fig:placement_solution_examples_appendix}
\end{figure}

\begin{figure}[!tbp]
    \centering 
    \includegraphics[width=\columnwidth]{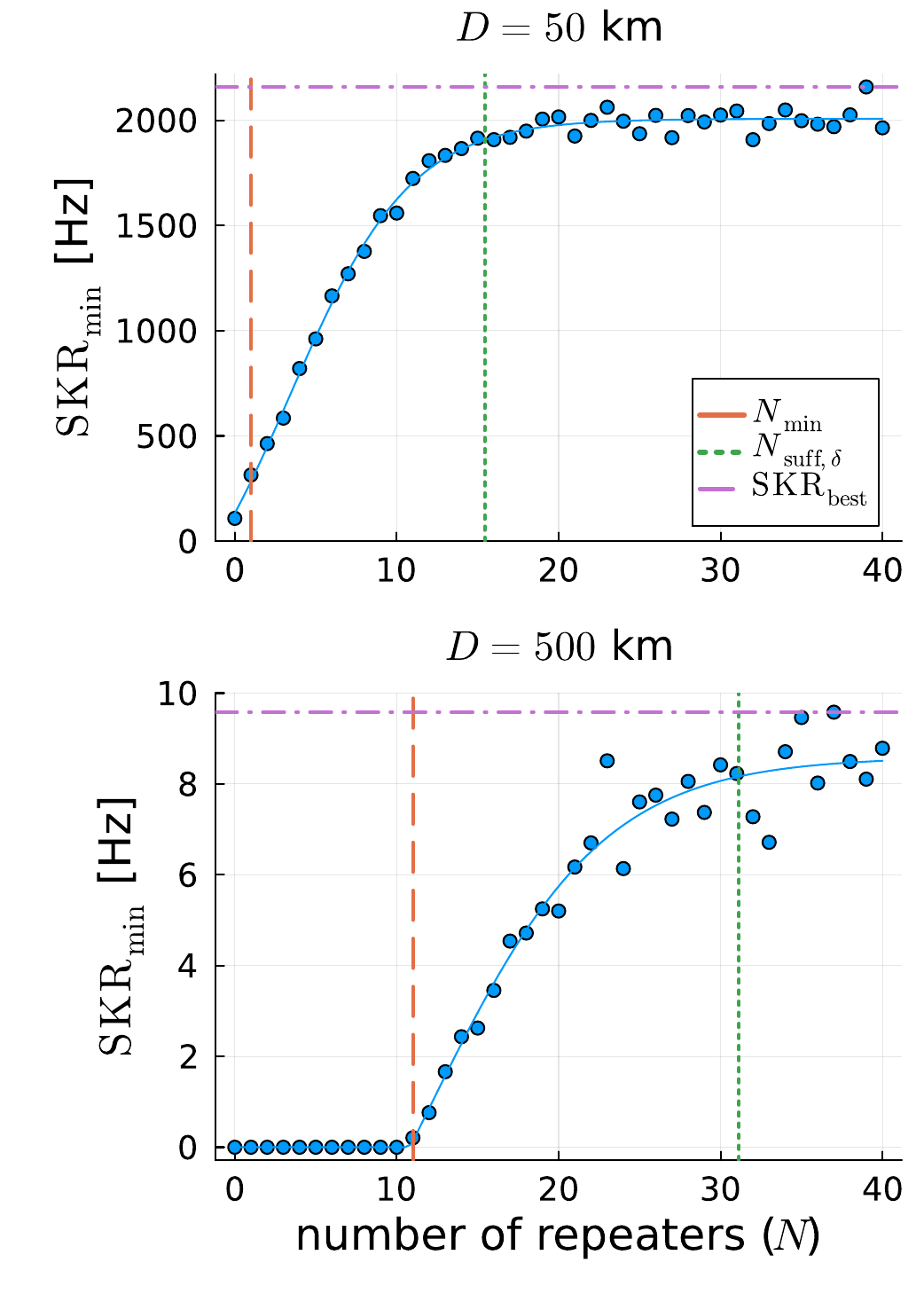}
    \caption{
    Dependence of the guaranteed minimal quality of service, $\text{SKR}_\text{min}$ as defined in Eq. \eqref{eq:skr_min}, on $N$ for $D = 50$ km and $D = 500$ km.
    Shown are the optimization results, a fit function as described by Eq. \eqref{eq:fit_function}, the minimal number of repeaters ($N_\text{min}$, defined in Eq. \eqref{eq:N_min}), the sufficient number of repeaters ($N_{\text{suff, }\delta}$, defined in Eq. \eqref{eq:N_suff}, with $\delta = 0.05$), and the best secret-key rate ($\text{SKR}_\text{best}$, defined in Eq. \eqref{eq:best_skr}).
    The error bars in the data are smaller than the marker size and have been suppressed.
    }
\label{fig:placement_features_appendix}
\end{figure} 

In this appendix, we show some additional solutions for the repeater-placement optimization discussed in Section~\ref{sec:repeater_placement}.
In Figure~\ref{fig:placement_solution_examples}, we have shown the best repeater placements we have found for $N = 4, 5, 6, 7$ and $D = 300$ km.
Here, in Figure~\ref{fig:placement_solution_examples_appendix}, we give some examples of how solutions look for larger values of $N$.
We observe that as the number of repeaters grows, the solutions become more irregular and less symmetric.
This is likely the case because the optimization problem is more difficult than for a smaller number of repeaters and hence the optimization has not had the chance to converge as well within the 25 epochs of 15 iterations each.
Better results would require more computational resources or more efficient optimizations (as discussed in Appendix~\ref{app:pathfinding}, pathfinding is currently the main bottleneck in the optimizations, especially when the number of repeaters is large).
Moreover, we observe that as $N$ increases, an increasing number of repeaters is not part of any of the optimal paths between the end nodes.
Therefore, the network utility ($\text{SKR}_\text{min}$, as defined in Eq. \eqref{eq:skr_min}) would not decrease in case those repeaters were removed.
This is in line with our observation that the utility stops growing with the number of repeaters for large values of $N$: at some point, adding more repeaters does not allow one to increase the network utility, as introducing them to the paths would reduce the end-to-end fidelity for only a small gain in entangling rate.
\\

In Figure~\ref{fig:placement_features_example} we have shown the dependence of the optimized network utility function $\text{SKR}_\text{min}$ on the number of repeaters $N$ for $D = 300$ km.
Here, in Figure~\ref{fig:placement_features_appendix}, we show what this dependence looks like for a few different values of $D$.
For $D = 500$ km, while it is clearly visible that the rate at which $\text{SKR}_\text{min}$ increases with $N$ slows down for larger values of $N$, it still appears to be growing at the largest value, $N = 40$.
For this reason, the estimates of $N_{\text{suff}, \delta}$ and $\text{SKR}_\text{best}$ are perhaps a bit less reliable than those for smaller values of $D$, as a significant part of the function may still be hidden beyond $N = 40$.
Obtaining significant data for $D > 500$ km would probably require solving the optimization for larger values of $N$, which may become computationally expensive.

\end{document}